\documentclass[12pt,a4paper]{article}

\usepackage[T1]{fontenc}
\usepackage[utf8]{inputenc}
\usepackage{microtype}
\usepackage{sourcesanspro}

\usepackage[a4paper,margin=1in]{geometry}
\usepackage{setspace}
\usepackage{amsmath,amsthm}

\usepackage{amssymb}
\usepackage{mathtools}
\usepackage{bbm}
\mathtoolsset{showonlyrefs}

\usepackage{booktabs}
\usepackage{tabularx}
\usepackage{subcaption}
\usepackage{graphicx}
\usepackage{tikz}
\usepackage{multirow}
\usepackage{xcolor}
\definecolor{mdlblue}{HTML}{173F5F}
\definecolor{modelrust}{HTML}{8F3B2D}
\definecolor{neutralgray}{HTML}{666666}
\usepackage{gensymb}
\usepackage{pgfplots}
\pgfplotsset{compat=1.18}

\usepackage[english]{babel}
\usepackage[round,authoryear]{natbib}

\usepackage[colorlinks=true,linkcolor=black,citecolor=blue,urlcolor=blue]{hyperref}

\usepackage{titlesec}

\titleformat{\section}
{\normalfont\large\bfseries}
{\thesection}{1em}{}

\titleformat{\subsection}
{\normalfont\normalsize\bfseries}
{\thesubsection}{1em}{}

\titleformat{\subsubsection}
{\normalfont\normalsize\bfseries}
{\thesubsubsection}{1em}{}

\usepackage{fancyhdr}
\usepackage{epigraph}
\newtheorem{remark}{Remark}

\usepackage{authblk}

\title{Theory as Data Compression}

\author[1]{Carlos Cueva\thanks{Email: \href{mailto:carlos.cueva@ua.es}{\texttt{carlos.cueva@ua.es}}. I am particularly grateful to Mat\'u\v{s} Teji\v{s}\v{c}\'ak and Pedro Albarr\'an for extensive discussions throughout the development of this project. I also thank Georgios Gerasimou and Miguel Costa-Gomes for valuable early discussions. Finally, I thank the authors of the three empirical studies for making their datasets available (Miettinen, Kosfeld, Fehr, and Weibull, 2020; Bruhin, Fehr-Duda, and Epper, 2010; Abdellaoui, Attema, and Bleichrodt, 2010).}}

\affil[1]{Departamento de Fundamentos del An\'alisis Econ\'omico (FAE)\\ Universidad de Alicante}

\date{August 2026}

\begin{document}

\maketitle

\thispagestyle{empty}

\vspace{2em}

\begin{abstract}

Comparing economic models requires balancing fit against flexibility. Yet standard selection criteria often proxy flexibility using parameter counts, overlooking differences due to functional form and experimental design. This paper introduces a data compression approach for evaluating economic models. Following the Minimum Description Length principle, models are interpreted as codes and evaluated by how effectively they compress data. This perspective yields compression-based analogues of Selten's \textit{predictive success} and Fudenberg et al.'s \textit{completeness} and \textit{restrictiveness}. Unlike Selten’s measure, its analogue unifies deterministic and stochastic models in a likelihood framework; unlike Fudenberg et al.’s measures, the proposed framework provides a complete selection criterion. Applications to social preferences, risky choice, and intertemporal choice illustrate the relevance of the approach. In simulations, accounting for flexibility differences due to functional form and experimental design improves model recovery relative to AIC, BIC, and cross-validation. In empirical reanalyses, MDL changes model rankings in favor of more parsimonious specifications.

\end{abstract}


\pagebreak

\section{Introduction}

In his defense of formal statistical inference in macroeconomics, \citet{Sims1996} describes advances in science as discoveries of ways to compress data with minimal loss of information. Kepler's laws compressed Tycho Brahe's observations of planetary motion, and Newton's inverse-square law compressed them further while extending the same structure to new domains. The same aspiration is present in economics. A useful model is one that describes economic phenomena with little information loss while remaining as simple as possible.

Traditionally, Occam's razor expresses this ideal of parsimony by favoring the simplest model among those with equal explanatory power. Modern statistics and machine learning offer a predictive perspective, framing model selection as a tradeoff between goodness of fit and model complexity, understood as flexibility in accommodating different patterns in the data. Penalizing complexity helps prevent overfitting, which occurs when a flexible model captures idiosyncratic variation in the observed sample rather than patterns that generalize beyond it.

Widely used methods such as AIC \citep{Akaike1974} and BIC \citep{Schwarz1978} implement this tradeoff by penalizing models according to the number of free parameters they contain. The problem with this approach is that parameter counts can be poor proxies for flexibility: two models with the same number of parameters but different functional forms may accommodate very different ranges of data patterns. Economists have long recognized this limitation, yet routinely rely on these methods for model selection.\footnote{For instance, in their seminal comparison of generalized expected utility theories, \citet{Harless1994} note that ``the number of patterns a theory allows does not necessarily correspond to the number of free parameters'' (p.~1259), yet their preferred model selection criterion ultimately penalizes models based on parameter counts. In a more recent example, \citet{Apesteguia2021} note that adapting their goodness-of-fit measure for model selection would require penalizing models according to their size, but since standard volume measures are uninformative in their setting, they too fall back on parameter counts.} A popular alternative is to use resampling methods such as cross-validation (CV), which avoid explicit parameter-counting penalties by evaluating out-of-sample prediction directly. However, because CV provides no explicit measure of complexity, it can also be misleading when differences in flexibility are central to the comparison.

This paper approaches model comparison through the lens of data compression, providing a way to compare models whose flexibility is not well captured by parameter counts. To illustrate the problem, consider two canonical theories of choice under risk, each with two free parameters. The first is Expected Utility (EU), specified with power utility over final wealth, $u(x)=(\omega+x)^\alpha$. The second is a parametric version of Yaari's \citeyearpar{Yaari1987} Dual Theory (DT), specified with linear utility and \cite{Prelec1998} probability weighting, $w(p)=\exp[-\delta(-\ln p)^\gamma]$. Despite their identical parameter counts, the models may differ in flexibility, making them a useful test case for comparing selection criteria. The comparison uses repeated simulations of noisy certainty equivalents for 50 binary lotteries under each specification. Table \ref{tab:hyp_risk1} reports how often each criterion selects the data-generating model over the alternative. Because AIC and BIC penalize only parameter count, they assign EU and DT identical complexity penalties and reduce model selection to a comparison of fit. When the data are generated by EU, both criteria perform no better than chance, recovering the true model in only 48\% of simulations. Cross-validation also struggles, successfully recovering EU in just 68\% of cases. By contrast, when the data are generated by DT, all criteria perform extremely well, recovering the true model in 99\% of simulations.

The reason why AIC and BIC fail to recover EU is that equal parameter counts conceal large differences in complexity. In EU, $\alpha$ and $\omega$ jointly determine the degree of concavity of the utility function, interacting in ways that restrict the range of observable choice patterns the model can generate. In DT, by contrast, $\gamma$ and $\delta$ govern distinct geometric features of the probability-weighting function---curvature and elevation---allowing the model to accommodate a much wider range of patterns. As a result, DT can often fit noisy EU-generated data better than EU itself, leading AIC and BIC to incorrectly select DT when EU is in fact the true model. The reverse error is much less likely: because EU is too restrictive to reproduce the broader range of patterns generated by DT, AIC and BIC almost always select DT when it is in fact the true model.\footnote{Cross-validation exhibits a similar asymmetry in model recovery, but the reason is more subtle. When EU generates the data, its weakly identified parameters produce unstable out-of-sample predictions, allowing DT to perform better despite being misspecified; see Section~\ref{sec:risk}.}

\begin{table}[t]
	\centering
	\caption{Model recovery rates: Expected Utility vs. Dual Theory}
	\begin{tabular*}{0.99\textwidth}{l @{\extracolsep{\fill}} c c c c c}
		\toprule
		\textbf{True Model} & \textbf{AIC} & \textbf{BIC} & \textbf{CV} & \textbf{MDL} & \textbf{COMP} \\
		\midrule
		EU  & 48\% & 48\% & 68\% & 99\% & 1.88 \\
		DT  & \makebox[0pt][r]{$>$}99\% & \makebox[0pt][r]{$>$}99\% & \makebox[0pt][r]{$>$}99\% & 99\% & 5.12 \\
		\bottomrule
	\end{tabular*}
	\par\medskip
	\parbox{0.99\textwidth}{\footnotesize \textit{Notes:} Selection frequencies are based on 10{,}000 simulated datasets per model. Each dataset contains certainty equivalents for 50 binary lotteries, formed by combining $p\in\{.1,.3,.5,.7,.9\}$ with all payoff pairs $x>z$ from $\{0,.25,.5,.75,1\}$. Parameters are drawn uniformly with $\alpha,\gamma\in[0.2,0.8]$ and $\omega,\delta\in[0.5,1.5]$. Noise is drawn from $\mathcal N(0,[.1(x-z)]^2)$. The criteria are $\mathrm{AIC}=-2\log\mathcal L(\hat\theta)+2k$, $\mathrm{BIC}=-2\log\mathcal L(\hat\theta)+k\log n$, and $\mathrm{MDL}=-\log\mathcal L(\hat\theta)+\mathrm{COMP}$; CV is five-fold. The Online Appendix provides simulation details, robustness checks, and COMP calculations.}
	\label{tab:hyp_risk1}
\end{table}

The last two columns of Table \ref{tab:hyp_risk1} show how the same comparison is evaluated using the Minimum Description Length (MDL) principle \citep{Rissanen1978,Rissanen1996,Grunwald2007}. The MDL column reports how often MDL selects the true model, while COMP reports the complexity penalty it assigns to each model. In MDL, a model is viewed less as a literal data-generating process than as a language or \textit{code} for describing data. The better the model captures structure in the data, the shorter the description of the observed dataset, measured by the number of bits needed to encode it. MDL treats model selection as a problem of data compression, selecting the model that yields the shortest description length of the data. This length combines a fit term, which is shorter when the model fits the observed data well, with the complexity term COMP, which is larger when the model can fit a wider range of possible datasets. Unlike AIC and BIC, MDL imposes a much larger complexity penalty for DT than for EU ($5.12$ versus $1.88$), reflecting DT's greater flexibility. By taking this difference into account, MDL recovers the true model in 99\% of simulations under both data-generating processes.

Building on the MDL principle, this paper’s first contribution is to construct compression-based analogues of measures from two influential approaches to model evaluation in economics. The first is Selten’s measure of predictive success \citep{Selten1991}. The second comprises the completeness and restrictiveness measures of \citet{Fudenberg2022,Fudenberg2026}. Both approaches move beyond naïve parameter counting to capture model flexibility, but each has limitations that the compression-based analogues address.

Starting with \citet{Selten1991}, predictive success measures the performance of a theory as the fraction of observations it correctly predicts (its hit rate $r$) minus the relative size of the outcome set it allows (its area $a$). Because this approach treats all misses alike rather than assigning probabilities to alternative outcomes, it discards information about the severity of violations and is difficult to reconcile with the stochastic models and likelihood-based estimation methods that dominate empirical work. Recasting Selten's ``area theories'' as codes yields an analogue of predictive success interpretable as compression gain relative to a zero-compression benchmark. The resulting measure preserves Selten's basic tradeoff, increasing in the hit rate $r$ and decreasing in the area $a$, but rather than imposing the linear difference $r-a$, it derives the functional form from coding-theoretic principles. This coding formulation places area theories and stochastic extensions, such as logit-augmented choice models, within a common likelihood-based framework. Both can then be estimated using standard maximum-likelihood methods and compared directly.

Turning to \citet{Fudenberg2022,Fudenberg2026}, completeness measures fit as the share of the possible reduction in expected prediction error achieved by a model, whereas restrictiveness measures flexibility as the model’s expected discrepancy from a predefined eligible set of synthetic datasets. Together, these measures identify a Pareto frontier, ruling out models that are both less complete and less restrictive than some alternative, but leaving unresolved the common case in which the more complete model is also less restrictive. Replacing prediction error by description length yields a compression-based analogue of completeness. This new measure is interpretable as the fraction of available compression achieved by the model, where the maximum compression attainable is determined by the entropy of the source. The analogue of restrictiveness replaces discrepancies by likelihoods, yielding a measure whose logarithm is equivalent to MDL’s complexity penalty, up to a constant (see Remark \ref{rem:restricted_complexity}). Crucially, the two analogues permit a direct MDL-based tradeoff between fit and flexibility, yielding a complete model selection criterion rather than the partial ordering implied by Fudenberg et al.’s Pareto frontier.

The paper next applies MDL’s data-compression perspective to three canonical domains in behavioral economics: social preferences, risk preferences, and time preferences. Behavioral models often give rise to non-nested and nonlinear specifications for which parameter counts are especially crude proxies for complexity, particularly in the small samples typical of laboratory experiments. Together, these features make behavioral economics a particularly promising area for MDL.

\paragraph{Application 1: Social Preferences.}

The first application studies social-preference models from the sequential Prisoner’s Dilemma experiment in \citet{Miettinen2020}, a discrete-choice setting that permits a direct comparison of Selten’s predictive success with its compression analogue. Across these models, Selten’s area and MDL complexity both reveal that parameter counts are poor proxies for model flexibility. Unlike Selten’s original framework, the coding formulation permits area theories to be evaluated alongside random-utility specifications of the same underlying models. This comparison shows that the stochastic specifications are slightly more complex than their area-theory counterparts, but that their improvement in fit more than offsets the additional complexity. The compression interpretation provides a further insight: although the best area theory achieves positive predictive success, it yields negative compression gain. Only the random-utility specifications achieve positive compression.

\paragraph{Application 2: Risk Preferences.}

This application revisits the initial risky-choice example to show how model complexity depends on the interaction between functional form and experimental design. Utility-curvature and probability-weighting models are compared across designs that vary payoffs and probabilities. When payoff variation is narrow, changes in utility curvature have little effect on predicted certainty equivalents, lowering the complexity of utility-curvature models. Conversely, when probability variation is narrow, changes in probability weighting have little effect on predicted certainty equivalents, lowering the complexity of probability-weighting models. Standard selection methods do not explicitly adjust for these design-induced differences in complexity. In the narrow designs, this leads AIC, BIC, and CV to favor the more complex specification, producing asymmetric recovery rates. MDL accounts for these differences and substantially reduces the asymmetries, achieving consistently high recovery across designs. This also suggests a role for MDL as an ex ante diagnostic tool, revealing when an experimental design is likely to bias standard selection methods toward particular specifications. We then reanalyze the certainty-equivalent data of \citet{Bruhin2010}. Compared with these standard methods, MDL favors parsimonious probability-weighting models over richer prospect-theory variants.

\paragraph{Application 3: Time Preferences.}

The final application compares exponential time discounting with two popular generalizations, hyperbolic discounting and constant sensitivity. Although both generalizations have two parameters, constant sensitivity is more flexible: it captures the decreasing-impatience patterns of hyperbolic discounting while also accommodating increasing impatience. In simulations, this allows constant sensitivity to mimic hyperbolic data closely, leading AIC, BIC, and CV to select constant sensitivity too often. MDL instead penalizes the additional flexibility of constant sensitivity, selecting hyperbolic discounting for decreasing-impatience data and constant sensitivity only for increasing-impatience data. These differences matter empirically. In a reanalysis of the experiment of \citet{Abdellaoui2010}, aggregate choices display decreasing impatience, yet standard criteria select constant sensitivity on the basis of its marginal fit advantage. MDL instead selects hyperbolic discounting because its lower complexity more than offsets that advantage.

The rest of the paper is organized as follows. Section \ref{sec:MDL} introduces the MDL principle and its foundations in lossless data compression. Section \ref{sec:mdl_econ} constructs compression-based analogues of Selten’s measure of predictive success and Fudenberg et al.’s completeness and restrictiveness measures. Sections \ref{sec:app_social}--\ref{sec:time} present the three applications. Section \ref{sec:conclusion} concludes.

\section{Models as Codes and the MDL Principle}
\label{sec:MDL}

Let $\mathcal X$ denote the sample space for a single outcome. $\mathcal X$ may be finite, countably infinite, or a connected subset of $\mathbb R^l$ for some $l\geq 1$.\footnote{For a comprehensive treatment of MDL and its coding-theoretic foundations, see \citet{Grunwald2007}. Throughout, formulas are written for the discrete case; in continuous settings, probabilities are replaced by densities and sums by integrals.} A sample of size $n$ is denoted $x^n=(x_1,\ldots,x_n)\in\mathcal X^n$, where $\mathcal X^n$ is the $n$-fold Cartesian product of $\mathcal X$. Let $\mathcal X^*=\bigcup_{n\geq 0}\mathcal X^n$ be the set of finite samples of arbitrary length, with $\mathcal X^0=\{x^0\}$ and $x^0$ denoting the empty sample. A probabilistic source is a function\footnote{The range $[0,\infty)$ allows the same notation to cover continuous sample spaces, where $P(x^n)$ denotes a density value and may exceed one. In finite or countable sample spaces, $P(x^n)$ is a probability mass and lies in $[0,1]$.} $P:\mathcal X^*\to[0,\infty)$ satisfying $P(x^0)=1$ and $\sum_{z\in\mathcal X}P(x^n,z)=P(x^n)$ for all $n\geq 0$ and $x^n\in\mathcal X^n$.

Let $\mathcal P$ denote the set of probabilistic sources on $\mathcal X$. A model is a subset $\mathcal M\subseteq\mathcal P$. A parametric model is one that can be written as $\mathcal M=\{P_\theta:\theta\in\Theta\}$, where $\Theta$ is a connected subset of $\mathbb R^k$ for some finite $k$ and $P_\theta(x^n)$ varies continuously with $\theta$ for every finite sample $x^n$. Given an observed sample $x^n$, let
$\hat\theta(x^n)\in\operatorname*{arg\,max}_{\theta\in\Theta}P_\theta(x^n)$ denote the maximum-likelihood estimate within $\mathcal M$. Economic models often specify the conditional distribution of a variable of interest given covariates. In such applications, write the likelihood as $P_\theta(y^n\mid z^n)$ and its maximizer as $\hat\theta(y^n\mid z^n)$, where $y^n$ denotes the observed values of the variable of interest and $z^n$ the observed covariates.

A \emph{code} is a rule that assigns each possible sample $x^n\in\mathcal X^n$ a \emph{codeword}, meaning a finite string of symbols drawn from a finite alphabet. The simplest and most common choice is the binary alphabet $\{0,1\}$. The length of the codeword assigned to $x^n$ is denoted by $L_C(x^n)$. A code is {uniquely decodable} if any concatenated sequence of codewords can be parsed back into the original sequence of samples without ambiguity. Suppose $X^n\sim P$. Shannon's source coding theorem \citep{Shannon1948} establishes that the expected length of any uniquely decodable code is bounded below by the entropy of $P$, denoted $H(P)$:

\[
\mathbb E_P[L_C(X^n)] \geq -\sum_{x^n\in\mathcal X^n}P(x^n)\log P(x^n) \equiv H(P).
\]

Here and throughout, $\log$ denotes the natural logarithm, so code lengths are expressed in nats rather than bits. Abstracting from integer constraints, this lower bound is attained by assigning to each sample $x^n\in\mathcal X^n$ its Shannon information content:

\begin{equation}
L_P(x^n)=-\log P(x^n). \label{eq:ideal_code}
\end{equation}

If the data are instead described using the ideal code induced by an alternative distribution $Q\neq P$, the expected description length exceeds the entropy bound by exactly the Kullback--Leibler divergence:
\[
\mathbb E_P[L_Q(X^n)] =\mathbb E_P[-\log Q(X^n)] =H(P)+D_{\mathrm{KL}}(P\,\|\,Q).
\]

Thus, minimizing expected description length is equivalent to minimizing KL divergence from $P$.

Of course, in practice the ``true'' data-generating probabilistic source is unknown.\footnote{The quotation marks reflect the fact that in this framework we need not presume the existence of a literally true data-generating probability distribution. Probabilities are instead treated as useful abstractions for data compression.} The analyst instead postulates a model $\mathcal M$, consisting of a set of hypothesized probabilistic sources. Each source $P\in\mathcal M$ induces an ideal code length through \eqref{eq:ideal_code}, but the model itself does not define a code until one specifies how the sources in $\mathcal M$ are used. One simple possibility is a two-part code. The first part identifies a source $P\in\mathcal M$, telling the receiver which code to use, and the second encodes the sample using that source. A more refined one-part alternative is the normalized maximum likelihood (NML) code, which turns the entire model into a single distribution, known as the \emph{Shtarkov} or \emph{NML distribution}, by assigning probability proportional to the likelihood of the best-fitting source for each possible sample:
\begin{equation}
P_{\mathrm{NML}}(x^n,\mathcal M) = \frac{P_{\hat\theta(x^n)}(x^n)} {\sum_{u^n\in\mathcal X^n}P_{\hat\theta(u^n)}(u^n)}. \label{eq:shtarkov}
\end{equation}
Provided the normalizing constant is finite, NML is minimax optimal in the sense that it minimizes the largest possible excess code length relative to the best-fitting source in hindsight, a quantity known as the code's regret. Its corresponding description length is
\begin{align}
L_{\mathrm{NML}}(x^n,\mathcal M)
&=
-\log P_{\hat\theta(x^n)}(x^n)
+
\log\sum_{u^n\in\mathcal X^n}
P_{\hat\theta(u^n)}(u^n).
\label{eq:nml}
\end{align}
The last term in \eqref{eq:nml} is independent of the observed sample $x^n$ and gives the NML code its constant regret. This term is known as the \textit{model complexity}, or, for parametric models, \textit{parametric complexity}:
\begin{equation}
\operatorname{COMP}_n(\mathcal M) = \log\sum_{u^n\in\mathcal X^n}P_{\hat\theta(u^n)}(u^n). \label{eq:comp}
\end{equation}

For conditional models, the counterparts of \eqref{eq:nml} and \eqref{eq:comp} are
\begin{equation}
L_{\mathrm{NML}}(y^n,\mathcal M\mid z^n) = -\log P_{\hat\theta(y^n\mid z^n)}(y^n\mid z^n) + \operatorname{COMP}_n(\mathcal M\mid z^n), \label{eq:nml_conditional}
\end{equation}
and
\begin{equation}
\operatorname{COMP}_n(\mathcal M\mid z^n) = \log\sum_{u^n\in\mathcal Y^n} P_{\hat\theta(u^n\mid z^n)}(u^n\mid z^n), \label{eq:comp_conditional}
\end{equation}
where $\hat\theta(u^n\mid z^n)$ maximizes the likelihood of $u^n$ given the covariate sequence $z^n$.

According to the MDL principle, the preferred model is the one that yields the shortest description length for the observed data. Given a set of candidate models $\{\mathcal M_j\}_{j=1}^J$, MDL therefore selects
\begin{equation}
\mathcal M^*(x^n) \in \operatorname*{arg\,min}_{\mathcal M\in\{\mathcal M_1,\ldots,\mathcal M_J\}} L_{\mathrm{NML}}(x^n,\mathcal M). \label{eq:mdl_choice}
\end{equation}

MDL admits several common implementations, including two-part, prequential plug-in, Bayesian, and NML codes \citep[see][]{Grunwald2007}. Among them, NML occupies a central place because of its minimax regret property and because its complexity term depends only on the model as a set of distributions, rather than on an additional choice of prior or parameterization. This paper therefore uses NML throughout.

\subsubsection*{A simple discrete example}

\begin{figure}[tb]
    \centering
    \begin{minipage}[t]{0.23\textwidth}
        \centering
        \includegraphics[width=\textwidth]{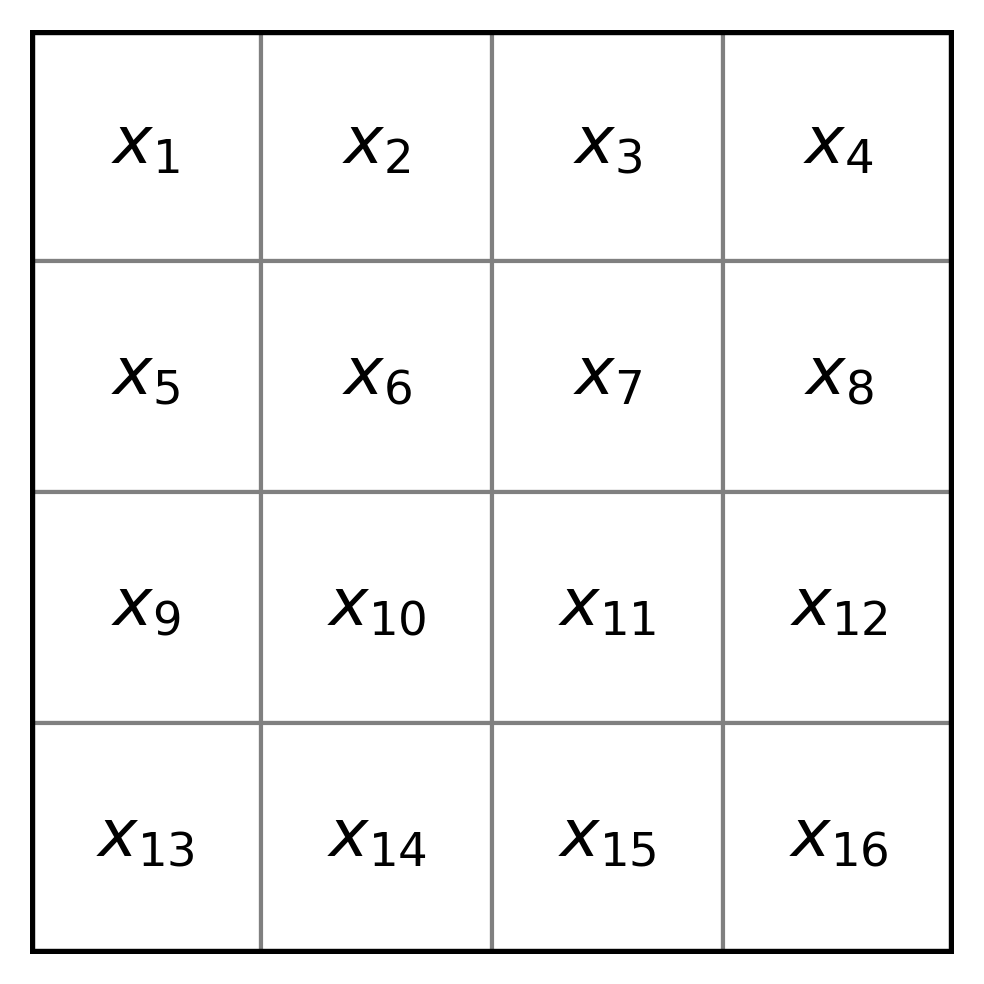}
        \\{\small (a) $\mathcal X$}
    \end{minipage}\hfill
    \begin{minipage}[t]{0.23\textwidth}
        \centering
        \includegraphics[width=\textwidth]{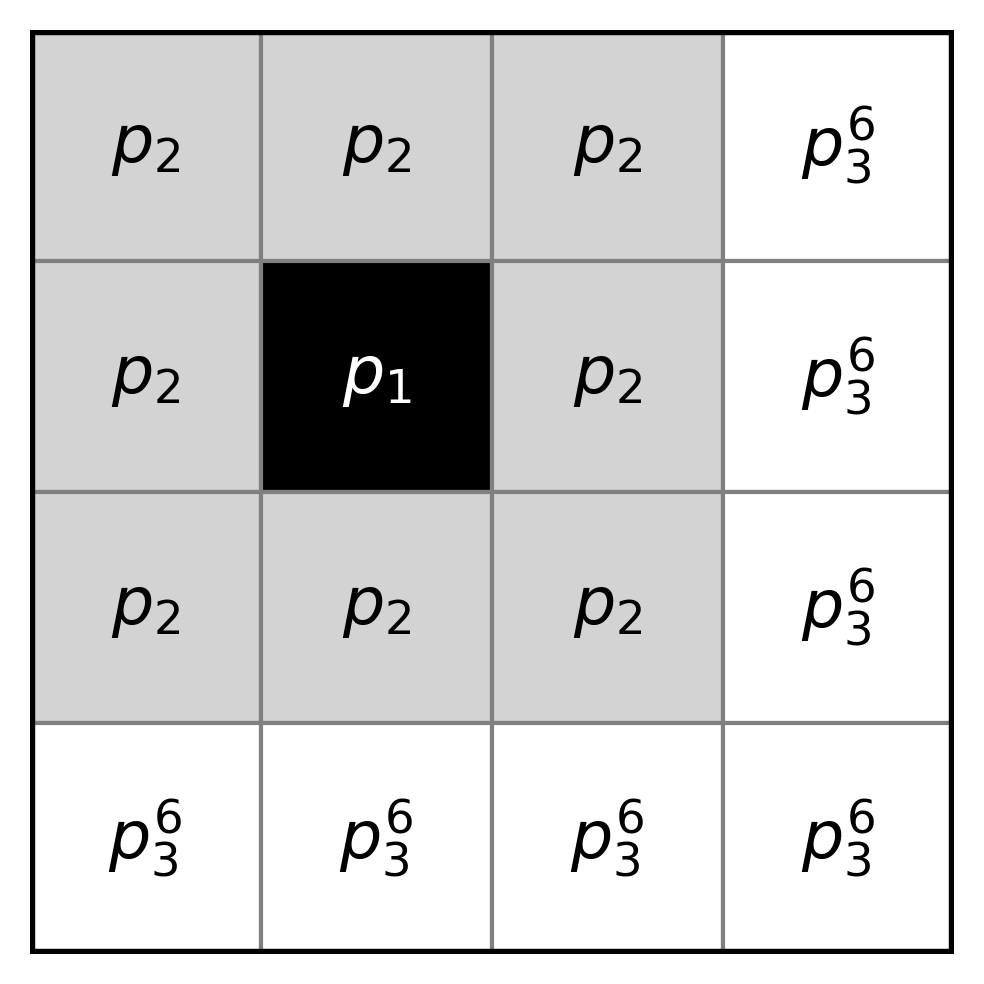}
        \\ {\small (b) $P_6$}
    \end{minipage}\hfill
    \begin{minipage}[t]{0.23\textwidth}
        \centering
        \includegraphics[width=\textwidth]{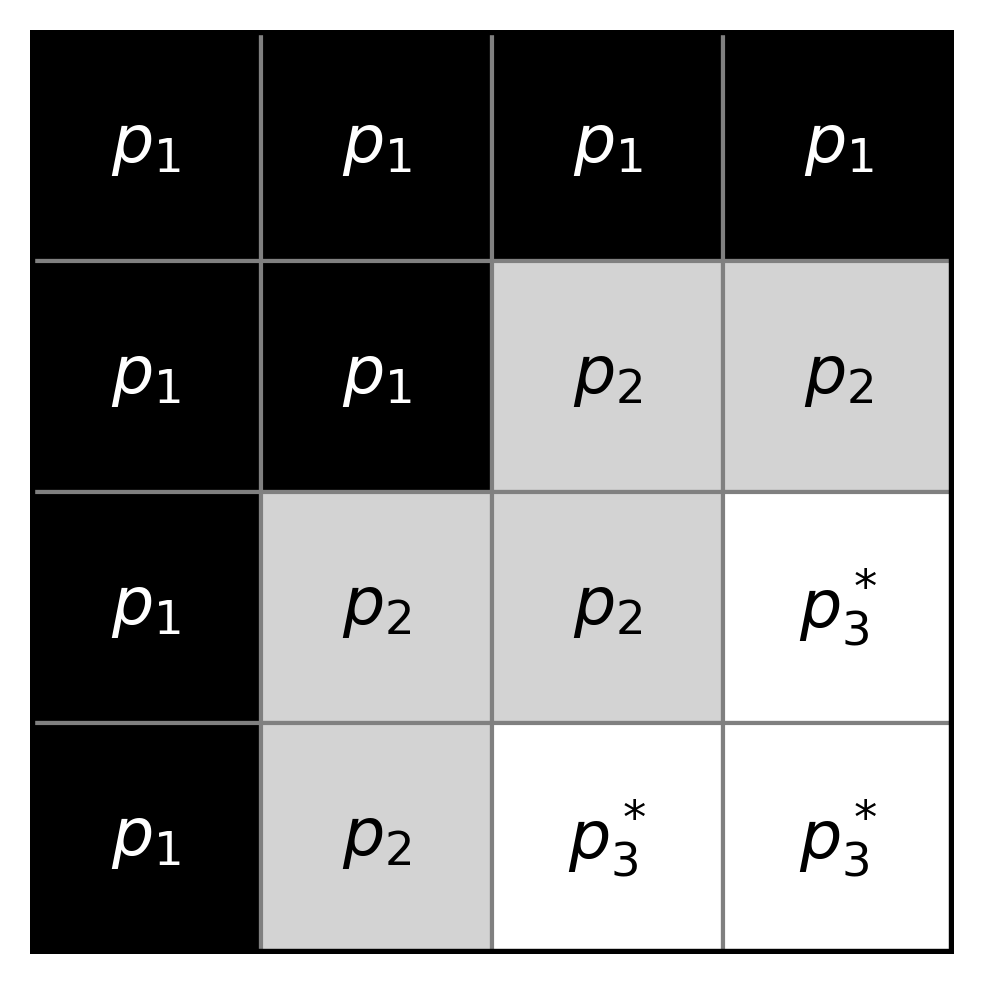}
        \\ {\small (c) $\max\limits_{P\in\mathcal M_1}P(x)$}
    \end{minipage}\hfill
    \begin{minipage}[t]{0.23\textwidth}
        \centering
        \includegraphics[width=\textwidth]{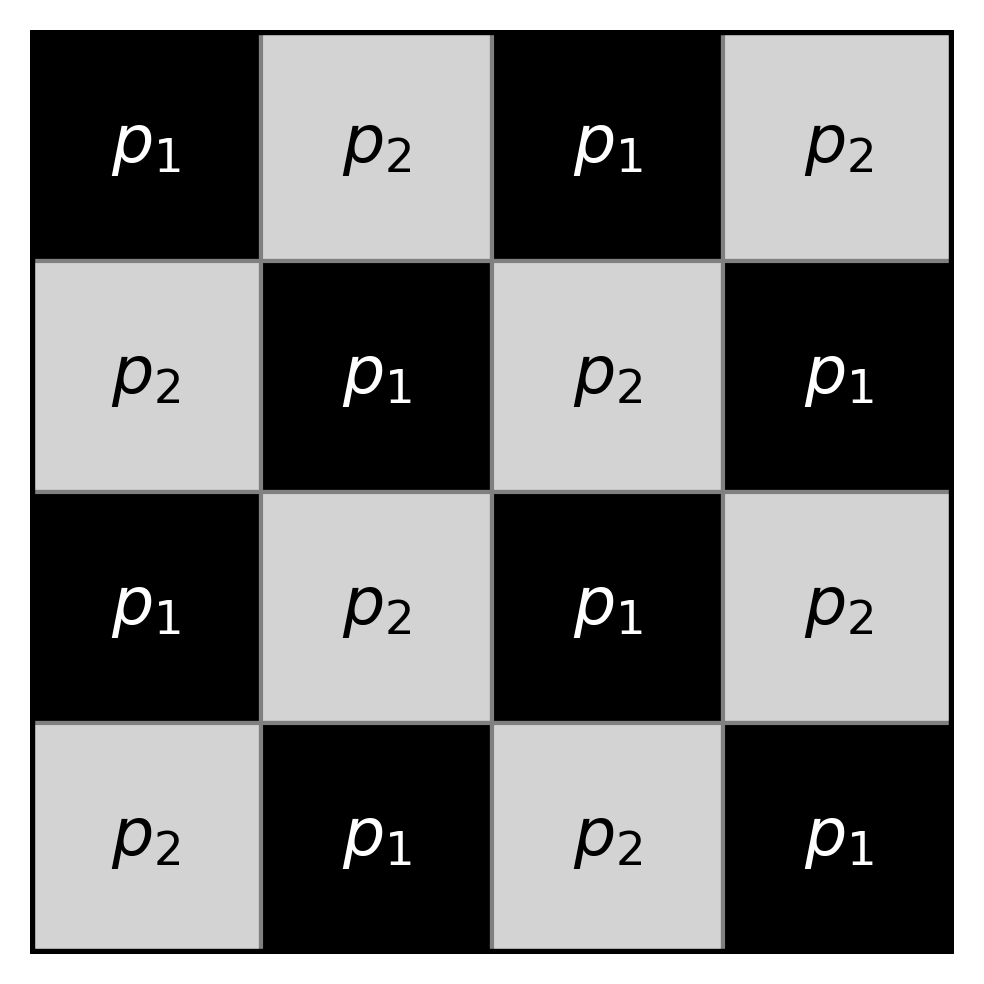}
        \\ {\small (d) $\max\limits_{P\in\mathcal M_2}P(x)$}
    \end{minipage}
    \caption{A simple discrete example. Panel (a) defines the sample space, and Panel (b) shows a probability distribution over $\mathcal X$ peaking at $x_6$ (darker shades represent larger probabilities). Panels (c) and (d) plot the maximized likelihood of each possible outcome under models $\mathcal M_1$ and $\mathcal M_2$, respectively. Both models contain $P_6$, but their remaining seven distributions are centered differently.} \label{fig:grids}
\end{figure}

The following example applies MDL to a finite sample space and two simple non-parametric models. Consider the 16-element sample space represented by the $4\times4$ grid in Figure~\ref{fig:grids}(a). Its elements may be interpreted either as one-observation samples, with $n=1$ and $\mathcal X=\{x_1,\ldots,x_{16}\}$, or as four-observation samples, with $n=4$ and $\mathcal X=\{0,1\}$. To simplify notation, we take $n=1$ and suppress the sample-size superscript, writing $x$ and $\mathcal X$ in place of $x^n$ and $\mathcal X^n$.

For each $j\in\{1,\ldots,16\}$, let $P_j$ be a probability distribution over $\mathcal X$ centered at $x_j$. Specifically, $P_j$ assigns probability $p_1$ to $x_j$, probability $p_2$ to each adjacent cell, and residual probability $p_3^j$ to every remaining cell, with $p_1>p_2>p_3^j$ and $\sum_{x\in\mathcal X}P_j(x)=1$. Figure~\ref{fig:grids}(b) plots $P_6$. Since $p_1$ and $p_2$ are fixed across $j$, the residual probability is largest for corner-centered distributions; denote this value by $p_3^*$.

Adapting the graphical example in \citet{Fudenberg2026}, consider two models, $\mathcal M_1$ and $\mathcal M_2$, each consisting of eight such distributions. In $\mathcal M_1$, the distributions are centered on or above the $45^\circ$ diagonal:
\[
\mathcal M_1=\{P_1,P_2,P_3,P_4,P_5,P_6,P_9,P_{13}\}.
\]
In $\mathcal M_2$, the centers are arranged in a checkerboard pattern:
\[
\mathcal M_2=\{P_1,P_3,P_6,P_8,P_9,P_{11},P_{14},P_{16}\}.
\]

Even though the two models contain the same number of distributions, they differ in the range of outcomes they can fit well. Panels (c) and (d) plot the maximized likelihood achieved by each model at every possible outcome, $\max_{P\in\mathcal M}P(x)$. Both models attain the maximum probability $p_1$ at eight cells, since each contains eight centered distributions. The difference arises away from those centers. In $\mathcal M_2$, the checkerboard arrangement ensures that every remaining cell is adjacent to one of its centers and therefore receives maximized likelihood $p_2$. In $\mathcal M_1$, by contrast, $x_{12}$, $x_{15}$, and $x_{16}$ are not adjacent to any center in the model; for each of these outcomes, $\operatorname*{arg\,max}_{P\in\mathcal M_1}P(x)=\{P_1,P_4,P_{13}\}$, so the maximized likelihood is $p_3^*<p_2$. Therefore, the sum of maximized likelihoods is larger for $\mathcal M_2$, and hence $\operatorname{COMP}(\mathcal M_2)>\operatorname{COMP}(\mathcal M_1)$.

Suppose that a single outcome $x\in\mathcal X$ is observed and the task is to select between $\mathcal M_1$ and $\mathcal M_2$. Because $\mathcal M_2$ is more complex, MDL selects it only when its advantage in fit is large enough to offset the difference in complexity. For any outcome on or above the $45^\circ$ diagonal, $\mathcal M_1$ fits at least as well as $\mathcal M_2$, so MDL selects $\mathcal M_1$. For outcomes below the diagonal for which $\mathcal M_2$ attains a higher maximized likelihood, MDL selects $\mathcal M_2$ if and only if
\[
\log \max_{P\in\mathcal M_2}P(x) - \log \max_{P\in\mathcal M_1}P(x) > \operatorname{COMP}(\mathcal M_2)-\operatorname{COMP}(\mathcal M_1).
\]

For comparison, consider Bayesian model selection, which chooses the model with the highest posterior probability. Unlike MDL, this procedure requires priors over both the models and the distributions within each model. With equal model priors, the posterior odds equal the Bayes factor, defined as the ratio of the models’ marginal likelihoods. Under a uniform prior over the eight distributions within each model, the marginal likelihood of $\mathcal M_j$ is $\frac{1}{8}\sum_{P\in\mathcal M_j}P(x)$, so Bayes selects the model with the higher average likelihood. Under these priors, it selects $\mathcal M_1$ for every outcome on or above the diagonal. Below the diagonal, it selects $\mathcal M_2$ except at $x_7$ and $x_{10}$, where the two models have equal posterior probability. MDL selects $\mathcal M_1$ at these ties and selects $\mathcal M_2$ among the remaining six outcomes only when the condition above is satisfied. Thus, under uniform priors, MDL sets a higher threshold than Bayes for selecting the more flexible model in this example. Bayesian model selection, of course, depends on the assumed priors, so different prior weights may produce different rankings.

\subsubsection*{Approximating model complexity}

The exact complexity term in \eqref{eq:comp} is readily computed in the finite example above, where the sample space contains only 16 elements. In many applications, however, summing maximized likelihoods over the entire sample space is infeasible. For parametric models, this problem can often be avoided using the following second-order asymptotic approximation \citep{Rissanen1996}:\footnote{The approximation requires standard regularity conditions. In particular, the maximum-likelihood estimate must remain sufficiently far from the boundary of the parameter space as $n$ grows, and the Fisher information integral $\int_{\Theta}\sqrt{\det I(\theta)}\,d\theta$ must be finite.}
\begin{equation}
\operatorname{COMP}_n(\mathcal M) \approx \frac{k}{2}\log\frac{n}{2\pi} + \log\int_\Theta\sqrt{\det I(\theta)}\,d\theta, \label{eq:rissanen}
\end{equation}
where $k$ is the dimension of $\theta$, $n$ is the sample size, and $I(\theta)$ is the Fisher information matrix, with typical element $I_{ij}(\theta)=-(1/n)\mathbb E_\theta[\partial^2\log P_\theta(X^n)/(\partial\theta_i\partial\theta_j)]$.

Substituting this approximation back into the objective function \eqref{eq:nml} yields the following asymptotic approximation to the NML description length:
\begin{equation}
L_{\mathrm{NML}}(x^n,\mathcal M) \approx -\log P_{\hat\theta(x^n)}(x^n) + \frac{k}{2}\log\frac{n}{2\pi} + \log\int_\Theta\sqrt{\det I(\theta)}\,d\theta. \label{eq:nml_approx}
\end{equation}

Beyond making complexity tractable, \eqref{eq:nml_approx} also connects MDL to BIC and Bayesian model selection. As $n\to\infty$, the Fisher term and the constant contribution from $\log(2\pi)$ in \eqref{eq:nml_approx} become negligible relative to the maximized log-likelihood and the leading $\frac{k}{2}\log n$ penalty. Dropping these terms yields BIC,\footnote{BIC is usually written as $-2\log P_{\hat\theta(x^n)}(x^n)+k\log n$, twice the expression obtained from \eqref{eq:nml_approx} after dropping the sample-size-independent terms.} so when the true model is among the candidates, MDL inherits BIC's consistency under the usual regularity conditions. If these terms are retained, however, \eqref{eq:nml_approx} is asymptotically equivalent to the negative log marginal likelihood under Jeffreys' prior. Consequently, MDL model selection based on this approximation coincides asymptotically with model selection via Bayes factors under this non-informative prior \citep[see][]{Grunwald2007}.

The same approximation also has a geometric interpretation. In exponentiated form, the complexity penalty in \eqref{eq:rissanen} represents a ``volume'' in which the size of a set of distributions is proportional to the number of statistically distinguishable distributions it contains \citep{Myung2000}. This perspective conceptually relates MDL's complexity term to Selten's notion of a theory's area \citep{Selten1991}. The next section examines the relationship between MDL complexity and Selten's area measure directly.

\section{Alternatives to Parameter Counting in Economics}
\label{sec:mdl_econ}

\subsection{Predictive Success} \label{sec:selten}

The Selten--Krischker (SK) measure of \textit{predictive success}, introduced by \citet{Selten1983} and axiomatized by \citet{Selten1991}, has been highly influential in experimental economics. This section reviews the original framework and presents a compression-based analogue that reformulates area theories as codes and reinterprets predictive success as compression gain with respect to a zero-compression benchmark.

\subsubsection*{The Selten--Krischker Measure}

Consider a finite outcome space $\mathcal X=\{1,\dots,I\}$. An \textit{area theory} specifies a subset $\mathcal A\subseteq\mathcal X$ of predicted outcomes. The \textit{area} of the theory is defined as $a=|\mathcal A|/I$. Given a sample $x^n=(x_1,\dots,x_n)\in\mathcal X^n$, the \textit{hit rate} is defined as the proportion of observations falling within the predicted subset, $r=\frac{1}{n}\sum_{i=1}^n \mathbbm{1}(x_i\in\mathcal A)$, where $\mathbbm{1}(\cdot)$ denotes the indicator function. Predictive success is defined as the linear difference $m=r-a$.

\cite{Selten1991} justifies this specific functional form by examining the structure of its ``unimprovable theories'' and by providing an axiomatic characterization. For a given probability distribution $P$ over the outcome space $\mathcal X$, an unimprovable theory maximizes the expected value of the measure. Under $m = r - a$, this theory includes exactly those outcomes that occur with greater-than-average probability, $\mathcal{A}^* = \{x \in \mathcal X : P(x) > 1/I\}$. Selten evaluates other plausible functional forms, but rejects them because they yield unimprovable theories that are either extremely permissive or extremely restrictive.\footnote{In particular, the ratio measure $m = r/a$ includes only the single most probable outcome, whereas the outside ratio measure $m = (r-a)/(1-a)$ includes every outcome except the single least probable one.} In this sense, the linear difference $r-a$ occupies a balanced middle ground.

Axiomatically, Selten requires that $m(r,a)$ be strictly increasing in the hit rate (Axiom 1), strictly decreasing in the area (Axiom 2), and evaluate the trivial theories $\mathcal{A}=\mathcal X$ and $\mathcal{A}=\emptyset$ as equally unsuccessful, meaning $m(1,1) = m(0,0)$ (Axiom 5). He then establishes two distinct mathematical characterizations of the $r-a$ functional form by introducing stronger linearity requirements. In his first theorem, adding continuity (Axiom 3) alongside the requirement that comparisons between theories depend only on the differences $r_1 - r_2$ and $a_1 - a_2$ (Axiom 4) characterizes the measure \textit{ordinally}. In his second theorem, dropping Axioms 3 and 4 but requiring that the measure for a pooled sample be the weighted average of its subsample measures, $m(\alpha r_1 + (1-\alpha) r_2, \alpha a_1 + (1-\alpha) a_2) = \alpha m(r_1,a_1) + (1-\alpha)m(r_2,a_2)$ (Axiom 6), characterizes the measure \textit{cardinally} (up to positive affine transformations).

\subsubsection*{A Compression Analogue of Predictive Success}

A defining feature of the SK framework is that it does not assign probabilities to outcomes within or outside a theory's area. This makes the approach difficult to apply directly to the stochastic models and likelihood-based estimation methods that dominate much of empirical economics. To bridge this gap, this section presents a compression-based analogue of predictive success that reformulates area theories as codes.

In the SK framework, area theories are defined over the outcome space $\mathcal X$, rather than over the sample space $\mathcal X^n$. Our reformulation accordingly treats an area theory as a code for individual outcomes and evaluates its average description length over the observed sample. For each area theory $\mathcal A\subseteq\mathcal X$, define the associated area model as
\[
\mathcal M_{\mathcal A} = \{P_y:y\in\mathcal A\}\cup\{P_0\},
\]
where, for each $y\in\mathcal A$, $P_y$ assigns probability one to $y$ and zero to every other outcome, and $P_0$ assigns probability $1/C$ to each outcome outside $\mathcal A$, distributing its remaining probability mass uniformly over $\mathcal A$. The constant $C$ governs the probability assigned to unpredicted outcomes and must satisfy $C\geq I-|\mathcal A|$ for $P_0$ to be a probability distribution.\footnote{For the boundary case $\mathcal A=\emptyset$, take $C=I$, so that $P_0$ is uniform over $\mathcal X$.}

The maximized likelihood under $\mathcal M_{\mathcal A}$ for any outcome $x\in\mathcal X$ is
\begin{equation}
\widehat P_{\mathcal A}(x)
\equiv
\max_{P\in\mathcal M_{\mathcal A}}P(x)
=
\begin{cases}
1 & \text{if } x\in\mathcal A,\\[3pt]
\frac{1}{C} & \text{if } x\notin\mathcal A.
\end{cases}
\label{eq:P_A}
\end{equation}
Thus, $\mathcal M_{\mathcal A}$ preserves the defining feature of an area theory: it distinguishes predicted from unpredicted outcomes, but makes no distinction among outcomes within either group.

To illustrate the representation, consider a standard rational-choice model over a fixed collection of finite choice sets. An outcome $x \in \mathcal X$ is a choice profile, recording one selected alternative from each choice set, and $\mathcal A \subseteq \mathcal X$ is the subset of choice profiles that can be rationalized by a strict preference relation. Each strict preference relation generates a unique profile by selecting the most-preferred alternative from each choice set. Hence, for every rationalizable profile $y\in\mathcal A$, there exists a preference relation that predicts $y$ with certainty, represented in $\mathcal M_{\mathcal A}^{0}$ by the degenerate distribution $P_y$. Accordingly, $\widehat P_{\mathcal A}(y)=1$ for every $y\in\mathcal A$. Conversely, no strict preference relation generates a profile outside $\mathcal A$, so such a profile is assigned likelihood $1/C$ under $P_0$, and its maximized likelihood is $\widehat P_{\mathcal A}(x)=1/C$.

Using the maximized likelihood function in \eqref{eq:P_A}, both the complexity and goodness-of-fit components of the MDL criterion can be calculated directly. The complexity of $\mathcal M_{\mathcal A}$ is
\begin{equation}
\operatorname{COMP}(\mathcal M_{\mathcal A})=\log\sum_{x\in\mathcal X}\widehat P_{\mathcal A}(x)=\log\left(|\mathcal A|+\frac{I-|\mathcal A|}{C}\right)=\log I+\log \tilde a. \label{eq:COMP_A}
\end{equation}
Here, $\tilde a=a+(1-a)/C$ denotes the error-adjusted area. Crucially, note that for fixed $C>1$ and $I$, $\operatorname{COMP}(\mathcal M_{\mathcal A})$ is strictly increasing in Selten's area $a$ and therefore ranks area theories in the same order.

Given a sample of observations $x^n$, goodness of fit is measured by the average log-likelihood per observation
\begin{equation}
\frac{1}{n}{\log\mathcal L}_{\mathcal A}(x^n)= \frac{1}{n}\sum_{i=1}^n\log\widehat P_{\mathcal A}(x_i)=r\log 1+(1-r)\log(1/C)=-(1-r)\log C. \label{eq:L_A}
\end{equation}

From equation \eqref{eq:nml}, the average description length per observation is then
\begin{equation}
\bar L(r,\mathcal A)\equiv\frac{1}{n}L_{\mathrm{NML}}(x^n,\mathcal M_{\mathcal A})=-\frac{1}{n}{\log\mathcal L}_{\mathcal A}(x^n)+\operatorname{COMP}(\mathcal M_{\mathcal A})=(1-r)\log C+\log I+\log \tilde a. \label{eq:mdl_a}
\end{equation}
An analogue of predictive success is obtained by expressing this description length as a \textit{compression gain} relative to a benchmark code. The uniform code over $\mathcal X$ provides a natural zero-compression benchmark, requiring $\log I$ nats per observation. Thus, compression gain is defined as
\begin{equation}
g(r,\tilde a)=\frac{\log I-\bar L(r,\mathcal A)}{\log I} =\frac{r\log C-\log \tilde a-\log C}{\log I}. \label{eq:m_ra}
\end{equation}

Here, $g=0$ means that the model achieves no compression relative to the uniform code, while $g=0.10$, for example, means that it reduces average description length by 10\% relative to the uniform code.

This leaves the specification of $C$, which determines the likelihood assigned to misses. Two natural alternatives can be mapped to different assumptions about the miss distribution $P_0$. Setting $C=I$ corresponds to a miss distribution that assigns uniform probability to all possible outcomes in $\mathcal X$. Conversely, setting $C=I-|\mathcal A|$ corresponds to a miss distribution that assigns uniform probability only over the unpredicted region $\mathcal X\setminus\mathcal A$. Both modeling choices are common in economics. For instance, in the context of behavioral game theory, \citet{Healy2023} estimate level-$k$ and cognitive hierarchy models using error structures that parallel these two specifications, which they term ``double-counting'' ($C=I$) and ``single-counting'' ($C=I-|\mathcal A|$).

For the present purpose, $C=I$ is the more natural formulation. First, it ensures that all models incur the same penalty for misses regardless of their area, in line with the original SK approach. This preserves Selten's monotonicity requirement that predictive success decreases with the area holding the hit rate fixed. Under single-counting, expanding the predicted set also reduces the penalty for remaining misses, so this monotonicity property need not hold. Also, the error structure implied by $C=I$ maps naturally to standard discrete choice models: as the sensitivity parameter of a logit function approaches zero, choice probabilities converge to a uniform distribution over all outcomes in $\mathcal X$.

We therefore set $C=I$, so that the adjusted area is given by $\tilde a=a+(1-a)/I$ and the compression gain expression in \eqref{eq:m_ra} becomes
\begin{equation}
g(r,\tilde a)=r-\frac{\log(I\tilde a)}{\log I}. \label{eq:m_anml}
\end{equation}

To see the parallel between compression gain and predictive success, let $\tilde A=I\tilde a$ denote the effective number of predicted outcomes after accounting for the miss state. Then $g=r-\frac{\log \tilde A}{\log I}$, whereas $m=r-\frac{|\mathcal A|}{I}$. Thus, both measures reward hit rates in the same way, but predictive success divides the number of predicted outcomes by the total number of possible outcomes, whereas compression gain takes the ratio of their logarithms. Furthermore, rewriting (\ref{eq:m_anml}) as $g=r-1-\frac{\log\tilde a}{\log I}$, shows that compression gain imposes a decreasing marginal penalty on the adjusted area $\tilde a$, and that this marginal penalty decreases as the outcome space expands. Intuitively, when $I$ is larger, a miss entails a greater loss of information, so $g(r,\tilde a)$ becomes more permissive about a model's area as the outcome space expands.

With $C=I$, both predictive success and compression gain take values in $(-1,1)$, and both measures assign zero to the two trivial theories, $\mathcal A=\emptyset$ and $\mathcal A=\mathcal X$. The interpretation of zero differs, however. In SK, $m=0$ when $r=a$, which is the expected score of a theory that selects its predicted outcomes at random. In MDL, $m=0$ instead means that the theory achieves no compression relative to the uniform code. Moreover, for any nontrivial theory, $m(r,a)>g(r,\tilde a)$. A theory can therefore achieve positive predictive success while still yielding negative compression gain. Application 1 illustrates an example of this.

Two further comparisons with Selten's framework are useful. The first concerns the unimprovable theories implied by compression gain; the second concerns Selten's axioms. Recall that minimizing description length is mathematically equivalent to minimizing KL divergence from the empirical distribution to the NML distribution induced by the code. Thus, an area theory yields a positive compression gain ($g>0$) if and only if the KL divergence from the empirical distribution to the theory's NML distribution is smaller than the KL divergence from the empirical to the uniform distribution. More generally, for any distribution $P$ over $\mathcal X$, an unimprovable theory in terms of compression gain is one that minimizes the KL divergence between $P$ and the distribution implied by its NML code. By contrast, an unimprovable theory in terms of predictive success includes exactly those outcomes that occur with above-average probability.

Turning to the axioms, compression gain preserves all of Selten's axioms except the two stronger requirements that impose linearity. It satisfies the monotonicity Axioms 1 and 2, continuity Axiom 3, and Axiom 5 on the equal treatment of trivial theories. However, because it imposes logarithmic penalties to capture information loss, it violates Axiom 4, which requires comparisons to depend only on the differences $r_1-r_2$ and $a_1-a_2$, and Axiom 6, which requires the pooled measure to be a convex combination of its subsample measures. Section~3 of the Online Appendix provides proofs.

\subsection{Completeness and Restrictiveness}
	\label{sec:comp_restr}

	\citet{Fudenberg2022,Fudenberg2026} propose two complementary measures for evaluating economic models. Completeness measures how much of the predictable variation in the data a model captures. Restrictiveness measures a model's flexibility by asking how well it can approximate synthetic datasets drawn from a predefined eligible set.

	Consider a prediction problem in which each observation is a pair $(z,y)$, where $z$ belongs to a finite covariate set $\mathcal Z$ and $y\in\mathcal Y$, with $\mathcal Y$ a compact subset of a finite-dimensional Euclidean space. A prediction rule is a mapping $f:\mathcal Z\to\mathcal Y$. Let $\overline{\mathcal F}=\mathcal Y^{|\mathcal Z|}$ denote the set of all prediction rules, let $\mathcal F_\Theta=\{f_\theta\}_{\theta\in\Theta}\subseteq\overline{\mathcal F}$ denote a parameterized family of prediction rules, and let $f_b$ denote a baseline prediction rule.

	Let $P$ denote the joint distribution of covariates and outcomes, and let $\ell:\mathcal Y\times\mathcal Y\to\mathbb R_+$ be a loss function. Define the expected prediction error of a prediction rule $f$ by $e_P(f)=\mathbb E_P[\ell(f(Z),Y)]$, and let $f^*\in\operatorname*{arg\,min}_{f\in\overline{\mathcal F}}e_P(f)$ denote a best possible prediction rule. The completeness of $\mathcal F_\Theta$ is
	\begin{equation}
\kappa(\mathcal F_\Theta) = \frac{e_P(f_b)-\inf_{f_\theta\in\mathcal F_\Theta}e_P(f_\theta)}{e_P(f_b)-e_P(f^*)}. \label{eq:fudenberg_completeness}
\end{equation}
	Completeness measures the share of possible reduction in expected prediction error achieved by the model. The numerator is the model's improvement over the baseline. The denominator is the largest possible improvement.

	Restrictiveness evaluates flexibility over a compact eligible set of prediction rules $\mathcal F\subseteq\overline{\mathcal F}$. Let $d:\overline{\mathcal F}\times\overline{\mathcal F}\to\mathbb R_+$ be a nonnegative discrepancy function, with $d(f,f')=0$ if and only if $f=f'$. For a target rule $f\in\mathcal F$, define the closest discrepancy attainable by $\mathcal F_\Theta$ as $d(\mathcal F_\Theta,f)=\inf_{f_\theta\in\mathcal F_\Theta}d(f_\theta,f)$. The restrictiveness of $\mathcal F_\Theta$ with respect to $\mathcal F$ is
	\begin{equation}
r(\mathcal F_\Theta,\mathcal F) = \frac{\mathbb E_{\lambda_{\mathcal F}}\!\left[d(\mathcal F_\Theta,f)\right]} {\mathbb E_{\lambda_{\mathcal F}}\!\left[d(f_b,f)\right]}, \label{eq:fudenberg_restrictiveness}
\end{equation}
	where $\lambda_{\mathcal F}$ is the uniform distribution over the eligible set and $f_b$ is a baseline prediction rule. The better $\mathcal F_\Theta$ is able to approximate a wide range of synthetic prediction rules, the lower its restrictiveness.

	\subsubsection*{Compression Analogues of Completeness and Restrictiveness}

	Fudenberg et al.’s completeness and restrictiveness measures evaluate predictive performance and flexibility separately, but their framework does not specify how the two should be traded off when they favor different models. The compression analogues developed here, by contrast, yield a complete model selection criterion. As shown below, maximizing the compression analogue of completeness is equivalent to minimizing description length, while the logarithm of the compression analogue of restrictiveness equals model complexity up to a constant common to all candidate models.

    To construct compression analogues of completeness and restrictiveness, the preceding framework is first translated into the notation used throughout this paper. The risky-choice experiment presented in the introduction provides an illustration. The experiment comprises $J$ lotteries. Each lottery is represented by a covariate vector $z$ containing its probability and payoff attributes, and $\mathcal Z$ denotes the finite set of the $J$ lottery vectors used in the experiment. Let $\mathcal Y\subset\mathbb R$ be the compact outcome space for a single lottery, consisting of its possible reported certainty equivalents. Let $\{(z_i,y_i)\}_{i=1}^n$ denote an empirical sample, where $z_i\in\mathcal Z$ and $y_i\in\mathcal Y$, and write $z^n=(z_1,\ldots,z_n)$ and $y^n=(y_1,\ldots,y_n)$ for the corresponding covariate and outcome sequences.

    Let $\mathcal X=\mathcal Y^J$ denote the space of complete outcome profiles, where each $x\in\mathcal X$ is a $J$-vector containing one certainty equivalent for each lottery. A prediction rule $f:\mathcal Z\to\mathcal Y$ induces the complete outcome profile $x_f=(f(z))_{z\in\mathcal Z}\in\mathcal X$. Conversely, every complete outcome profile $x\in\mathcal X$ defines a prediction rule on $\mathcal Z$, establishing a one-to-one correspondence between prediction rules $f\in\overline{\mathcal F}$ and complete outcome profiles $x\in\mathcal X$. Under this correspondence, the eligible set $\mathcal F\subseteq\overline{\mathcal F}$ induces the compact set $\mathcal S=\{x_f:f\in\mathcal F\}\subseteq\mathcal X$, and the parameterized family $\mathcal F_\Theta=\{f_\theta\}_{\theta\in\Theta}$ induces $\mathcal X_\Theta=\{x_{f_\theta}:f_\theta\in\mathcal F_\Theta\}\subseteq\mathcal X$. Thus, $\mathcal X_\Theta$ is a set of point predictions over $\mathcal X$, analogous to an area theory in Selten's framework.

    To turn the parameterized family of prediction rules $\mathcal F_\Theta$ into a parametric model $\mathcal M_\Theta$, associate each prediction rule $f_\theta$ with a conditional probabilistic source $P_\theta(y^n\mid z^n)$.\footnote{A natural way to do this is to use the loss function $\ell$ to define the conditional likelihood via $P_\theta(y\mid z)=\exp\{-\beta\ell(f_\theta(z),y)\}/Z_\theta(z,\beta)$, where $\beta>0$ is a fixed scaling factor and $Z_\theta(z,\beta)$ normalizes the distribution over $\mathcal Y$ \citep[see, e.g.,][]{Grunwald1999}. Under conditional independence, $P_\theta(y^n\mid z^n)=\prod_{i=1}^nP_\theta(y_i\mid z_i)$.} Let $\mathcal M_\Theta=\{P_\theta:\theta\in\Theta\}$ denote the resulting model. The compression analogue of completeness is defined as
\begin{equation}
\mathcal C(y^n,\mathcal M_\Theta\mid z^n) = \frac{ \bar L(y^n,\mathcal M^b\mid z^n) - \bar L(y^n,\mathcal M_\Theta\mid z^n) }{ \bar L(y^n,\mathcal M^b\mid z^n) - \widehat H(y^n\mid z^n) }, \label{eq:mdl_completeness_sample}
\end{equation}

    where $\bar L(y^n,\mathcal M\mid z^n)=\frac{1}{n}L_{\mathrm{NML}}(y^n,\mathcal M\mid z^n)$ denotes conditional NML code length per observation, with $L_{\mathrm{NML}}$ defined in \eqref{eq:nml_conditional}; $\mathcal M^b=\{P_\phi:\phi\in\Phi\}$ denotes the baseline model; and $\widehat H(y^n\mid z^n)$ is an estimator of the conditional entropy per observation, $\frac{1}{n}H(Y^n\mid Z^n=z^n)$.

    Like completeness, $\mathcal C$ evaluates predictive performance relative to a baseline and a best attainable benchmark. Its numerator is the reduction in conditional NML code length relative to the baseline. Its denominator estimates the largest achievable reduction, measured by the gap between the baseline code length and the conditional-entropy lower bound. Thus, $\mathcal C$ measures the fraction of available compression achieved by the model.

    For restrictiveness, the relevant objects are instead prediction rules or, equivalently, complete outcome profiles. Under the correspondence established above, each eligible prediction rule $f\in\mathcal F$ is represented by a complete outcome vector $x_f\in\mathcal S\subseteq\mathcal X$. The compression analogue of restrictiveness is defined as \begin{equation}
F(\mathcal M_\Theta,\mathcal S) = \frac{ \mathbb E_{\lambda_{\mathcal S}} [P_{\hat\theta(S)}(S)] }{ \mathbb E_{\lambda_{\mathcal S}} [P_{\hat\phi(S)}(S)] }. \label{eq:likelihood_flexibility}
\end{equation}

	Here $S\sim\lambda_{\mathcal S}$ is a complete outcome profile drawn uniformly from $\mathcal S$. To simplify notation, $P_\theta(S)$ and $\hat\theta(S)$ stand for $P_\theta(S\mid z^J)$ and $\hat\theta(S\mid z^J)$, respectively, where $z^J$ denotes a sequence containing each element of $\mathcal Z$ once.

    Like restrictiveness, $F$ normalizes the model's average best attainable fit over a predefined evaluation set by the baseline model's average best attainable fit. Because $F$ measures fit by maximized likelihood rather than minimized discrepancy, larger values indicate better fit to outcomes drawn from the eligible set. For this reason, $F(\mathcal M_\Theta,\mathcal S)$ will be called \emph{likelihood flexibility}.\footnote{Note that even if $P_\theta$ is constructed from the discrepancy function as $P_\theta(s)\propto\exp\{-\beta d(x_{f_\theta},s)\}$, likelihood flexibility and restrictiveness need not rank models identically because the transformation is nonlinear.}

    The following remarks link the MDL criterion to the proposed completeness and restrictiveness analogues.

\begin{remark}
    \label{rem:completeness_mdl}
    Let $\{(z_i,y_i)\}_{i=1}^n$ denote an empirical sample. Fix a baseline model $\mathcal M^b$ and a conditional entropy estimator $\widehat H(y^n\mid z^n)$. Provided $\bar L(y^n,\mathcal M^b\mid z^n) > \widehat H(y^n\mid z^n)$, MDL selects the model that maximizes $\mathcal C(y^n,\mathcal M_\Theta\mid z^n)$.

    \noindent\textit{Proof.} The result follows by rearranging the definition of $\mathcal C$: MDL code length is an affine, decreasing function of $\mathcal C$, with a positive coefficient common to all candidate models. \qed
\end{remark}

\medskip

\begin{remark}
\label{rem:logF_COMP}
Let $\operatorname{COMP}(\mathcal M_\Theta,\mathcal S) = \log\sum_{s\in\mathcal S}P_{\hat\theta(s)}(s)$ denote the complexity of $\mathcal M_\Theta$ evaluated over $\mathcal S$. Then
\begin{equation}
\log F(\mathcal M_\Theta,\mathcal S) = \operatorname{COMP}(\mathcal M_\Theta,\mathcal S) - \operatorname{COMP}(\mathcal M^b,\mathcal S). \label{eq:logF_COMP}
\end{equation}

\noindent\textit{Proof.} Since $\lambda_{\mathcal S}$ is uniform, its common normalizing constant cancels from the ratio defining $F$; taking logarithms then yields the difference between the two complexity terms. \qed
\end{remark}

\medskip

\begin{remark}
\label{rem:restricted_complexity}
Let $L_{\mathrm{NML}}(x,\mathcal M;\mathcal S)$ denote the restricted NML code length obtained by taking $\mathcal S$ as the sample space. When MDL is implemented using this restricted code, it selects, for any $x\in\mathcal S$, the model that minimizes
$$
-\log P_{\hat\theta(x)}(x)+\log F(\mathcal M_\Theta,\mathcal S).
$$

\noindent\textit{Proof.} By Remark~\ref{rem:logF_COMP}, restricted NML code length equals the stated objective plus the baseline complexity, which is common to all candidate models. \qed
\end{remark}

\medskip

Remark \ref{rem:completeness_mdl} concerns an empirical sample $\{(z_i,y_i)\}_{i=1}^n$, whereas Remark \ref{rem:restricted_complexity} concerns one complete outcome profile $x \in \mathcal S$. The latter result extends to a sample of $n$ independent complete outcome profiles, $x^n\in\mathcal S^n$, by taking $\mathcal S^n$ as the sample space. MDL then selects the model that minimizes $-\log P_{\hat\theta(x^n)}(x^n)+\log F(\mathcal M_\Theta,\mathcal S^n)$.

Fudenberg et al.\@ define completeness and restrictiveness at the population level, but also provide finite-sample estimators of both measures. The most direct comparison is therefore between these estimators and our compression analogues, which are defined for observed data; Table~\ref{tab:mdl_analogues} compares the quantities entering each.

\begin{table}[tb]
\centering
\small
\setlength{\belowcaptionskip}{0.5em}
\caption{Sample quantities entering Fudenberg et al.'s measures and their compression analogues.}
\label{tab:mdl_analogues}
\renewcommand{\arraystretch}{1.25}
\begin{tabular}{@{}p{0.225\textwidth}p{0.30\textwidth}@{\hspace{0.5em}}p{0.43\textwidth}@{}}
    \hline
    & \textit{Fudenberg et al.}
    & \textit{Compression Analogue} \\
    \hline
    \textit{Completeness} & & \\
    \quad Baseline
    & $\widehat e_{\mathrm{CV}}(f_b)$
    & $\bar L(y^n,\mathcal M^b\mid z^n)$ \\
    \quad Model
    & $\widehat e_{\mathrm{CV}}(\mathcal F_\Theta)$
    & $\bar L(y^n,\mathcal M_\Theta\mid z^n)$ \\
    \quad Best attainable
    & $\widehat e_{\mathrm{CV}}(\overline{\mathcal F})$
    & $\widehat H(y^n\mid z^n)$ \\
    \textit{Restrictiveness} & & \\
    \quad Model
    & $\frac{1}{M}\sum_{m=1}^M d(\mathcal X_\Theta,s_m)$
    & $\frac{1}{M}\sum_{m=1}^M
       P_{\hat\theta(s_m)}(s_m)$ \\
    \quad Baseline
    & $\frac{1}{M}\sum_{m=1}^M d(x_{f_b},s_m)$
    & $\frac{1}{M}\sum_{m=1}^M
       P_{\hat\phi(s_m)}(s_m)$ \\
    \hline
\end{tabular}
\par\smallskip\footnotesize\raggedright
\textit{Note:}
$\widehat e_{\mathrm{CV}}$ denotes cross-validated prediction error. The quantity $\bar L(y^n,\mathcal M\mid z^n)=\frac{1}{n}L_{\mathrm{NML}}(y^n,\mathcal M\mid z^n)$ denotes conditional NML code length per observation, and $\widehat H(y^n\mid z^n)$ is an estimator of the conditional entropy per observation, $\frac{1}{n}H(Y^n\mid Z^n=z^n)$. The set $\mathcal S$ is the eligible set of complete outcome profiles, $\lambda_{\mathcal S}$ is the uniform distribution over it, and $s_1,\ldots,s_M\sim\lambda_{\mathcal S}$ are $M$ independent draws.
\par
\end{table}

    \medskip
    \noindent\textbf{Probabilistic Prediction Rules}\par\smallskip

	The preceding discussion interprets prediction rules as deterministic point predictions. However, Fudenberg et al.'s framework also accommodates probabilistic prediction rules, which can be incorporated into the compression framework directly. To illustrate, consider an experiment consisting of a fixed collection of $J$ menus, $\mathcal Z=(Z_1,\ldots,Z_J)$, where each menu contains $K$ alternatives, $Z_j=\{a_{j1},\ldots,a_{jK}\}$. A single outcome in this experiment is a complete choice profile $x=(x_1,\ldots,x_J)\in\mathcal X$, where $x_j\in Z_j$ and $\mathcal X=Z_1\times\cdots\times Z_J$.

	A probabilistic prediction rule is a probability mass function $f:\mathcal X\to[0,1]$, where $f(x)$ is the probability of choice profile $x$ and $\sum_{x\in\mathcal X}f(x)=1$. Thus, the set of all prediction rules is $\overline{\mathcal F}=\Delta(\mathcal X)$, and a parameterized family of prediction rules is itself a probabilistic model, $\mathcal F_\Theta=\mathcal M_\Theta=\{P_\theta:\theta\in\Theta\}$. For a sample $x^n$ of independent outcomes, each prediction rule $P_\theta$ therefore supplies the likelihood $P_\theta(x^n)$ directly, and the same compression analogues of completeness and restrictiveness apply.

\section{Application 1: Social Preferences}
\label{sec:app_social}

Our first application examines models of social preferences in a simple strategic environment with a finite, discrete outcome space. As in the introductory example, the analysis shows how the number of parameters, $k$, can be a poor proxy for model complexity. At the same time, this setting provides an ideal illustration of Selten's measure of predictive success and how it relates to its analogue, compression gain. While the former measure applies exclusively to deterministic area theories, precluding its use in standard maximum likelihood estimation, compression gain can also be applied to error-augmented stochastic versions of these models. Comparing both types of models under this unified framework reveals that error augmentation yields substantial gains in fit that more than offset a modest increase in complexity.

\subsection{Experimental Setup and Models}

We analyze data from \cite{Miettinen2020}, who conduct a sequential Prisoner's Dilemma experiment to evaluate various models of social preferences. Subjects are anonymously matched in pairs. The first mover chooses whether to cooperate ($\mathrm{C}$) or defect ($\mathrm{D}$). The second mover observes this choice and responds with either $\mathrm{C}$ or $\mathrm{D}$. Decisions are elicited using the strategy method. Prior to role assignment, each subject $i$ specifies one full plan of action, represented by the vector $x_i=(x_{i1},x_{i2},x_{i3})\in\{\mathrm{C},\mathrm{D}\}^3$, where $x_{i1}$ is the first-mover choice, $x_{i2}$ is the second-mover response to $\mathrm{C}$, and $x_{i3}$ is the second-mover response to $\mathrm{D}$. Thus, each subject contributes one strategy $x_i$, and the outcome space consists of $I=8$ distinct strategies. For example, $\mathrm{CCD}$ represents cooperation as a first mover followed by conditional cooperation as a second mover. The sample consists of 96 subjects.

We evaluate five models of social preferences drawn from \cite{Miettinen2020}, summarized in Table \ref{tab:models_summary}.\footnote{We omit the Reciprocity model because it is observationally equivalent to the Conditional Welfare (CW) model in our analysis.} Consider the benchmark model \textit{Homo Oeconomicus} (HO). Since HO is a model of selfish preferences, it always plays $\mathrm{D}$ as second mover. For a first mover, however, HO may predict $\mathrm{C}$ or $\mathrm{D}$, depending on the agent's beliefs about second mover behavior. For instance, if the agent believes the second mover is a conditional cooperator (who plays $\mathrm{C}$ in response to $\mathrm{C}$ and $\mathrm{D}$ in response to $\mathrm{D}$), then playing $\mathrm{C}$ maximizes the expected material payoff of a first mover, so HO prescribes $\mathrm{C}$.

In their experiment, Miettinen et al.\ elicit subjects' beliefs and derive belief-contingent predictions for each model. To simplify the exposition, however, we assume first movers believe that a second mover cooperates with probability $0.47$ after first-mover cooperation and $0.16$ after first-mover defection, which correspond exactly to the empirical frequencies of second-mover cooperation in the sample.

As noted in Section \ref{sec:selten}, Selten's framework is limited to area theories, which is a natural way to formulate the deterministic models considered in this section. However, a much more common approach is to augment these models with a stochastic error structure and to estimate parameters via maximum likelihood. In this vein, Miettinen et al. propose a standard random utility specification in which choice probabilities at any node of the game are given by the logistic function
\begin{equation}
\label{eq:logit} P(X=C)=\frac{e^{\tau u_C}}{e^{\tau u_C}+e^{\tau u_D}},
\end{equation}
where $u_C$ and $u_D$ are the DM's utility from playing $\mathrm{C}$ and $\mathrm{D}$ at that node, respectively, and $\tau > 0$ is an additional parameter governing the noisiness or sensitivity of the choice function.

\begin{table}[tb]
\caption{Overview of the social preference models in \cite{Miettinen2020}.}
\centering
\resizebox{\textwidth}{!}{
\begin{tabular}{@{}lllc@{}}
\toprule
Model & Abbr. & Utility Function & Parameter Restrictions \\
\midrule
Homo Oeconomicus & HO & $U(s) = \pi_i(s)$ & -- \\
\addlinespace[1.5ex]
Altruism & AL & $U(s) = \pi_i(s) + \theta \cdot \pi_j(s)$ & $\theta \in (0, 1)$ \\
\addlinespace[1.5ex]
Homo Moralis & HM & $U(x, y) = (1 - \kappa)\pi(x, y) + \kappa\pi(x, x)$ & $\kappa \in [0, 1]$ \\
\addlinespace[1.5ex]
Inequity Aversion & IA & $U(s) = \begin{cases} \makebox[12.5em][l]{$\pi_i(s) - \alpha \cdot [\pi_j(s) - \pi_i(s)]$} & \text{if } \pi_i(s) \le \pi_j(s) \\ \makebox[12.5em][l]{$\pi_i(s) - \beta \cdot [\pi_i(s) - \pi_j(s)]$} & \text{if } \pi_i(s) > \pi_j(s) \end{cases}$ & $\alpha \ge \beta \ge 0$ \\
\addlinespace[1.5ex]
Conditional Welfare & CW & $U(s) = \begin{cases} \makebox[12.5em][l]{$(1 - \rho)\pi_i(s) + \rho\pi_j(s)$} & \text{if } \pi_i(s) \ge \pi_j(s) \\ \makebox[12.5em][l]{$(1 - \sigma)\pi_i(s) + \sigma\pi_j(s)$} & \text{if } \pi_i(s) < \pi_j(s) \end{cases}$ & $\begin{array}{@{}c@{}} 0 \le \sigma \le 1/2 \\ \sigma \le \rho \le 1 \end{array}$ \\
\bottomrule
\end{tabular}
}
\label{tab:models_summary}
\par
\raggedright
\footnotesize \textit{Note:} $\pi_i(s)$ denotes the standard material payoff given the pure strategy profile $s=(s_1, s_2)$. For the Homo Moralis model, $\pi(x, y)$ represents the ex-ante expected material payoff in the symmetrized game if a player uses strategy $x$ against an opponent using $y$.
\end{table}

\subsection{Results}\label{sec:social_results}

Recall that predictive success evaluates a model by the difference between its hit rate and its area. The area is the fraction of the outcome space compatible with the model, while the hit rate is the proportion of observations that fall into its area. For example, HO is only compatible with 1 of the 8 possible outcomes ($\mathrm{CDD}$), so its area is $1/8$. For this model, a subject that chooses $\mathrm{CDD}$ counts as a hit, while any other outcome counts as a miss.

\begin{table}[tb]
\caption{Predictive Success and Compression Gain of social preferences models.}
\centering
\resizebox{\textwidth}{!}{
\begin{tabular}{@{} l c@{\hspace{4pt}}c@{\hspace{4pt}}c@{\hspace{10pt}} c@{\hspace{4pt}}c@{\hspace{4pt}}c@{\hspace{10pt}} c@{\hspace{4pt}}c@{\hspace{4pt}}c@{\hspace{10pt}} c@{\hspace{4pt}}c@{\hspace{4pt}}c@{\hspace{10pt}} c@{\hspace{4pt}}c@{\hspace{4pt}}c @{}}
\toprule
 & \multicolumn{3}{c@{\hspace{10pt}}}{HO ($k=0$)} & \multicolumn{3}{c@{\hspace{10pt}}}{AL ($k=1$)} & \multicolumn{3}{c@{\hspace{10pt}}}{HM ($k=1$)} & \multicolumn{3}{c@{\hspace{10pt}}}{IA ($k=2$)} & \multicolumn{3}{c}{CW ($k=2$)} \\
\cmidrule(lr){2-4} \cmidrule(lr){5-7} \cmidrule(lr){8-10} \cmidrule(lr){11-13} \cmidrule(lr){14-16}
Outcome & Hit & $\widehat{P}_\mathcal{A}$ & $P_{\hat{\theta}}$ & Hit & $\widehat{P}_\mathcal{A}$ & $P_{\hat{\theta}}$ & Hit & $\widehat{P}_\mathcal{A}$ & $P_{\hat{\theta}}$ & Hit & $\widehat{P}_\mathcal{A}$ & $P_{\hat{\theta}}$ & Hit & $\widehat{P}_\mathcal{A}$ & $P_{\hat{\theta}}$ \\
\midrule
CCC ($n=7$)  & 0 & 0.125 & 0.125 & 1 & 1.000 & 1.000 & 1 & 1.000 & 0.500 & 0 & 0.125 & 0.125 & 1 & 1.000 & 1.000 \\
CCD ($n=33$) & 0 & 0.125 & 0.125 & 0 & 0.125 & 0.125 & 1 & 1.000 & 1.000 & 0 & 0.125 & 0.221 & 1 & 1.000 & 1.000 \\
CDC ($n=1$)  & 0 & 0.125 & 0.171 & 1 & 1.000 & 1.000 & 0 & 0.125 & 0.171 & 0 & 0.125 & 0.171 & 1 & 1.000 & 1.000 \\
CDD ($n=13$) & 1 & 1.000 & 1.000 & 1 & 1.000 & 1.000 & 1 & 1.000 & 1.000 & 1 & 1.000 & 1.000 & 1 & 1.000 & 1.000 \\
DCC ($n=2$)  & 0 & 0.125 & 0.125 & 0 & 0.125 & 0.125 & 0 & 0.125 & 0.190 & 0 & 0.125 & 0.125 & 0 & 0.125 & 0.154 \\
DCD ($n=3$)  & 0 & 0.125 & 0.125 & 0 & 0.125 & 0.125 & 0 & 0.125 & 0.125 & 1 & 1.000 & 1.000 & 0 & 0.125 & 0.187 \\
DDC ($n=5$)  & 0 & 0.125 & 0.164 & 0 & 0.125 & 0.166 & 0 & 0.125 & 0.468 & 0 & 0.125 & 0.164 & 0 & 0.125 & 0.166 \\
DDD ($n=32$) & 0 & 0.125 & 0.410 & 0 & 0.125 & 0.410 & 0 & 0.125 & 0.410 & 1 & 1.000 & 1.000 & 0 & 0.125 & 0.410 \\
\midrule
Sum & 1 & 1.875 & 2.244 & 3 & 3.625 & 3.951 & 3 & 3.625 & 3.864 & 3 & 3.625 & 3.805 & 4 & 4.500 & 4.917 \\
\midrule
\multicolumn{16}{@{}l}{\textbf{Overall} ($n=96$)} \\ [1ex]
${\operatorname{COMP}}$ & -- & 0.629 & 0.808 & -- & 1.288 & 1.374 & -- & 1.288 & 1.352 & -- & 1.288 & 1.336 & -- & 1.504 & 1.593 \\
Area & 0.125 & 0.234 & -- & 0.375 & 0.453 & -- & 0.375 & 0.453 & -- & 0.375 & 0.453 & -- & 0.500 & 0.562 & -- \\
Hit rate & 0.135 & 0.135 & -- & 0.219 & 0.219 & -- & 0.552 & 0.552 & -- & 0.500 & 0.500 & -- & 0.562 & 0.562 & -- \\
$m$ & 0.010 & -- & -- & -0.156 & -- & -- & \textbf{0.177} & -- & -- & 0.125 & -- & -- & 0.062 & -- & -- \\
$g$ & -- & -0.167 & -0.055 & -- & -0.401 & -0.244 & -- & \textbf{-0.067} & \textbf{0.107} & -- & -0.119 & -0.040 & -- & -0.161 & 0.002 \\ [0.5ex]
$\bar{L}$ & -- & 2.426 & 2.193 & -- & 2.912 & 2.588 & -- & \textbf{2.219} & \textbf{1.857} & -- & 2.328 & 2.163 & -- & 2.414 & 2.075 \\
\bottomrule
\end{tabular}
}
\par
\raggedright
\footnotesize \textit{Note:} The top panel reports compatibility (Hit) alongside likelihoods for the area-based ($\widehat{P}_\mathcal{A}$) and logit-augmented ($P_{\hat{\theta}}$) specifications for each observed choice pattern. The row ``Sum'' aggregates these values over the eight patterns. The bottom panel evaluates performance over the $96$ subjects. Area represents $a$ under Selten's definition and $\tilde{a}$ under the compression analogue, both obtained by dividing the corresponding column sum by 8. Predictive success ($m$) applies only to the Selten framework, while compression gain ($g$), complexity (${\operatorname{COMP}}$), and average description length ($\bar{L}$) apply to the compression framework. Bold values indicate the preferred model for the respective selection criterion.
\label{tab:social_results}
\end{table}

Table \ref{tab:social_results} displays the results. For each model, the column ``Hit'' takes value 1 if an outcome is compatible with the model and 0 otherwise. Since HO is only compatible with $\mathrm{CDD}$, the first column has value 1 for that outcome and 0 for all others. AL, for instance, is compatible with 3 outcomes ($\mathrm{CCC}, \mathrm{CDC},\mathrm{CDD}$). Columns $\widehat{P}_\mathcal{A}$ and $P_{\hat{\theta}}$ present the maximized likelihoods under the area-based likelihood function \eqref{eq:P_A} (with miss penalty $C=I=8$) and the random utility specification, respectively. Note that $P_{\hat{\theta}}(x)=1$ for most choice patterns that count as hits ($x \in \mathcal{A}$). In these cases the random utility model achieves perfect fit by setting a sufficiently high sensitivity parameter $\hat{\tau}$, which essentially collapses it into a deterministic model.\footnote{The only exception to this is unconditional cooperation ($\mathrm{CCC}$) in HM. The ML parameter estimate for unconditional cooperation in HM is $\hat{\kappa} = 1$, but for this value of $\hat{\kappa}$ the DM is indifferent between $\mathrm{CCC}$ and $\mathrm{CCD}$, so the logit choice function assigns probability $0.5$ to $\mathrm{CCC}$.} By contrast, while all misses ($x \notin \mathcal{A}$) are equal in Selten's framework, and thus receive equal likelihood under $\widehat{P}_\mathcal{A}$, this is no longer the case for the logit-augmented models. For example, $\mathrm{CCC}$ counts as a much more severe miss for HO ($P_{\hat{\theta}}=0.125$) than $\mathrm{DDD}$ ($P_{\hat{\theta}}=0.410$).

The bottom panel of Table \ref{tab:social_results} displays the sample statistics used to compute predictive success and compression gain. The first row displays complexity, ${\operatorname{COMP}}$, which is simply the logarithm of the sum of the 8 likelihoods in the top panel. For each model, the first number is the complexity of the area-based specification and the second number is the complexity of the random utility specification. The second row displays the model's area: the first number corresponds to Selten's area, $a$, which equals the sum of outcomes compatible with the model divided by 8, and the second number corresponds to the error-adjusted area, $\tilde a$, which equals the sum of likelihoods $\widehat{P}_\mathcal{A}$, divided by 8. The third row displays the hit rate, $r$, the fraction of subjects whose observed choice is compatible with the model. For example, HO has a hit rate of $13/96=0.135$ because 13 of the 96 subjects play $\mathrm{CDD}$, the only outcome compatible with HO.

The last three rows display predictive success, $m = r - a$, compression gain, $g$ and average description length per observation, $\bar{L} = \frac{1}{n}L_{\text{NML}}$. For the area-based specification, compression gain is $g=r-\log(I\tilde a)/\log I$, as defined in \eqref{eq:m_anml}. For the random utility specification, $g = (\log I - \bar L)/\log I$.

There are a few results worth noting from Table \ref{tab:social_results}. First, and consistent with the motivating example in Table \ref{tab:hyp_risk1}, the number of free parameters $k$ is a poor proxy for model complexity. AL and HM have only one parameter, yet they can accommodate the same number of outcomes as the two-parameter IA model, which in turn admits fewer outcomes than CW despite having the same number of parameters. Second, predictive success and compression gain agree in selecting HM as the winning model for this sample. This is not too surprising: HM combines a near-maximal hit rate with the smallest area among the models that fit more than one outcome. However, while HM's positive predictive success ($m=0.177$) may suggest good model performance, the compression gain of its area-based analogue is negative ($g=-0.067$), indicating that HM leads to worse compression than the zero-compression uniform code. In statistical terms, this means that a uniform-random distribution provides a closer approximation to the empirical data, in terms of KL divergence, than the NML distribution induced by the area-based HM model.

While the area-based models consistently fail to beat the zero-compression benchmark, this is no longer true for the random utility specifications. Even though these models are slightly more complex than their area counterparts, the increase in complexity is more than compensated for by improvements in fit. Specifically, every random utility model achieves a lower description length (and therefore a higher compression gain) than its area counterpart. The improvements are substantial: while all area models obtain negative compression gain, the random utility specification achieves a $\approx10\%$ compression gain for HM and a marginal compression gain for CW ($\approx0.2\%$).

\begin{figure}[tb]
    \centering
    \includegraphics[width=\textwidth]{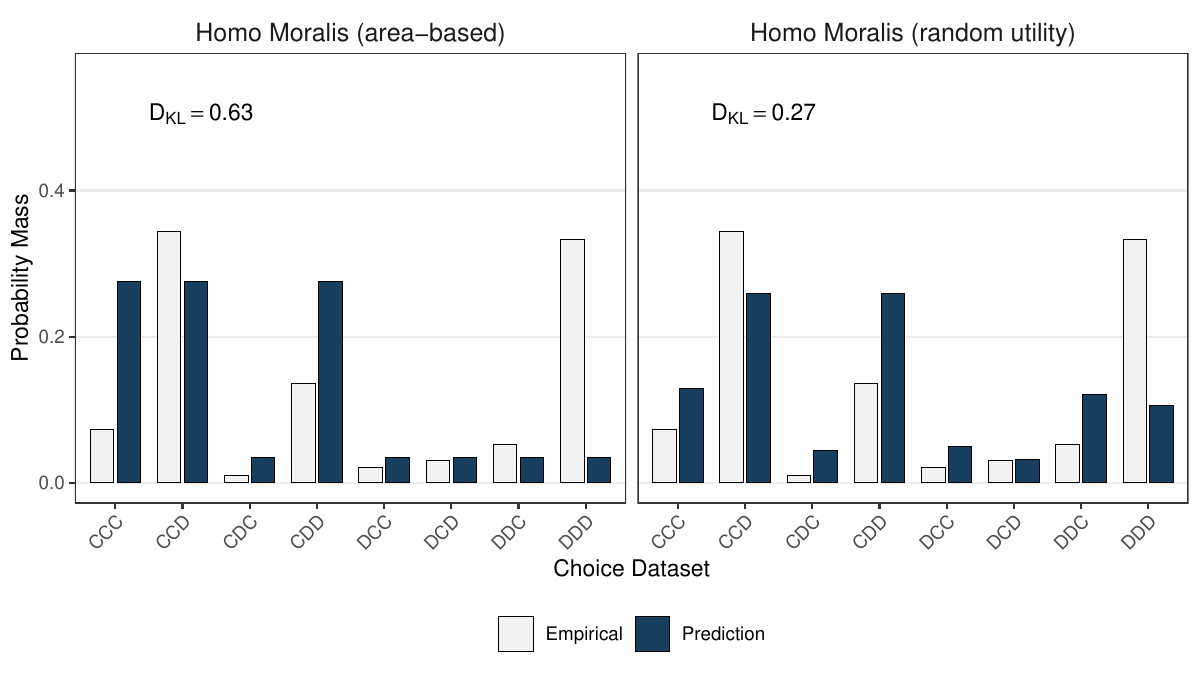}
    \caption{Comparison of empirical choice frequencies and the NML distributions induced by the area-based (left) and logit-augmented (right) specifications of the Homo Moralis model. The values reported inside each panel are the KL divergences from the empirical distribution to the corresponding NML distribution.}    \label{fig:HM_comparison}
\end{figure}

To illustrate this result, Figure \ref{fig:HM_comparison} displays the NML distributions induced by the area-based and random utility specifications of HM alongside the empirical frequencies. As noted earlier, the area-based HM model fits the empirical data rather poorly, as indicated by its negative compression gain. This is reflected in a large KL divergence from the empirical distribution (left panel). By contrast, the random utility HM specification achieves a positive compression gain, which is reflected in a lower KL divergence from the empirical distribution (right panel).

Overall, the application highlights the broader advantage of data compression as a framework for model selection. Selten's approach improves on parameter-counting criteria by measuring flexibility in terms of the outcomes a model can accommodate, rather than the number of free parameters. At the same time, it imposes a coarse measure of goodness of fit by treating all violations equally. Likelihood-based approaches have the opposite strength: they preserve information about the severity of misses, but are usually paired with complexity penalties based on parameter counts, as in AIC or BIC. The compression approach retains the strengths of both methods by accounting for both model complexity and goodness of fit in a unified likelihood framework. Moreover, by reformulating models as codes, the compression approach can be applied to standard stochastic models as well as Selten-style area theories, making the two directly comparable.

\section{Application 2: Risk Preferences}
\label{sec:risk}

This section returns to the motivating example from the introduction and studies models of choice under risk. The application emphasizes two sources of complexity. The first, already familiar, is functional form: models with the same number of parameters can generate very different ranges of observable choice patterns. The second is experimental design: the flexibility implied by a given functional form depends on the set of observations over which it is evaluated. Accounting for this design dependence makes it possible to diagnose experimental imbalances that may produce asymmetric model recovery and to adjust model selection accordingly.

\subsection{Model Framework}

Following the example in Table \ref{tab:hyp_risk1}, consider a standard setting in which a decision maker reports certainty equivalents (CEs) for binary lotteries $L=(p,x;1-p,z)$ with $x>z$. Preferences are represented by a value function $v(\cdot)$ and a probability weighting function $w(\cdot)$, yielding the predicted certainty equivalent
\begin{equation}
\hat{ce}(L)=v^{-1}\left[w(p)v(x)+(1-w(p))v(z)\right]. \label{eq:ce_hat}
\end{equation}
Observed CEs are assumed to satisfy $ce(L)\sim \mathcal{N}(\hat{ce}(L),[\sigma(x-z)]^2)$.

Most of the models considered in this section are nested within a common parameterization of Cumulative Prospect Theory (CPT), combining a piecewise power value function with a two-parameter weighting function. The value function is
\begin{equation}
v(y)=
    \begin{cases}
        y^\alpha & \text{if } y\geq 0,\\
        -(-y)^\beta & \text{if } y<0,
    \end{cases}
    \label{eq:power_value}
\end{equation}
where $\alpha,\beta\in[0,1]$ govern curvature for gains and losses. Since neither the hypothetical experiments nor the empirical application include mixed lotteries, we omit the usual loss-aversion parameter $\lambda$, which is not identified in this setting. The loss-curvature parameter $\beta$ is irrelevant for the gain-only simulations below, but is retained for the empirical application, which includes both gain-only and loss-only lotteries.

Probability weighting follows the two-parameter \cite{Prelec1998} form
\begin{equation}
w(p)=\exp\left(-\delta(-\ln p)^\gamma\right), \label{eq:prelec}
\end{equation}
where $\delta>0$ and $\gamma\in[0,1]$ govern elevation and curvature, respectively. The unrestricted model is indexed by $\theta=(\alpha,\beta,\gamma,\delta)$.

\subsection{Simulations}
\label{sec:risk_sim}

This analysis illustrates how model complexity is jointly determined by functional form and experimental design. In the simulations below, changing the design can even reverse the complexity ranking of competing models. Let experiment $i$ be defined by the tuple $(P_i,Q_i)$, which generates the set of lotteries obtained by combining every probability $p\in P_i$ with every pair of outcomes $x,z\in Q_i$ such that $x>z$. The three designs each yield $50$ lotteries:

	\begin{itemize}
		\item {Experiment 1 (Balanced):} $Q_1=\{0, .25, .5, .75, 1\}$ and $P_1=\{.1, .3, .5, .7, .9\}$.

    	\item {Experiment 2 (Narrow Payoffs):} $Q_2=\{.3, .4, .5, .6, .7\}$ and $P_2=P_1$.
\item {Experiment 3 (Narrow Probabilities):} $Q_3=Q_1$ and $P_3=\{.32, .34, .37, .40, .42\}$.
\end{itemize}

Experiment 1 is a balanced baseline spanning the full range of payoffs and probabilities, and corresponds to the design used in Table \ref{tab:hyp_risk1}. Experiment 2 keeps the probability set fixed but restricts payoffs to a narrower interval. This makes choices less sensitive to changes in utility curvature, limiting the range of patterns generated by different values of $\alpha$. Experiment 3 mirrors this logic for probability weighting: probabilities are concentrated around $p=1/e\approx 0.37$, the fixed point of the one-parameter Prelec specification obtained when $\delta=1$, where variation in $\gamma$ has little effect on choices.

We compare two one-parameter restrictions of the general CPT model. Since payoffs are always non-negative, the loss-curvature parameter $\beta$ is irrelevant. The first is \textit{Expected Utility}, $\mathcal{M}_{\alpha}$, obtained by setting $\delta=\gamma=1$, so that $w(p)=p$ and risk attitudes are driven solely by utility curvature, $v(x)=x^\alpha$. The second is the \textit{Dual Theory}, $\mathcal{M}_{\gamma}$, obtained by setting $\alpha=\delta=1$, so that utility is linear and risk attitudes are driven solely by Prelec I probability weighting through $\gamma$.

For each experiment-model pair, we simulate 10{,}000 datasets, drawing the free parameter of the true model uniformly from $[0.2,0.8]$ and fixing $\sigma=0.1$. Table \ref{tab:risk_sim} reports the results. The third column displays the asymptotic complexity approximation (COMP), computed using \eqref{eq:rissanen} (for details on the COMP calculation and robustness checks, see Sections~1.2 and~2.2 of the Online Appendix). As expected, complexity varies sharply across designs. In the balanced design (Experiment 1), the two models have similar complexity. In Experiment 2, $\mathcal{M}_\alpha$ is much less complex than $\mathcal{M}_\gamma$, while the reverse holds in Experiment 3. These differences translate directly into model selection performance.

\begin{table}[tb]
    \centering
    \caption{Model selection performance across simulated risky-choice experiments}
    \label{tab:risk_sim}
    \begin{tabular*}{\textwidth}{@{\extracolsep{\fill}}lcccccc@{}}
        \toprule
        & & & \multicolumn{4}{c}{\textbf{Recovery Rates}} \\
        \cmidrule(l){4-7}
        \textbf{Experiment} & \textbf{True $\mathcal{M}$} & \textbf{COMP}
        & \textbf{MDL} & \textbf{AIC} & \textbf{BIC} & \textbf{CV} \\
        \midrule
        \multirow{2}{*}{\textbf{1. Balanced}}
        & $\mathcal{M}_{\alpha}$ & 2.50 & 95\% & 98\% & 98\% & 97\% \\
        & $\mathcal{M}_{\gamma}$ & 2.18 & \makebox[0pt][r]{$>$}99\% & 98\% & 98\% & 97\% \\
        \addlinespace
        \multirow{2}{*}{\textbf{2. Narrow Payoffs}}
        & $\mathcal{M}_{\alpha}$ & 0.19 & 99\% & 88\% & 88\% & 56\% \\
        & $\mathcal{M}_{\gamma}$ & 2.18 & 94\% & 96\% & 96\% & 97\% \\
        \addlinespace
        \multirow{2}{*}{\textbf{3. Narrow Probabilities}}
        & $\mathcal{M}_{\alpha}$ & 1.97 & 96\% & 98\% & 98\% & 99\% \\
        & $\mathcal{M}_{\gamma}$ & 0.04 & \makebox[0pt][r]{$>$}99\% & 88\% & 88\% & 49\% \\
        \bottomrule
    \end{tabular*}
    \begin{flushleft}
    \footnotesize
    \textit{Note:} COMP reports model complexity computed using the asymptotic
    approximation in \eqref{eq:rissanen}, integrating over
    $\alpha,\gamma\in[0,1]$. Recovery rates are the percentage of simulated
    datasets in which each criterion selects the true model, based on 10{,}000
    simulations for each experiment-model pair. CV is five-fold.
    \end{flushleft}
\end{table}

In Experiment 1, where complexity differences are small, all methods recover the true model with high accuracy. In Experiments 2 and 3, however, recovery under AIC and BIC becomes asymmetric. Because both models have one parameter, these criteria impose equal penalties and rank them entirely by fit. When the less complex model is true, the more flexible rival is incorrectly selected in approximately 12\% of simulations, whereas the reverse error occurs in 2--4\% of cases.

One might expect CV to reduce this asymmetry, since it guards against overfitting by evaluating out-of-sample prediction. In these designs, however, the asymmetry becomes substantially more pronounced: CV recovers the less complex model in only 56\% and 49\% of the simulations in each of the narrow experiments, compared with a 97--99\% recovery when the more complex model is true. This pattern can be understood through the bias--variance tradeoff. When the less complex model is true, it has no misspecification bias but is weakly identified because the design contains little variation along the dimension governed by its parameter. Its estimates are therefore highly sensitive to noise in the training sample, generating high-variance predictions. By contrast, the more complex but misspecified rival can exploit the richer variation along the other dimension to produce more stable predictions. In these simulations, this variance advantage often outweighs its misspecification bias, leading CV to select the rival model.

By explicitly accounting for design-induced differences in complexity, MDL substantially reduces asymmetric model recovery, selecting the true model in at least 94\% of simulations across all designs. Beyond improving selection, COMP helps explain and predict the failures of the other methods. Large complexity differences predict both the emergence and direction of asymmetric recovery, with AIC, BIC, and CV tending to favor the more complex specification. Because COMP can be calculated before collecting data, it provides an ex ante diagnostic for potential experimental design imbalance, indicating when standard methods are likely to produce asymmetric model selection.

\subsection{Real Data}
\label{sec:risk_real}

This section performs a similar model selection exercise using real choice data from \citet{Bruhin2010} (henceforth BFE).\footnote{The analysis uses the \textit{Zurich '03} wave, the largest and most comprehensive subsample in the study.} The sample consists of 179 subjects, each providing certainty equivalents for 50 binary lotteries (25 gain-framed and 25 loss-framed). Models are estimated and compared separately for each subject.

\subsubsection*{Model Specifications}

Two families of models are considered. The first is Expected Utility, denoted $\mathcal{M}_{\alpha}^{\text{EU}}$, which combines a power utility function over final wealth, $u(w)=w^\alpha$, with linear probability weighting. This model assumes asset integration and therefore ignores the gain/loss framing of lotteries. The second family consists of reference-dependent Prospect Theory models, in which gains and losses are evaluated relative to a status quo. The maximal model is $\mathcal{M}_{\alpha,\beta,\delta,\gamma}$, with two value-function parameters ($\alpha$ for gains and $\beta$ for losses) and two probability-weighting parameters ($\delta$ for elevation and $\gamma$ for sensitivity).\footnote{Throughout the analysis, payoffs are normalized to the $[0,1]$ interval for gains and to the $[0,-1]$ interval for losses.} We evaluate this unrestricted model together with the relevant nested submodels, obtained either by fixing parameters to unity (e.g., $\mathcal{M}_{\gamma}$ imposes $\alpha=\beta=\delta=1$) or by imposing equality restrictions (e.g., $\mathcal{M}_{\alpha=\beta}$). The fully restricted case, $\alpha=\beta=\delta=\gamma=1$, corresponds to Expected Value, $\mathcal{M}^{\text{EV}}$, which serves as our minimal baseline.

In the simulations of Section \ref{sec:risk_sim}, the noise parameter $\sigma$ was fixed. In empirical applications, however, it is typically estimated jointly with the structural parameters. This poses a complication for MDL because allowing $\sigma$ to become arbitrarily small makes model complexity infinite. With certainty equivalents and parameter values treated as continuous, a model can place an arbitrarily narrow probability density around continuously varying predictions. The resulting set of distinguishable fitted distributions becomes unbounded as $\sigma$ approaches zero. A finite complexity penalty therefore requires a positive lower bound on $\sigma$.

The BFE elicitation procedure provides a natural empirical basis for this lower bound. Certainty equivalents are elicited by choosing among 20 equally spaced sure amounts between the lower and upper payoff of each lottery, so observations are recorded in increments of size $\Delta=(x-z)/20$. It is therefore not meaningful to attribute to models a precision beyond this resolution.\footnote{In other applications, the bound might instead be dictated by intrinsic limits of the data-generating process rather than by the limited precision of the recording device. For instance, in sensory domains, physiological constraints on discrimination (e.g., just-noticeable differences in pitch or luminance) impose a natural limit on precision.} Treating the latent certainty equivalent as uniformly distributed within the chosen interval implies a standard deviation of $\Delta/\sqrt{12}$. Matching this quantity to the Gaussian noise term, $ce(L)\sim\mathcal{N}(\hat{ce}(L),[\sigma(x-z)]^2)$, yields $\underline{\sigma}=(20\sqrt{12})^{-1}\approx 0.014$ (for details, see Section~1.3.1 of the Online Appendix).

\subsubsection*{Results}

\begin{figure}[tb]
    \centering
    \begin{tikzpicture}
        \begin{axis}[
            width=0.82\textwidth,
            height=0.52\textwidth,
            xmin=-0.1, xmax=4.2,
            ymin=0, ymax=15.5,
            ytick={0,5,10,15},
            xlabel={Structural Parameters ($k$)},
            ylabel={Effective Penalty (relative to EV)},
            xtick={0,1,2,3,4},
            ytick={0,5,10,15,20},
            axis lines=left,
            ymajorgrids=true,
            grid style={dashed, gray!25},
            legend style={
                draw=none,
                fill=none,
                font=\small,
                at={(0.03,0.97)},
                anchor=north west
            },
            tick label style={font=\small},
            label style={font=\small},
            clip=false
        ]

\addplot[domain=0:4.2, black, dashed, thick] {x};
        \addlegendentry{AIC}

        \addplot[domain=0:4.2, black!60, dotted, thick] {x * ln(50) / 2};
        \addlegendentry{BIC}

\addplot[
            only marks,
            mark=*,
            mark size=2.2pt,
            mdlblue
        ] coordinates {
            (0,0)
            (1,3.10)
            (1,3.10)
            (1,2.80)
            (2,6.23)
            (2,6.29)
            (2,7.02)
            (3,9.41)
            (3,10.65)
            (4,14.30)
        };

\node[font=\scriptsize, anchor=south] at (axis cs:0.1,0.3)
            {$\mathcal{M}_{\mathrm{EV}}$};

\node (riskAlphaEU) [font=\scriptsize, anchor=south east] at (axis cs:0.90,3.55)
            {$\mathcal{M}_{\alpha}^{\mathrm{EU}}$};
        \draw[black, thin, shorten <=2.2pt] (axis cs:1,3.10) -- (riskAlphaEU);
        \node (riskAlphaBetaEq) [font=\scriptsize, anchor=south west] at (axis cs:1.12,3.55)
            {$\mathcal{M}_{\alpha=\beta}$};
        \draw[black, thin, shorten <=2.2pt] (axis cs:1,3.10) -- (riskAlphaBetaEq);
        \node (riskGamma) [font=\scriptsize, anchor=east] at (axis cs:0.88,2.98)
            {$\mathcal{M}_{\gamma}$};
        \draw[black, thin, shorten <=2.2pt] (axis cs:1,2.80) -- (riskGamma);

\node (riskAlphaGamma) [font=\scriptsize, anchor=west] at (axis cs:2.14,5.98)
            {$\mathcal{M}_{\alpha,\gamma}$};
        \draw[black, thin, shorten <=2.2pt] (axis cs:2,6.23) -- (riskAlphaGamma);
        \node (riskAlphaBeta) [font=\scriptsize, anchor=east] at (axis cs:1.90,6.36)
            {$\mathcal{M}_{\alpha,\beta}$};
        \draw[black, thin, shorten <=2.2pt] (axis cs:2,6.29) -- (riskAlphaBeta);
        \node[font=\scriptsize, anchor=south west] at (axis cs:2.04,7.08)
            {$\mathcal{M}_{\delta,\gamma}$};

        \node[font=\scriptsize, anchor=west] at (axis cs:3.04,9.41)
            {$\mathcal{M}_{\alpha,\beta,\gamma}$};
        \node[font=\scriptsize, anchor=south east] at (axis cs:2.96,10.72)
            {$\mathcal{M}_{\alpha,\delta,\gamma}$};

        \node[font=\scriptsize, anchor=east] at (axis cs:4,14.30)
            {$\mathcal{M}_{\alpha,\beta,\delta,\gamma}$};

        \end{axis}
    \end{tikzpicture}
    \caption{Complexity penalties in the \citet{Bruhin2010} experiment, normalized by subtracting the EV penalty. Points show MDL complexity; dashed and dotted lines show the corresponding AIC and BIC penalties after division by two (AIC: $k$; BIC: $\tfrac{k}{2}\log 50$). The horizontal axis reports the number $k$ of structural parameters; all models additionally include the noise parameter $\sigma$.}    \label{fig:risk_pen}
\end{figure}
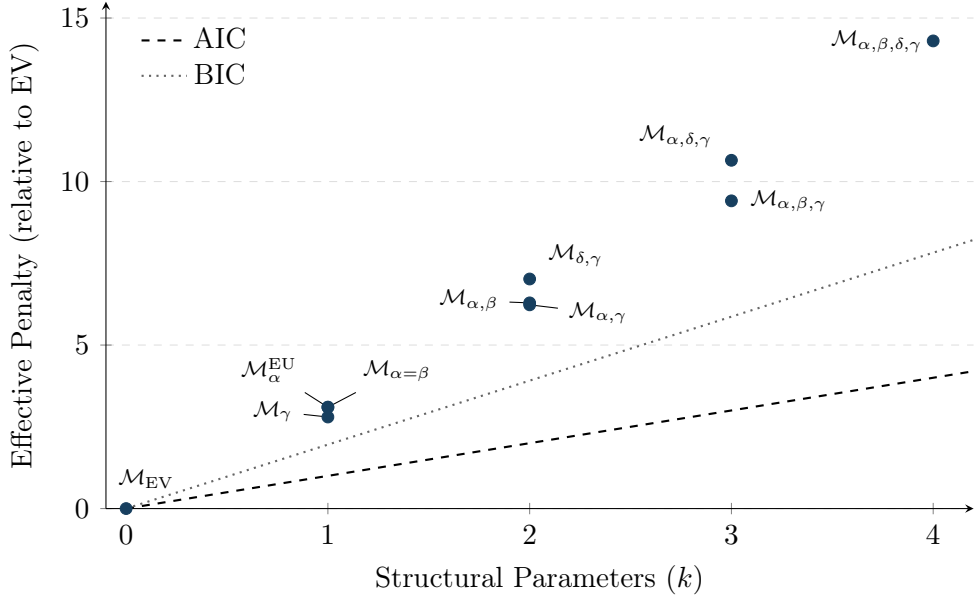

Figure \ref{fig:risk_pen} compares the complexity penalties implied by MDL, AIC, and BIC, normalized relative to the EV baseline. As expected, AIC and BIC impose penalties that are linear in parameter count, whereas MDL imposes substantially larger penalties and, crucially, distinguishes between models with the same number of parameters.

In this dataset, adding the elevation parameter $\delta$ to the probability-weighting function incurs a larger complexity penalty than adding a loss-specific curvature parameter $\beta$ to the value function. This indicates that variation in probability-weighting elevation affords greater flexibility than variation in loss-specific utility curvature under the experimental design considered here.

Table \ref{tab:risk_sbj} shows how these complexity differences translate into subject-level model selection. Across all criteria, most subjects are assigned to models with probability weighting parameters, consistent with the broader experimental evidence that probability weighting often plays a larger role than value curvature in risky choice. The distribution across specific models nevertheless varies considerably. MDL is the most conservative, assigning the largest share of subjects to the one-parameter weighting model $\mathcal{M}_{\gamma}$. AIC and CV are the most permissive, both favoring the fully parameterized prospect theory variant, while BIC occupies an intermediate position, most often selecting the two-parameter weighting model $\mathcal{M}_{\delta,\gamma}$.

Taken together, these results show that a fuller account of model complexity can shift model selection toward simpler specifications, favoring parsimonious probability-weighting models over more flexible Prospect Theory variants that also incorporate value-function curvature.

\begin{table}[tb]
    \centering
    \caption{Subject-level model classifications in the \citet{Bruhin2010} experiment}
    \label{tab:risk_sbj}
    \resizebox{\textwidth}{!}{\begin{tabular}{lcccccccccc}
            \toprule
            \textbf{Model} & $\mathcal{M}^{\text{EV}}$ & $\mathcal{M}_{\gamma}$ & $\mathcal{M}_{\alpha=\beta}$ & $\mathcal{M}_{\alpha}^{\text{EU}}$ & $\mathcal{M}_{\alpha, \gamma}$ & $\mathcal{M}_{\alpha, \beta}$ & $\mathcal{M}_{\delta, \gamma}$ & $\mathcal{M}_{\alpha, \beta, \gamma}$ & $\mathcal{M}_{\alpha, \delta, \gamma}$ & $\mathcal{M}_{\alpha, \beta, \delta, \gamma}$ \\
            \midrule
            \textbf{COMP} & 2.83 & 5.63 & 5.93 & 5.93 & 9.06 & 9.12 & 9.85 & 12.24 & 13.48 & 17.13 \\            \addlinespace
            \textit{\# Subjects} & & & & & & & & & & \\
            MDL & 9 & \textbf{86} & 0 & 11 & 0 & 1 & 56 & 0 & 1 & 15 \\
            AIC & 1 & 29 & 0 & 5 & 0 & 1 & 53 & 7 & 7 & \textbf{76} \\
            BIC & 4 & 49 & 0 & 9 & 0 & 0 & \textbf{70} & 1 & 2 & 44 \\
            CV & 4 & 26 & 0 & 7 & 2 & 0 & 43 & 9 & 10 & \textbf{78} \\
            \bottomrule
        \end{tabular}}
    \begin{flushleft}
    \footnotesize \textit{Note:} COMP reports model complexity computed using the asymptotic approximation in \eqref{eq:rissanen}, with parameter ranges $\alpha,\beta,\gamma\in[0,1]$, $\delta\in[0,\infty)$, and $\sigma\in[\underline{\sigma},1]$. The Fisher-information integral for $\delta$ is evaluated over the positive half-line, with finite truncation used only as a numerical fallback. The remaining rows report the number of subjects, out of 179, assigned to each model by each criterion. CV uses five folds within each subject. Numbers in bold indicate the most frequently selected model for each criterion.
    \end{flushleft}
\end{table}

\section{Application 3: Time Preferences}\label{sec:time}

Models of intertemporal choice provide a clean illustration of how specifications with the same number of parameters can differ substantially in the range of patterns they generate. Whereas complexity differences in the risky-choice simulations of Section \ref{sec:risk} were largely induced by the experimental design, here they arise from intrinsic differences in flexibility across functional forms. As in Application 2, this point is first demonstrated through simulations and then examined using real data. The empirical application shifts from individual-level estimation to a pooled representative-agent analysis.

\subsection{Model Framework}

Models of \textit{discounted utility} assume that preferences over temporal payment profiles $(x_0,\ldots,x_T)$ can be represented by a utility function of the form $U(x_0,\ldots,x_T)=\sum_{t=0}^T \lambda(t)u(x_t)$, where $u(x_t)$ is instantaneous utility and $\lambda(t)$ is the discount function.

We consider a standard time-discounting experiment that elicits subjects' present values for a given monetary payment at different delays. Formally, the present value $\hat{x}(t)$ of reward $x_t$ solves $u(\hat{x}(t))=\lambda(t)u(x_t)$. We consider three alternative specifications of the discount function:

\begin{alignat}{2}
    \lambda_1(t) & = (1 + \delta)^{-t}, & \qquad \delta > 0 \tag{Exponential} \label{eq:exp} \\
    \lambda_2(t) & = (1 + \gamma t)^{-\alpha/\gamma}, & \qquad \alpha, \gamma > 0 \tag{Hyperbolic} \label{eq:hyp} \\
    \lambda_3(t) & = e^{-(at)^b}, & \qquad a, b > 0 \tag{Constant sensitivity} \label{eq:cs}
\end{alignat}
\noeqref{eq:exp, eq:hyp, eq:cs}

The hyperbolic parameterization follows \citet{Loewenstein1992}, whereas the constant-sensitivity specification was introduced by \citet{Ebert2007}. Exponential discounting imposes a constant discount rate. Hyperbolic discounting allows decreasing impatience and converges to exponential discounting as $\gamma \to 0$. Constant sensitivity is more flexible: it reduces to exponential discounting when $b=1$, allows decreasing impatience for $b<1$, and increasing impatience for $b>1$. Thus, despite having the same number of parameters, constant sensitivity spans a broader range of discounting patterns than hyperbolic discounting.

\begin{figure}[tbp]
    \centering
    \includegraphics[width=\textwidth]{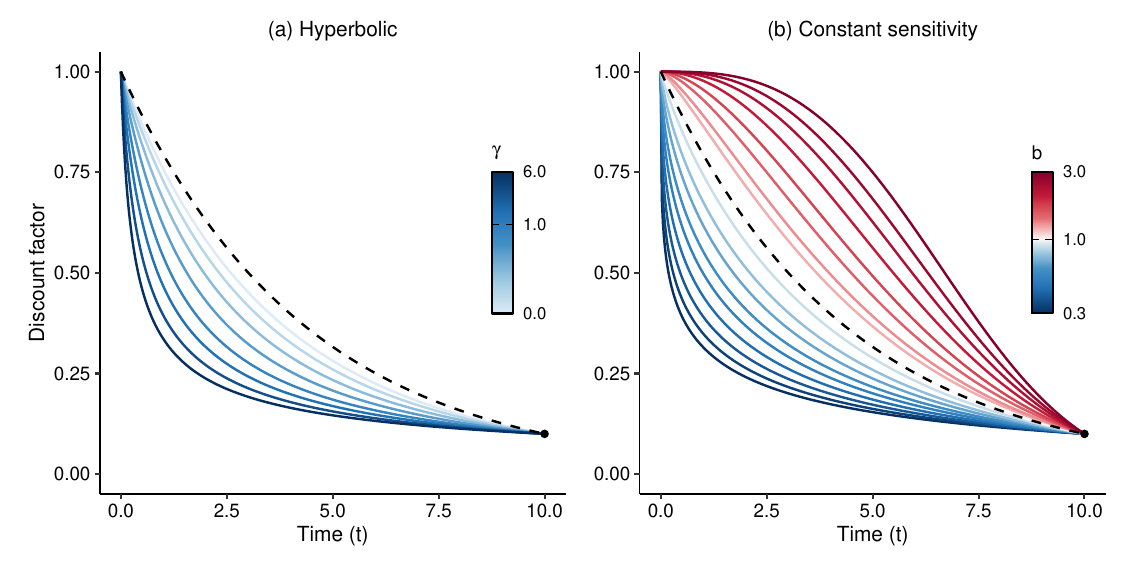}
    \caption{Functional forms of discounting models. Panel (a) displays the hyperbolic discount function for different values of $\gamma$. Panel (b) displays the constant sensitivity discount function for different values of $b$. In both panels, the dashed line represents the exponential benchmark, and all curves are anchored to a discount factor of $0.10$ at $t=10$.}
    \label{fig:discount_shapes}
\end{figure}

Figure \ref{fig:discount_shapes} illustrates this difference. Holding fixed the discount factor at $0.10$ for $t=10$, hyperbolic discounting spans only convex profiles associated with decreasing impatience, whereas constant sensitivity spans both convex and concave profiles, corresponding respectively to decreasing and increasing impatience.

\subsection{Simulations}

We simulate present values from the hyperbolic and constant sensitivity models and perform model selection using AIC, BIC, CV and MDL. For simplicity, we assume linear instantaneous utility and $x_t=1$ for all $t$, so that the present value is $\hat{x}(t)=\lambda(t)$. Observed valuations are generated for $t\in\{1,\dots,10\}$ by adding Gaussian noise, $x(t)=\lambda(t)+\epsilon$, where $\epsilon\sim\mathcal{N}(0,\sigma^2)$ with $\sigma=0.05$. To generate distinct but comparable patterns, we calibrate the hyperbolic model to produce moderate decreasing impatience and the constant sensitivity model to produce moderate increasing impatience, choosing the remaining parameters so that both imply the same discount factor at $t=10$.\footnote{Specifically, we set $\gamma=0.2$ and $b=1.4$. We then fix $\alpha=0.219$ and $a=0.114$ so that $\lambda_2(\alpha,\gamma,T)=\lambda_3(a,b,T)=0.3$.} Robustness checks varying $\gamma$, $b$, and $\sigma$ are reported in Section~2.3 of the Online Appendix.

Table \ref{tab:hyp_csen} reports the simulation results. As expected, constant sensitivity is more complex than hyperbolic discounting, even though both models have two parameters. Since AIC and BIC impose the same penalty on both models, their rankings are driven entirely by fit. When the data are generated by the hyperbolic model, constant sensitivity can mimic the true model closely by choosing $b<1$. Because AIC, BIC, and CV all reward this close fit without accounting for differences in flexibility, they frequently select the more complex constant sensitivity model, recovering the true model only about 60\% of the time. MDL, by contrast, penalizes the additional flexibility of constant sensitivity and identifies the simpler hyperbolic model almost always. When the data are generated by the constant sensitivity model with increasing impatience ($b>1$), however, the hyperbolic model is incapable of matching the resulting concave profile. In this case, the fit advantage of constant sensitivity outweighs its higher complexity, and all criteria recover the true model with high accuracy.

\begin{table}[tbp]
    \centering
    \caption{Model recovery in simulated time-discounting experiment}
    \label{tab:hyp_csen}
    \begin{tabular*}{\textwidth}{@{\extracolsep{\fill}}lcccc@{}}
        \toprule
        \textbf{True Model} & \textbf{COMP} & \textbf{MDL} & \textbf{AIC/BIC} & \textbf{CV} \\
        \midrule
        Hyperbolic           & 4.54 & \makebox[0pt][r]{$>$}99\% & 62\% & 58\% \\
        Constant sensitivity & 5.95 & 90\%  & \makebox[0pt][r]{$>$}99\% & 90\% \\
        \bottomrule
    \end{tabular*}
    \begin{flushleft}
    \footnotesize \textit{Note:} Simulated datasets contain noisy present values at delays $t\in\{1,\ldots,10\}$, with Gaussian noise $\sigma=0.05$. COMP reports model complexity computed using the asymptotic approximation in \eqref{eq:rissanen} (for details, see Section~1.2 of the Online Appendix). Recovery rates are the percentage of simulated datasets in which each criterion selects the true model, based on 1{,}000 simulations for each true model. CV is five-fold.
    \end{flushleft}
\end{table}

\subsection{Real Data}
\label{sec:time_real}

This empirical application uses data from \cite{Abdellaoui2010} to examine whether the complexity difference between hyperbolic and constant sensitivity discounting matters for model selection. The dataset elicits time weights, $\lambda(t)$, for 67 subjects at six delays ranging from 3 months to 4 years. The analysis focuses on the treatment with positive rewards. In contrast to the previous applications, model selection is conducted at the aggregate level, with all models estimated on the pooled dataset ($n=402$).

To apply MDL in this pooled setting, a lower bound must again be specified for the error standard deviation, $\underline{\sigma}$. As in the previous empirical application, this issue arises because $\sigma$ is estimated jointly with the structural parameters, so allowing it to approach zero would make the complexity term diverge. Here, however, measurement resolution is no longer the relevant constraint. A pooled model imposes a single prediction for all subjects at a given delay, treating between-subject heterogeneity as irreducible noise. The relevant lower bound is therefore the residual standard deviation of a model that perfectly fits the conditional mean at each delay. This floor is estimated using a saturated linear model with a dummy for each delay, yielding $\underline{\sigma}=\hat{\sigma}_s=0.21$ (for details, see Section~1.3.1 of the Online Appendix).

The pooled structure of the data also makes this application particularly well suited to illustrate the compression-based analogue of completeness introduced in Section~\ref{sec:comp_restr}. With repeated observations at each delay, a flexible benchmark for the conditional distribution can be estimated directly. Recall that completeness measures the reduction in description length achieved by a model relative to the largest attainable reduction, with conditional entropy providing the best-attainable code length. An estimate of this entropy is obtained from the same saturated model used to determine $\underline{\sigma}$. This model assigns a separate mean to each delay while retaining a common Gaussian error variance. Treating its fitted distribution as an estimate of the conditional source yields the entropy estimate $\widehat H_G=\frac{1}{2}\log(2\pi e\hat{\sigma}_s^2)$. This benchmark preserves the homoskedastic Gaussian error structure imposed by the candidate models while removing their restrictions on the conditional mean profile. Completeness therefore measures how much of the improvement in description length attainable by allowing mean present value to vary across delays is achieved by each discounting model. Because the benchmark and candidate models share the same homoskedastic Gaussian error specification, the measure does not evaluate their ability to capture heteroskedasticity, non-Gaussian residuals, or structured individual heterogeneity.

Table \ref{tab:time_selection} summarizes the results. The first column reports maximized log-likelihoods: both hyperbolic and constant sensitivity improve substantially on exponential discounting, while constant sensitivity attains only a modest additional gain over hyperbolic discounting. The second column reports MDL complexity. Constant sensitivity is substantially more complex than hyperbolic discounting ($8.25$ versus $7.31$), and both are more complex than exponential discounting ($6.03$). The third column reports the completeness analogue. All three models capture a large share of the available compression, but hyperbolic discounting achieves the highest completeness ($0.862$), followed by constant sensitivity ($0.850$) and exponential discounting ($0.819$). Thus, the small fit advantage of constant sensitivity is insufficient to offset its greater complexity. Consistent with this ranking, MDL selects hyperbolic discounting, whereas AIC, BIC, and CV select constant sensitivity.

\begin{table}[tb]
    \centering
    \caption{Model selection in the \citet{Abdellaoui2010} experiment}
    \label{tab:time_selection}
    \begin{tabular*}{\textwidth}{@{\extracolsep{\fill}}lccccccc@{}}
        \toprule
        \textbf{Model} & \textbf{LogL} & \textbf{COMP}
        & \textbf{$\mathcal C$} & \textbf{MDL}
        & \textbf{AIC} & \textbf{BIC} & \textbf{CV} \\
        \midrule
        Exponential  & 50.21 & 6.03 & 0.819 & --         & --         & --         & -- \\
        Hyperbolic   & 53.85 & 7.31 & 0.862 & \checkmark & --         & --         & -- \\
        Const. Sens. & 54.15 & 8.25 & 0.850 & --         & \checkmark & \checkmark & \checkmark \\
        \bottomrule
    \end{tabular*}
    \begin{flushleft}
    \footnotesize
    \textit{Note:} The table uses the positive-reward treatment from
    \citet{Abdellaoui2010}, pooled across subjects ($n=402$).
    LogL reports the maximized log-likelihood, and COMP reports MDL complexity
    computed using the asymptotic approximation in \eqref{eq:rissanen} over
    $\sigma\in[\hat\sigma_s,1]$.
    $\mathcal C$ denotes completeness relative to a constant-mean Gaussian
    baseline, using the saturated homoskedastic Gaussian model as the entropy
    benchmark. For CV, five folds are assigned at the subject level, so all
    observations from a held-out subject enter the same fold. Checkmarks
    indicate the model selected by each criterion.
    \end{flushleft}
\end{table}

\begin{figure}[tb]
    \centering
    \begin{subfigure}[b]{0.48\textwidth}
        \centering
        \begin{tikzpicture}
            \begin{axis}[
                width=\textwidth,
                height=.9\textwidth, xmin=0, xmax=4.2,
                ymin=0.55, ymax=1.00,
                xtick={0,1,2,3,4},
                ytick={0.6,0.7,0.8,0.9,1.0},
                xlabel={Delay ($t$)},
                ylabel={Discount Factor},
                axis lines=left,
                tick label style={font=\small},
                label style={font=\small},
                ymajorgrids=true,
                grid style={dashed,gray!20},
                legend style={
                    draw=none,
                    fill=none,
                    font=\scriptsize,
                    at={(0.97,0.96)},
                    anchor=north east
                },
                clip=false,
                domain=0:4,
                samples=200
            ]

\addplot[neutralgray, solid, thick]
                {(1 + 0.1399907048)^(-x)};
            \addlegendentry{Exponential}

\addplot[mdlblue, dashed, thick]
                {(1 + 0.6292091292*x)^(-0.2304554253/0.6292091292)};
            \addlegendentry{Hyperbolic}

\addplot[modelrust, dotted, thick]
                {exp(-(0.08674150097*x)^0.7113568361)};
            \addlegendentry{Const. sens.}

\addplot[
                only marks,
                mark=o,
                mark size=3.0pt,
                mark options={solid, draw=black, fill=white, line width=0.6pt},
                black
            ] coordinates {
                (0.25,0.9277614)
                (0.50,0.9036250)
                (1.00,0.8421868)
                (2.00,0.7507667)
                (3.00,0.6765255)
                (4.00,0.6265185)
            };

            \end{axis}
        \end{tikzpicture}
        \caption{}
        \label{fig:time_fit}
    \end{subfigure}
    \hfill
    \begin{subfigure}[b]{0.48\textwidth}
        \centering
        \begin{tikzpicture}
            \begin{axis}[
                width=\textwidth,
                height=0.9\textwidth,
                xmin=0.95, xmax=2.08,
                ymin=0, ymax=3.05,
                xtick={1,2},
                ytick={0,1,2,3},
                xlabel={Structural Parameters ($k$)},
                ylabel={Effective Penalty},
                axis lines=left,
                tick label style={font=\small},
                label style={font=\small},
                ymajorgrids=true,
                grid style={dashed,gray!20},
                legend style={
                    draw=none,
                    fill=none,
                    font=\scriptsize,
                    at={(0.06,0.96)},
                    anchor=north west
                },
                clip=false
            ]

\addplot[domain=1:2.04, black, dashed, thick] {x-1};
            \addlegendentry{AIC}

            \addplot[domain=1:2.04, black!60, dotted, thick] {(ln(402)/2)*(x-1)};
            \addlegendentry{BIC}

\addplot[
                only marks,
                mark=*,
                mark size=2.4pt,
                mdlblue
            ] coordinates {
                (1.00,0.00)   (2.00,1.28)   (2.00,2.22)   };

\node[font=\scriptsize, anchor=south] at (axis cs:1.00,0.00) {Exp};
            \node[font=\scriptsize, anchor=west] at (axis cs:2.00,1.28) {Hyp};
            \node[font=\scriptsize, anchor=west] at (axis cs:2.00,2.22) {CS};

            \end{axis}
        \end{tikzpicture}
        \caption{}
        \label{fig:time_pen}
    \end{subfigure}
    \caption{Model fit and complexity in the positive-reward treatment of the \citet{Abdellaoui2010} experiment, pooled across subjects ($n=402$). Panel (a) plots mean elicited discount weights together with the fitted exponential, hyperbolic, and constant sensitivity models. Panel (b) plots effective complexity penalties relative to the exponential baseline; points show MDL complexity, while dashed and dotted lines show the penalties implied by AIC and BIC.}
    \label{fig:time_fig}
\end{figure}
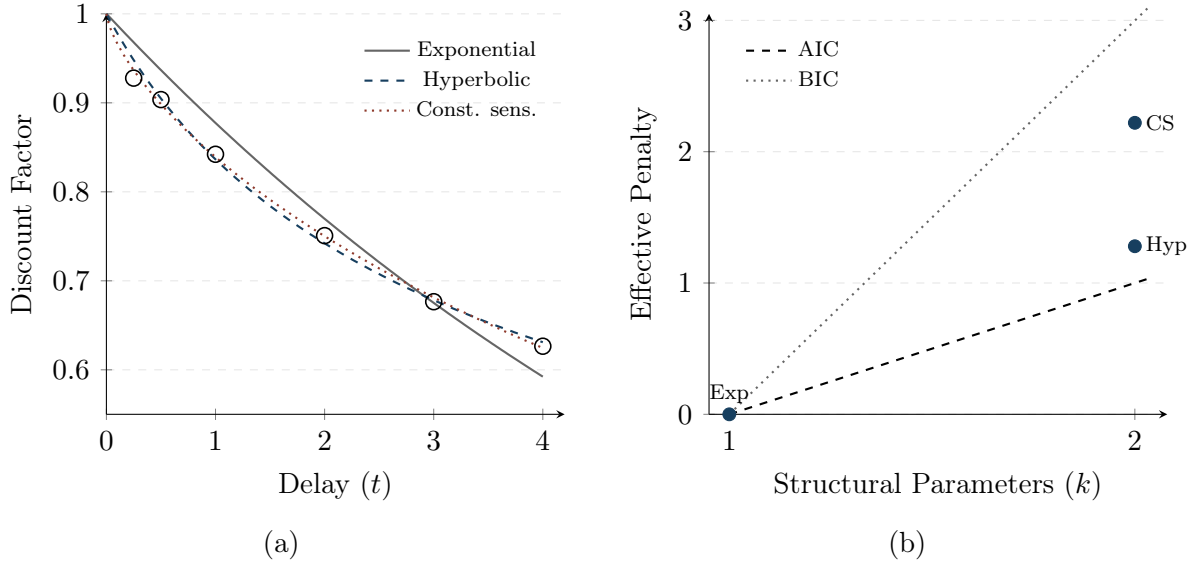

Figure \ref{fig:time_fig} illustrates the results. Panel (a) displays the mean elicited time weights together with the maximum-likelihood predictions of each model. Both hyperbolic and constant sensitivity improve substantially on exponential discounting, capturing the steep initial decline and later flattening of discount weights, but the two fitted curves are nearly indistinguishable from each other. Nevertheless, AIC, BIC, and CV all select constant sensitivity based on its marginal fit advantage over the hyperbolic model. Panel (b) plots the effective complexity penalties of each criterion. As usual, BIC is more conservative than AIC, imposing a larger (and of course equal) penalty on both constant sensitivity and hyperbolic discounting for their additional parameter. By contrast, MDL penalizes constant sensitivity much more steeply than the hyperbolic model. Indeed, the increase in complexity from hyperbolic to constant sensitivity is roughly as large as the increase from exponential to hyperbolic, illustrating how differences in functional form can matter as much as an additional structural parameter. Because the small improvement in fit achieved by constant sensitivity is not large enough to compensate for its added complexity, MDL selects the hyperbolic model, which yields the shortest description length for the data.

\section{Discussion}
\label{sec:conclusion}

This paper applies a data-compression perspective to the evaluation of economic models based on the MDL principle. By viewing models as data compression devices and framing model selection as a compression problem, this approach yields two main contributions.

First, it offers a way to reformulate under a common framework two influential approaches to measuring model flexibility in economics. Recasting Selten’s area theories as codes yields a compression-based analogue of predictive success that extends his approach to the standard stochastic models and maximum-likelihood techniques routinely used in applied work. Similarly, the compression analogues of Fudenberg et al.'s completeness and restrictiveness measures retain their core insight, but map directly into MDL to yield a complete model selection criterion.

Second, the applications to social, risk, and time preferences illustrate the practical relevance of this perspective by showing how functional form and experimental design shape model complexity in ways that parameter-count penalties or cross-validation do not capture. In simulations, complexity differences across models predict asymmetric recovery patterns under AIC, BIC, and CV, allowing the complexity measure to identify, ex ante, designs likely to favor particular specifications. By accounting for these differences, MDL substantially improves model recovery in the simulations considered, while in the empirical reanalyses, it often changes model rankings in favor of more parsimonious specifications.

\bibliographystyle{apalike}
\bibliography{entropy}

\end{document}


\maketitle

\section{Model Complexity Calculations}
\label{sec:oa_complexity}

This appendix describes how the MDL complexity term, denoted $\COMP$, is calculated in the paper. The goal is to make explicit the asymptotic approximation, its specialization to the Normal location models used in the simulations, and its extension to the empirical applications where the standard deviation parameter is also estimated.

\subsection{The general approximation}
\label{sec:oa_general_approximation}

Consider a parametric model $\mathcal M=\{p_\theta:\theta\in\Theta\}$ with $k$ free parameters. Observation $i$ consists of an outcome $y_i$ and fixed covariates or design variables $\mathbf z_i$. The conditional density is written $p_\theta(y_i\mid \mathbf z_i)$, and
\[
\log \mathcal L_i(\theta) = \log p_\theta(y_i\mid \mathbf z_i), \qquad \log \mathcal L(\theta) = \sum_{i=1}^n \log \mathcal L_i(\theta).
\]
The subscript $i$ indexes the observation or design point: in the risk application it indexes lotteries, while in the time application it indexes delays or subject-delay observations. The covariates $\mathbf z_i$ are treated as fixed throughout the complexity calculation.

For regular $k$-dimensional parametric models, \citet{Rissanen1996}'s Fisher information approximation is
\begin{equation}
\COMP(\mathcal M) \approx \frac{k}{2}\log\frac{n}{2\pi} + \log \int_{\Theta} \sqrt{\det I(\theta)}\,\dd\theta, \label{eq:fia_general}
\end{equation}
where $I(\theta)$ is the normalized Fisher information matrix,
\begin{equation}
I(\theta) = - \frac{1}{n}\sum_{i=1}^n \E_{\theta} \left[ \frac{\partial^2}{\partial\theta\,\partial\theta'} \log \mathcal L_i(\theta) \;\middle|\; \mathbf z_i \right]. \label{eq:fisher_hessian}
\end{equation}
Here $\E_\theta[\cdot\mid \mathbf z_i]$ denotes expectation with respect to the conditional distribution $p_\theta(y_i\mid \mathbf z_i)$.

In the Normal location models below, it is simpler to compute the Fisher information using products of first derivatives. The equivalence follows from the standard information matrix equality. Since
\[
\E_\theta\left[ \frac{\partial \log \mathcal L_i(\theta)}{\partial\theta} \;\middle|\; \mathbf z_i \right] = 0
\]
for all interior $\theta$, differentiating both sides with respect to $\theta'$ gives
\[
0 = \E_\theta \left[ \frac{\partial^2 \log \mathcal L_i(\theta)} {\partial\theta\,\partial\theta'} \;\middle|\; \mathbf z_i \right] + \E_\theta \left[ \frac{\partial \log \mathcal L_i(\theta)}{\partial\theta} \frac{\partial \log \mathcal L_i(\theta)}{\partial\theta'} \;\middle|\; \mathbf z_i \right].
\]
Hence
\[
- \E_\theta \left[ \frac{\partial^2 \log \mathcal L_i(\theta)} {\partial\theta\,\partial\theta'} \;\middle|\; \mathbf z_i \right] = \E_\theta \left[ \frac{\partial \log \mathcal L_i(\theta)}{\partial\theta} \frac{\partial \log \mathcal L_i(\theta)}{\partial\theta'} \;\middle|\; \mathbf z_i \right].
\]
Substituting this expression into \eqref{eq:fisher_hessian} gives the form used below:
\begin{equation}
I(\theta) = \frac{1}{n}\sum_{i=1}^n \E_{\theta} \left[ \frac{\partial \log \mathcal L_i(\theta)}{\partial\theta} \frac{\partial \log \mathcal L_i(\theta)}{\partial\theta'} \;\middle|\; \mathbf z_i \right]. \label{eq:fisher_score}
\end{equation}

\subsection{Normal location models with fixed sigma}
\label{sec:oa_fixed_sigma}

The simulations in the risk and time applications use Normal location models with fixed noise. Let $\phi\in\Phi\subseteq\R^k$ denote the structural parameters. Conditional on covariates $\mathbf z_i$, write the model as
\begin{equation}
y_i = m(\mathbf z_i;\phi)+\varepsilon_i, \qquad \varepsilon_i\sim N(0,\sigma_i^2), \label{eq:normal_location}
\end{equation}
where $\sigma_i$ is fixed. In what follows, write $m_i(\phi)=m(\mathbf z_i;\phi)$ for the prediction at observation $i$. The log-likelihood function is
\begin{equation}
\log \mathcal L_i(\phi) = -\frac{1}{2}\log(2\pi\sigma_i^2) - \frac{(y_i-m_i(\phi))^2}{2\sigma_i^2}.
\end{equation}
Let $J_i(\phi)=\partial m_i(\phi)/\partial\phi$ be the $k\times 1$ gradient of the model prediction. The score is
\begin{equation}
\frac{\partial \log \mathcal L_i}{\partial\phi} = \frac{y_i-m_i(\phi)}{\sigma_i^2}J_i(\phi),
\end{equation}
and, since $\E_\phi[(y_i-m_i(\phi))^2\mid\mathbf z_i]=\sigma_i^2$, the Fisher information is
\begin{equation}
I(\phi) = \frac{1}{n} \sum_{i=1}^n \frac{J_i(\phi)J_i(\phi)'}{\sigma_i^2}. \label{eq:normal_location_fisher}
\end{equation}
For a two-parameter model $\phi=(\phi_1,\phi_2)$, \eqref{eq:normal_location_fisher} is
\begin{equation}
I(\phi)
    =
    \frac{1}{n}
    \sum_{i=1}^n
    \frac{1}{\sigma_i^2}
    \begin{bmatrix}
        \left(\frac{\partial m_i(\phi)}{\partial \phi_1}\right)^2
        &
        \frac{\partial m_i(\phi)}{\partial \phi_1}
        \frac{\partial m_i(\phi)}{\partial \phi_2}
        \\
        \frac{\partial m_i(\phi)}{\partial \phi_1}
        \frac{\partial m_i(\phi)}{\partial \phi_2}
        &
        \left(\frac{\partial m_i(\phi)}{\partial \phi_2}\right)^2
    \end{bmatrix}.
    \label{eq:two_param_fisher}
\end{equation}
The corresponding COMP approximation is therefore
\begin{equation}
\COMP_n(\mathcal M) \approx \frac{k}{2}\log\frac{n}{2\pi} + \log \int_{\Phi} \sqrt{ \det\left[ \frac{1}{n} \sum_{i=1}^n \frac{J_i(\phi)J_i(\phi)'}{\sigma_i^2} \right] }\,\dd\phi . \label{eq:normal_location_fixed_comp}
\end{equation}

\subsubsection{Risk-preference simulations}

In the risk-preference simulations, each observation is a certainty equivalent for a binary lottery $L_i=(p_i,x_i;1-p_i,z_i)$, with $x_i>z_i$. The simulation noise is proportional to the lottery spread:
\begin{equation}
y_i = m_i(\phi)+\varepsilon_i, \qquad \varepsilon_i\sim N(0,\sigma_i^2), \qquad \sigma_i^2=s^2(x_i-z_i)^2. \label{eq:risk_noise}
\end{equation}

\subsubsubsection*{Two-parameter Expected Utility (Introduction)}

The first model in the motivating example is Expected Utility. Preferences over lotteries are represented by
\[
U(L_i;\alpha,\omega) = p_i(\omega+x_i)^\alpha+(1-p_i)(\omega+z_i)^\alpha.
\]
For lottery $L_i$, the predicted certainty equivalent satisfies
\[
(\omega+m_i(\alpha,\omega))^\alpha = U(L_i;\alpha,\omega).
\]
Then
\begin{equation}
m_i(\alpha,\omega) = U(L_i;\alpha,\omega)^{1/\alpha}-\omega. \label{eq:eu_ak_ce}
\end{equation}
Writing $U_i=U(L_i;\alpha,\omega)$ in the next display, the derivatives entering the information matrix are
\begin{align}
    \frac{\partial m_i}{\partial\alpha}
    &=
    U_i^{1/\alpha}
    \left[
    \frac{
        p_i(\omega+x_i)^\alpha\log(\omega+x_i)
        +(1-p_i)(\omega+z_i)^\alpha\log(\omega+z_i)
    }{\alpha U_i}
    -
    \frac{\log U_i}{\alpha^2}
    \right],
    \label{eq:eu_ak_dalpha}
    \\
    \frac{\partial m_i}{\partial\omega}
    &=
    U_i^{1/\alpha-1}
    \left[
        p_i(\omega+x_i)^{\alpha-1}
        +(1-p_i)(\omega+z_i)^{\alpha-1}
    \right]
    -1.
    \label{eq:eu_ak_domega}
\end{align}
Substituting \eqref{eq:eu_ak_dalpha}--\eqref{eq:eu_ak_domega} into \eqref{eq:two_param_fisher}, with $\sigma_i^2=s^2(x_i-z_i)^2$ as in \eqref{eq:risk_noise}, gives
\[
I(\alpha,\omega;s)
    =
    \frac{1}{n}
    \sum_{i=1}^n
    \frac{1}{s^2(x_i-z_i)^2}
    \begin{bmatrix}
        \left(\frac{\partial m_i}{\partial \alpha}\right)^2
        &
        \frac{\partial m_i}{\partial \alpha}
        \frac{\partial m_i}{\partial \omega}
        \\
        \frac{\partial m_i}{\partial \alpha}
        \frac{\partial m_i}{\partial \omega}
        &
        \left(\frac{\partial m_i}{\partial \omega}\right)^2
    \end{bmatrix}
    \equiv
    \frac{1}{s^2}I(\alpha,\omega).
\]
The common scale then separates from the square-root determinant:
\[
\sqrt{\det I(\alpha,\omega;s)} = s^{-2} \sqrt{\det I(\alpha,\omega)}.
\]
Thus, for the two-parameter EU model,
\begin{align*}
    \COMP_n(\mathcal M_{\mathrm{EU}})
    &\approx
    \log\frac{n}{2\pi}
    +
    \log
    \int_{\Phi_{\mathrm{EU}}}
    \sqrt{\det I(\alpha,\omega)}
    \,\dd\alpha\,\dd\omega
    -
    2\log s.
\end{align*}
The complexity calculation uses the parameter domain $\alpha\in(0,1)$ and $\omega\in(0,\infty)$.

\subsubsubsection*{Two-parameter Dual Theory (Introduction)}

The second model in the motivating example is Dual Theory with linear utility and two-parameter Prelec weighting. Preferences over lotteries are represented by
\[
U(L_i;\delta,\gamma) = w(p_i;\delta,\gamma)x_i + \left[1-w(p_i;\delta,\gamma)\right]z_i,
\]
where
\[
w(p;\delta,\gamma) = \exp[-\delta(-\log p)^\gamma].
\]
Since utility is linear, the predicted certainty equivalent satisfies
\[
m_i(\delta,\gamma) = U(L_i;\delta,\gamma).
\]
Then
\begin{equation}
m_i(\delta,\gamma) = w(p_i;\delta,\gamma)x_i + \left[1-w(p_i;\delta,\gamma)\right]z_i. \label{eq:dt_dg_ce}
\end{equation}
Let $w_i=w(p_i;\delta,\gamma)$. The derivatives entering the information matrix are
\begin{align}
    \frac{\partial m_i}{\partial\delta}
    &=
    -(x_i-z_i)(-\log p_i)^\gamma w_i,
    \label{eq:dt_dg_ddelta}
    \\
    \frac{\partial m_i}{\partial\gamma}
    &=
    -(x_i-z_i)\delta(-\log p_i)^\gamma\log(-\log p_i)w_i.
    \label{eq:dt_dg_dgamma}
\end{align}
Substituting \eqref{eq:dt_dg_ddelta}--\eqref{eq:dt_dg_dgamma} into \eqref{eq:two_param_fisher}, with $\sigma_i^2=s^2(x_i-z_i)^2$ as in \eqref{eq:risk_noise}, gives
\[
I(\delta,\gamma;s)
    =
    \frac{1}{n}
    \sum_{i=1}^n
    \frac{1}{s^2(x_i-z_i)^2}
    \begin{bmatrix}
        \left(\frac{\partial m_i}{\partial \delta}\right)^2
        &
        \frac{\partial m_i}{\partial \delta}
        \frac{\partial m_i}{\partial \gamma}
        \\
        \frac{\partial m_i}{\partial \delta}
        \frac{\partial m_i}{\partial \gamma}
        &
        \left(\frac{\partial m_i}{\partial \gamma}\right)^2
    \end{bmatrix}
    \equiv
    \frac{1}{s^2}I(\delta,\gamma).
\]
Thus, for the two-parameter Dual Theory model,
\[
\COMP_n(\mathcal M_{\mathrm{DT}}) \approx \log\frac{n}{2\pi} + \log \int_{\Phi_{\mathrm{DT}}} \sqrt{\det I(\delta,\gamma)} \,\dd\delta\,\dd\gamma - 2\log s.
\]
The complexity calculation uses the parameter domain $\delta\in(0,\infty)$ and $\gamma\in(0,1)$.

\subsubsubsection*{One-parameter Expected Utility (Application 2: Risk Preferences)}

The risk-preference simulations in Application 2 compare one-parameter restrictions of the CPT framework. For $\mathcal M_\alpha$, preferences over lotteries are represented by
\[
U(L_i;\alpha) = p_i x_i^\alpha+(1-p_i)z_i^\alpha.
\]
The predicted certainty equivalent satisfies
\[
m_i(\alpha)^\alpha = U(L_i;\alpha).
\]
Then
\begin{equation}
m_i(\alpha) = U(L_i;\alpha)^{1/\alpha}. \label{eq:malpha_ce}
\end{equation}
Writing $U_i=U(L_i;\alpha)$ in the next display, the derivative entering the information matrix is
\begin{equation}
\frac{\partial m_i}{\partial\alpha} = U_i^{1/\alpha} \left[ \frac{ p_i x_i^\alpha\log x_i +(1-p_i)z_i^\alpha\log z_i }{\alpha U_i} - \frac{\log U_i}{\alpha^2} \right], \label{eq:malpha_derivative}
\end{equation}
Terms of the form $y^\alpha\log y$ are set to zero at $y=0$, since $\lim_{y\downarrow 0} y^\alpha\log y=0$ for $\alpha>0$. Since this model has one parameter, substituting \eqref{eq:malpha_derivative} into \eqref{eq:normal_location_fisher}, with $\sigma_i^2=s^2(x_i-z_i)^2$ as in \eqref{eq:risk_noise}, gives
\[
I(\alpha;s) = \frac{1}{n}\sum_{i=1}^n \frac{1}{s^2(x_i-z_i)^2} \left( \frac{\partial m_i(\alpha)}{\partial\alpha} \right)^2 \equiv \frac{1}{s^2}I(\alpha).
\]
Thus, for $\mathcal M_\alpha$,
\[
\COMP_n(\mathcal M_\alpha) \approx \frac{1}{2}\log\frac{n}{2\pi} + \log \int_{\Phi_\alpha} \sqrt{I(\alpha)} \,\dd\alpha - \log s.
\]
The complexity calculation uses the parameter domain $\alpha\in(0,1)$.

\subsubsubsection*{One-parameter Dual Theory (Application 2: Risk Preferences)}

For $\mathcal M_\gamma$, preferences over lotteries are represented by
\[
U(L_i;\gamma) = w(p_i;\gamma)x_i + \left[1-w(p_i;\gamma)\right]z_i,
\]
where
\[
w(p;\gamma)=\exp[-(-\log p)^\gamma].
\]
Since utility is linear, the predicted certainty equivalent is
\begin{equation}
m_i(\gamma) = w(p_i;\gamma)x_i + \left[1-w(p_i;\gamma)\right]z_i, \label{eq:mgamma_ce}
\end{equation}
and the derivative entering the information matrix is
\begin{equation}
\frac{\partial m_i}{\partial\gamma} = -(x_i-z_i)(-\log p_i)^\gamma \log(-\log p_i) \exp[-(-\log p_i)^\gamma]. \label{eq:mgamma_derivative}
\end{equation}
Substituting \eqref{eq:mgamma_derivative} into \eqref{eq:normal_location_fisher}, with $\sigma_i^2=s^2(x_i-z_i)^2$ as in \eqref{eq:risk_noise}, gives
\[
I(\gamma;s) = \frac{1}{n}\sum_{i=1}^n \frac{1}{s^2(x_i-z_i)^2} \left( \frac{\partial m_i(\gamma)}{\partial\gamma} \right)^2 \equiv \frac{1}{s^2}I(\gamma).
\]
Thus, for $\mathcal M_\gamma$,
\[
\COMP_n(\mathcal M_\gamma) \approx \frac{1}{2}\log\frac{n}{2\pi} + \log \int_{\Phi_\gamma} \sqrt{I(\gamma)} \,\dd\gamma - \log s.
\]
The complexity calculation uses the parameter domain $\gamma\in(0,1)$.

\subsubsection{Time-preference simulations}

In the time-preference simulations, each observation is a present value or time weight at delay $t_i$:
\[
y_i=\lambda(t_i;\phi)+\varepsilon_i, \qquad \varepsilon_i\sim N(0,\sigma^2),
\]
with fixed $\sigma=0.05$. Equations \eqref{eq:normal_location_fisher} and \eqref{eq:normal_location_fixed_comp} therefore apply with $m_i(\phi)=\lambda(t_i;\phi)$ and $\sigma_i^2=\sigma^2$.

\subsubsubsection*{One-parameter Exponential Discounting}

For exponential discounting, the predicted time weight is
\[
\lambda(t;\delta)=(1+\delta)^{-t}.
\]
The derivative entering the information matrix is
\[
\frac{\partial\lambda}{\partial\delta} = -t(1+\delta)^{-t-1}.
\]
For this model,
\[
I(\delta;\sigma) = \frac{1}{n} \sum_{i=1}^n \frac{1}{\sigma^2} \left[ \frac{\partial\lambda(t_i;\delta)}{\partial\delta} \right]^2 \equiv \frac{1}{\sigma^2}I(\delta).
\]
Thus
\[
\COMP_n(\mathcal M_e) \approx \frac{1}{2}\log\frac{n}{2\pi} + \log \int_{\Phi_e} \sqrt{I(\delta)} \,\dd\delta - \log\sigma.
\]
The complexity calculation uses the parameter domain $\delta\in(0,\infty)$.

\subsubsubsection*{Two-parameter Hyperbolic Discounting}

For hyperbolic discounting, the predicted time weight is
\[
\lambda(t;\alpha,\gamma) = (1+\gamma t)^{-\alpha/\gamma}.
\]
The derivatives entering the information matrix are
\begin{align}
    \frac{\partial\lambda}{\partial\alpha}
    &=
    -\frac{\log(1+\gamma t)}{\gamma}\lambda(t;\alpha,\gamma),
    \label{eq:time_hyperbolic_dalpha}
    \\
    \frac{\partial\lambda}{\partial\gamma}
    &=
    \lambda(t;\alpha,\gamma)
    \left[
        \frac{\alpha\log(1+\gamma t)}{\gamma^2}
        -
        \frac{\alpha t}{\gamma(1+\gamma t)}
    \right].
    \label{eq:time_hyperbolic_dgamma}
\end{align}
For this model,
\[
I(\alpha,\gamma;\sigma)
    =
    \frac{1}{n}
    \sum_{i=1}^n
    \frac{1}{\sigma^2}
    \begin{bmatrix}
        \left(\frac{\partial\lambda(t_i;\alpha,\gamma)}{\partial\alpha}\right)^2
        &
        \frac{\partial\lambda(t_i;\alpha,\gamma)}{\partial\alpha}
        \frac{\partial\lambda(t_i;\alpha,\gamma)}{\partial\gamma}
        \\
        \frac{\partial\lambda(t_i;\alpha,\gamma)}{\partial\alpha}
        \frac{\partial\lambda(t_i;\alpha,\gamma)}{\partial\gamma}
        &
        \left(\frac{\partial\lambda(t_i;\alpha,\gamma)}{\partial\gamma}\right)^2
    \end{bmatrix}
    \equiv
    \frac{1}{\sigma^2}I(\alpha,\gamma).
\]
Thus
\[
\COMP_n(\mathcal M_h) \approx \log\frac{n}{2\pi} + \log \int_{\Phi_h} \sqrt{\det I(\alpha,\gamma)} \,\dd\alpha\,\dd\gamma - 2\log\sigma.
\]
The complexity calculation uses the parameter domain $\alpha\in(0,\infty)$ and $\gamma\in(0,\infty)$.

\subsubsubsection*{Two-parameter Constant Sensitivity}

For constant sensitivity, the predicted time weight is
\[
\lambda(t;a,b)=\exp[-(at)^b].
\]
The derivatives entering the information matrix are
\begin{align}
    \frac{\partial\lambda}{\partial a}
    &=
    -\frac{b}{a}(at)^b\lambda(t;a,b),
    \label{eq:time_cs_da}
    \\
    \frac{\partial\lambda}{\partial b}
    &=
    -(at)^b\log(at)\lambda(t;a,b).
    \label{eq:time_cs_db}
\end{align}
For this model,
\[
I(a,b;\sigma)
    =
    \frac{1}{n}
    \sum_{i=1}^n
    \frac{1}{\sigma^2}
    \begin{bmatrix}
        \left(\frac{\partial\lambda(t_i;a,b)}{\partial a}\right)^2
        &
        \frac{\partial\lambda(t_i;a,b)}{\partial a}
        \frac{\partial\lambda(t_i;a,b)}{\partial b}
        \\
        \frac{\partial\lambda(t_i;a,b)}{\partial a}
        \frac{\partial\lambda(t_i;a,b)}{\partial b}
        &
        \left(\frac{\partial\lambda(t_i;a,b)}{\partial b}\right)^2
    \end{bmatrix}
    \equiv
    \frac{1}{\sigma^2}I(a,b).
\]
Thus
\[
\COMP_n(\mathcal M_c) \approx \log\frac{n}{2\pi} + \log \int_{\Phi_c} \sqrt{\det I(a,b)} \,\dd a\,\dd b - 2\log\sigma.
\]
The complexity calculation uses the parameter domain $a\in(0,\infty)$ and $b\in(0,\infty)$. For numerical evaluation, the upper limit of the Fisher-volume integral is set to $b=600$; the integral is already stable by $b=100$, with the implied log volume changing by less than $5\times10^{-6}$ when the limit is increased from $100$ to $600$.

\subsection{Normal location models with estimated sigma}
\label{sec:oa_estimated_sigma}

When the noise scale is estimated along with the structural parameters, begin with the homoskedastic Normal location model
\begin{equation}
y_i=m_i(\phi)+\varepsilon_i, \qquad \varepsilon_i\sim N(0,\sigma^2). \label{eq:estimated_sigma_model}
\end{equation}
The full parameter vector is $\theta=(\phi,\sigma)$. In this subsection, $k$ denotes the total number of parameters, so $\phi$ has $k-1$ components.

The log-likelihood contribution is
\begin{equation}
\log \mathcal L_i(\phi,\sigma) = -\log \sigma-\frac{1}{2}\log(2\pi) - \frac{(y_i-m_i(\phi))^2}{2\sigma^2}.
\end{equation}
The score components are
\begin{equation}
\frac{\partial \log \mathcal L_i}{\partial\phi} = \frac{y_i-m_i(\phi)}{\sigma^2}J_i(\phi), \qquad \frac{\partial \log \mathcal L_i}{\partial \sigma} = -\frac{1}{\sigma} + \frac{(y_i-m_i(\phi))^2}{\sigma^3}.
\end{equation}
Let
\[
\frac{\partial \log \mathcal L_i(\theta)}{\partial\theta}
    =
    \begin{bmatrix}
        \partial \log \mathcal L_i/\partial\phi \\
        \partial \log \mathcal L_i/\partial\sigma
    \end{bmatrix},
    \qquad
    \theta=(\phi,\sigma).
\]
The normalized Fisher information matrix is the expected outer product of this vector,
\[
I(\theta) = \frac{1}{n} \sum_{i=1}^n \E_{\theta} \left[ \frac{\partial \log \mathcal L_i(\theta)}{\partial\theta} \frac{\partial \log \mathcal L_i(\theta)}{\partial\theta'} \mid \mathbf z_i \right],
\]
which gives
\begin{equation}
I(\phi,\sigma)
    =
    \begin{bmatrix}
        \sigma^{-2} I_{\phi}(\phi) & 0 \\
        0 & 2\sigma^{-2}
    \end{bmatrix},
    \label{eq:block_fisher}
\end{equation}
where
\[
I_{\phi}(\phi) = \frac{1}{n} \sum_{i=1}^n J_i(\phi)J_i(\phi)'.
\]
This is the specification used in the time-preference application. In the risk-preference application, the standard deviation is $\sigma(x_i-z_i)$. The same block form in \eqref{eq:block_fisher} then applies with
\[
I_{\phi}(\phi) = \frac{1}{n} \sum_{i=1}^n \frac{J_i(\phi)J_i(\phi)'}{(x_i-z_i)^2},
\]
as in Section~\ref{sec:oa_fixed_sigma}; the scale block and the power of $\sigma$ in the determinant are unchanged.
To see why the off-diagonal block is zero, write $e_i=y_i-m_i(\phi)$. For observation $i$, that block is
\begin{align*}
    I_i(\phi,\sigma)
    &=
    \E_{\phi,\sigma}
    \left[
        \frac{\partial \log \mathcal L_i}{\partial\phi}
        \frac{\partial \log \mathcal L_i}{\partial\sigma}
        \mid\mathbf z_i
    \right]
    \\
    &=
    \E_{\phi,\sigma}
    \left[
        \frac{e_i}{\sigma^2}J_i(\phi)
        \left(
            -\frac{1}{\sigma}
            +
            \frac{e_i^2}{\sigma^3}
        \right)
        \mid\mathbf z_i
    \right]
    \\
    &=
    \frac{J_i(\phi)}{\sigma^5}
    \E_{\phi,\sigma}
    \left[
        e_i(e_i^2-\sigma^2)
        \mid\mathbf z_i
    \right].
\end{align*}
Because $e_i\mid\mathbf z_i\sim N(0,\sigma^2)$, its first and third moments are zero. Hence
\[
\E_{\phi,\sigma} \left[ e_i(e_i^2-\sigma^2) \mid\mathbf z_i \right] = \E_{\phi,\sigma}[e_i^3\mid\mathbf z_i] - \sigma^2\E_{\phi,\sigma}[e_i\mid\mathbf z_i] = 0.
\]
Thus the determinant factorizes as
\begin{equation}
\det I(\phi,\sigma) = 2\sigma^{-2k}\det I_{\phi}(\phi),
\end{equation}
so
\begin{equation}
\sqrt{\det I(\phi,\sigma)} = \sqrt{2}\,\sigma^{-k} \sqrt{\det I_{\phi}(\phi)}. \label{eq:sigma_volume_element}
\end{equation}
If $\sigma\in[\underline{\sigma},\overline{\sigma}]$, substituting \eqref{eq:sigma_volume_element} into the COMP approximation \eqref{eq:fia_general} gives
\begin{equation}
\COMP_n(\mathcal M) \approx \frac{k}{2}\log\frac{n}{2\pi} + \log\left[ \sqrt{2}\, \int_{\Phi}\sqrt{\det I_{\phi}(\phi)}\,\dd\phi \int_{\underline{\sigma}}^{\overline{\sigma}}\sigma^{-k}\,\dd\sigma \right]. \label{eq:free_sigma_comp}
\end{equation}
The scale integral is
\begin{equation}
\int_{\underline{\sigma}}^{\overline{\sigma}}\sigma^{-k}\,\dd\sigma
    =
    \begin{cases}
        \log(\overline{\sigma}/\underline{\sigma}), & k=1,\\[
4pt]
        \dfrac{\underline{\sigma}^{-(k-1)}-\overline{\sigma}^{-(k-1)}}{k-1}, & k>1.
    \end{cases}
\end{equation}
For $k>1$, the resulting complexity can be written as
\begin{equation}
\begin{aligned}
    \COMP_n(\mathcal M)
    \approx{}&
    \frac{k}{2}\log\frac{n}{2\pi}
    +\log\int_{\Phi}\sqrt{\det I_{\phi}(\phi)}\,\dd\phi
    +(k-1)\log\frac{1}{\underline{\sigma}}
    \\
    &+\log\left[1-\left(\frac{\underline{\sigma}}{\overline{\sigma}}\right)^{k-1}\right]
    -\log(k-1)
    +\frac{1}{2}\log 2.
\end{aligned}
    \label{eq:free_sigma_decomposition}
\end{equation}
This decomposition shows why a lower bound on $\sigma$ is necessary. As $\underline{\sigma}\downarrow 0$, the scale integral in \eqref{eq:free_sigma_comp} diverges; for $k>1$, this is reflected in the term $(k-1)\log(1/\underline{\sigma})$. Intuitively, allowing the noise scale to approach zero lets the model distinguish arbitrarily fine differences in the data, producing an infinite integral.

In both empirical applications, outcomes and model predictions are normalized to an interval of width one, and we set $\overline{\sigma}=1$. This upper bound does not restrict the maximum-likelihood estimate for any eligible dataset. In the risk-preference application, observed and predicted certainty equivalents lie within each lottery's payoff range, so the absolute residual divided by the payoff spread cannot exceed one. In the time-preference application, observed time weights and predicted discount factors lie in $[0,1]$, yielding the same bound. Thus, $\overline{\sigma}=1$ admits every attainable maximum-likelihood estimate while excluding noise scales larger than the entire outcome range.

\subsubsection{Setting the sigma lower bound}
\label{sec:oa_sigma_lower_bound}

The previous derivation shows that the complexity of the normal location model is finite only if the standard deviation of the noise term, $\sigma$, is bounded away from zero. Lowering $\underline{\sigma}$ increases the scale integral in \eqref{eq:free_sigma_comp}. Because this contribution depends only on $k$ and the common scale bounds, $\underline{\sigma}$ has no effect on comparisons among models with the same number of parameters, but it can affect comparisons among models with different $k$: the smaller $\underline{\sigma}$ is, the more conservative MDL becomes about adding parameters. This makes it important to tie the lower bound, when possible, to objective features of the experimental design or of the data-generating process.

\subsubsubsection*{Risk-preferences experiment \citep{Bruhin2010}}

In the risk-preferences experiment, each subject provides 50 certainty equivalents, one for each lottery. The models are then estimated separately for each subject. For lottery $i$, the observed certainty equivalent is modeled as
\[
    y_i=m_i(\phi)+\varepsilon_i.
\]
Following \citet{Bruhin2010}, the noise term is modeled as
\[
\varepsilon_i\sim N(0,\sigma^2(x_i-z_i)^2),
\]
so its standard deviation is proportional to the lottery spread $x_i-z_i$.

The certainty equivalents are constructed from a finite elicitation grid. In each decision sheet, the subject chooses between the lottery and a sequence of guaranteed payoffs. The certainty equivalent is recorded as the midpoint between the two adjacent guaranteed payoffs at which the subject switches. If the grid has $Q$ equally spaced steps over the lottery spread, the distance between adjacent guaranteed payoffs is
\[
\delta_i=\frac{x_i-z_i}{Q}.
\]
This finite resolution creates a natural lower bound for $\sigma$. The recorded midpoint locates the certainty equivalent only up to an interval of width $\delta_i$. Assuming that the true certainty equivalent is uniformly distributed within the interval identified by the two adjacent switching payoffs, its variance around the recorded midpoint is
\[
\frac{1}{\delta_i} \int_{-\delta_i/2}^{\delta_i/2} u^2\,\dd u = \frac{\delta_i^2}{12}.
\]
The corresponding standard deviation is therefore $\delta_i/\sqrt{12}$. Requiring the model's smallest admissible noise standard deviation, $\underline{\sigma}(x_i-z_i)$, to be no smaller than this uncertainty gives
\[
\underline{\sigma}(x_i-z_i) = \frac{\delta_i}{\sqrt{12}} = \frac{x_i-z_i}{Q\sqrt{12}}.
\]
Canceling the spread yields
\[
\underline{\sigma} = \frac{1}{Q\sqrt{12}}.
\]
In \citet{Bruhin2010}, each decision sheet has 20 equally spaced steps, so $Q=20$ and
\[
\underline{\sigma}=\frac{1}{20\sqrt{12}}.
\]

\subsubsubsection*{Time-preferences experiment \citep{Abdellaoui2010}}

In this experiment, each subject provides 6 present values, or time weights, for delays ranging from 3 months to 4 years. In contrast to the risk-preferences experiment, where models are estimated separately for each subject, this analysis uses a representative-agent framework. For a given delay, a model assigns a single predicted time weight,
\[
y_i=\lambda(t_i;\phi)+\varepsilon_i, \qquad \varepsilon_i\sim N(0,\sigma^2).
\]
Here $\lambda(t_i;\phi)$ is the discounting model's predicted time weight at delay $t_i$. The error term captures both within-subject noise and between-subject heterogeneity. The lower bound for $\sigma$ is therefore not determined by a common elicitation-grid width. Instead, the relevant limitation comes from pooling: subjects differ in their measured time weights, and this within-delay heterogeneity cannot be eliminated by any representative-agent model that assigns one prediction to each delay.

The lower bound can therefore be anchored to the most flexible representative-agent benchmark available at the delay level. This saturated benchmark assigns a separate mean to each delay, so it removes all systematic variation that can be captured by a single prediction per delay. If $\hat\mu_t$ is the sample mean of all observations measured at delay $t$, then the remaining residual dispersion is
\begin{equation}
\underline{\sigma} = \left[ \frac{1}{n} \sum_i (y_i-\hat\mu_{t_i})^2 \right]^{1/2}.
\end{equation}
By construction, no representative-agent model that assigns one prediction to each delay can achieve an estimated $\sigma$ below this residual dispersion.\footnote{Unlike the bound in the risk-preference application, this bound is data-dependent and must therefore, strictly speaking, itself be transmitted. It can be treated as the first component of a two-part code. Because the associated codelength is common to all candidate models, this component does not affect their ranking.}

\section{Simulation details and Robustness Checks}
\label{sec:oa_simulation_robustness}

The paper uses simulations in 3 places: the introductory comparison between Expected Utility and Dual Theory, the risk-preference simulations in Application 2, and the time-preference simulations in Application 3. This section reports robustness checks that vary the parts of each simulation most likely to affect model recovery: parameter values, noise levels, and experimental designs. Throughout the simulation exercises in the paper and Online Appendix, CV uses five folds, randomly redrawn for each simulated dataset and held fixed across models; models are ranked by average held-out log-likelihood.

\subsection{Motivating example: Expected Utility vs. Dual Theory}
\label{sec:oa_intro_risk_robustness}

This subsection gives the simulation details behind the introductory comparison in the paper. The motivating example compares two-parameter versions of Expected Utility and Dual Theory. Each simulated dataset contains certainty equivalents for 50 binary lotteries $L_i=(p_i,x_i;z_i)$, where $p_i$ is the probability of payoff $x_i$ and $1-p_i$ is the probability of payoff $z_i$. Under Expected Utility, preferences over lotteries are represented by
\[
U_{\mathrm{EU}}(L_i;\alpha,\omega) = p_i(\omega+x_i)^\alpha+(1-p_i)(\omega+z_i)^\alpha,
\]
so the predicted certainty equivalent is
\[
m_i^{\mathrm{EU}}(\alpha,\omega) = U_{\mathrm{EU}}(L_i;\alpha,\omega)^{1/\alpha}-\omega.
\]
Under Dual Theory, preferences over lotteries are represented by
\[
U_{\mathrm{DT}}(L_i;\delta,\gamma) = w(p_i;\delta,\gamma)x_i + \left[1-w(p_i;\delta,\gamma)\right]z_i,
\]
where
\[
w(p;\delta,\gamma) = \exp[-\delta(-\log p)^\gamma],
\]
and the predicted certainty equivalent is
\[
m_i^{\mathrm{DT}}(\delta,\gamma) = U_{\mathrm{DT}}(L_i;\delta,\gamma).
\]
The lotteries are obtained by combining $p_i\in\{.1,.3,.5,.7,.9\}$ with all payoff pairs $x_i>z_i$ from $\{0,.25,.5,.75,1\}$. For each true model, parameters are drawn independently and uniformly, with $\alpha,\gamma\in[0.2,0.8]$ and $\omega,\delta\in[0.5,1.5]$. Let $m_i(\phi)$ denote the corresponding predicted certainty equivalent, with $\phi=(\alpha,\omega)$ under EU and $\phi=(\delta,\gamma)$ under DT. Simulated certainty equivalents are generated as
\[
y_i=m_i(\phi)+\varepsilon_i, \qquad \varepsilon_i\sim N(0,s^2(x_i-z_i)^2),
\]
with $s=0.1$ in the baseline. Both models are then re-estimated by maximum likelihood and compared using MDL, AIC, BIC, and CV.

\begin{figure}[!t]
    \centering
    \includegraphics[width=.82\textwidth]{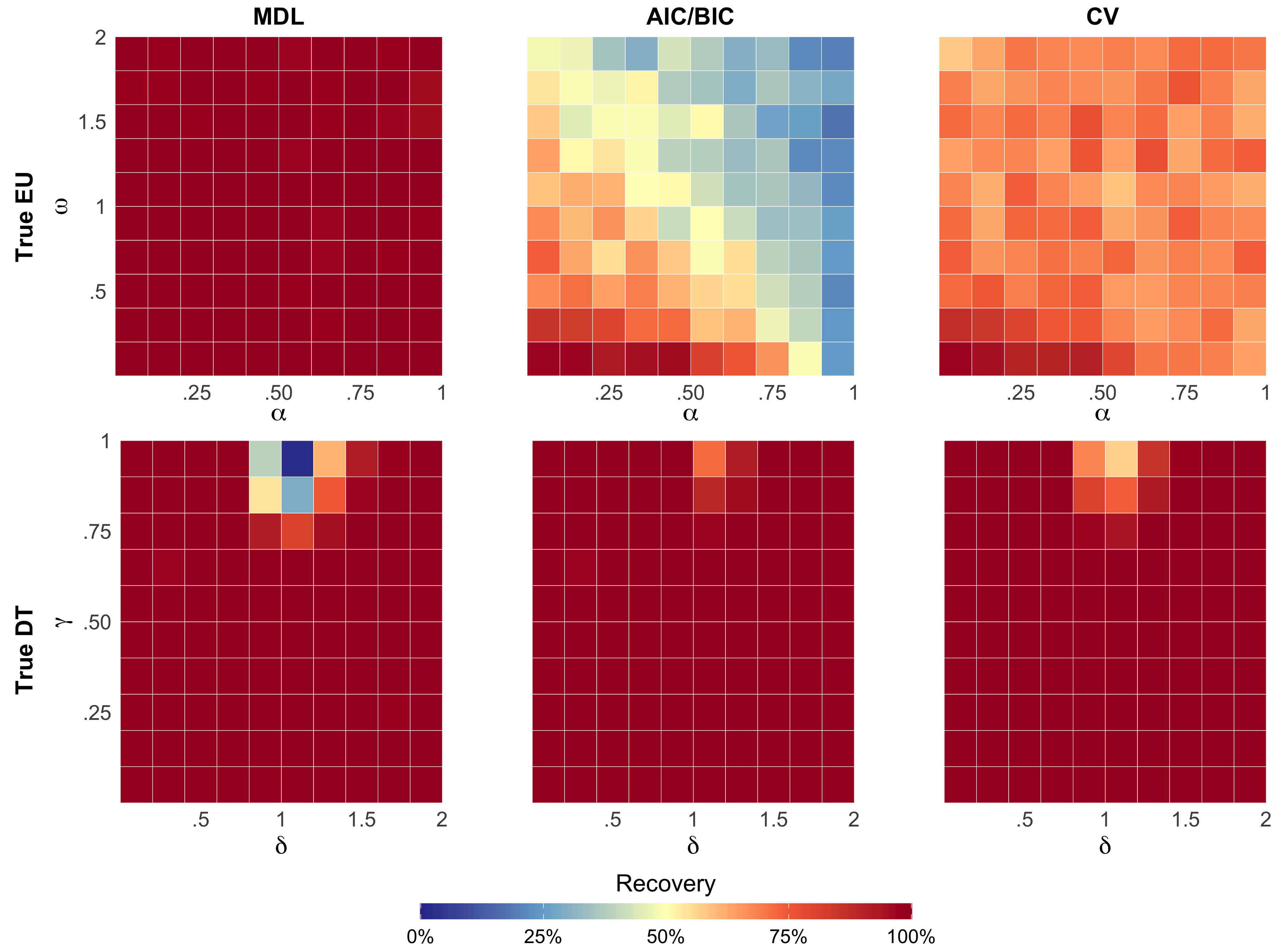}
    \caption{Model recovery over the parameter space in the introductory risk simulation. Each cell reports the fraction of 100 simulated datasets in which the criterion selects the data-generating model.}
    \label{fig:oa_intro_parameter_grid}
\end{figure}

The introductory simulation table in the main text reports near-perfect recovery for MDL under both true models. It also reports near-perfect recovery for all criteria when DT is true, but much weaker recovery when EU is true: about 50\% for AIC/BIC and 68\% for CV. The first robustness check examines how these averages vary across the parameter space by replacing random parameter draws with a $10\times 10$ grid over a broader parameter space. For EU-generated data, the grid varies $(\alpha,\omega)$ over $(0,1)\times(0,2)$. For DT-generated data, it varies $(\delta,\gamma)$ over $(0,2)\times(0,1)$. At each grid point, the data-generating parameter is set to the midpoint of the corresponding cell. The lottery set and noise specification are unchanged, 100 datasets are simulated, both models are estimated, and recovery is the fraction of simulated datasets in which the criterion selects the data-generating model.

Figure~\ref{fig:oa_intro_parameter_grid} shows that the averages reported in the introductory simulation table are replicated over most of the parameter space. MDL recovers EU in 96--100\% of simulations throughout the EU parameter grid, and it recovers DT almost everywhere except near $(\delta,\gamma)=(1,1)$. Note that at $(\delta,\gamma)=(1,1)$, DT is observationally equivalent to EU, so the fact that MDL chooses the simpler model is arguably the correct decision.

For AIC/BIC and CV, the weak average recovery of EU reflects poor performance over most of the EU parameter grid. The small bottom-left region, where $U$ becomes extremely convex, is the main area where these methods perform well; elsewhere recovery is low. AIC/BIC recovery is below 50\% in 55 of the 100 EU grid cells, and below 70\% in 86 cells. CV is more robust than AIC/BIC in this case, but still much less stable than MDL, with recovery rates ranging from 58--97\% and a middle-half range of 66--73\%.

The simulations in Figure~\ref{fig:oa_intro_parameter_grid} keep the noise scale fixed at $s=0.1$. The second robustness check asks how recovery changes as the data become more or less noisy. This check uses the same design as the introductory simulation table in the main text, but evaluates it at 10 values of $s$, with 1{,}000 simulated datasets for each true model and value of $s$. The values of $s$ range from 0.01 to 0.75 and are evenly spaced on a logarithmic scale. The lottery set, parameter ranges, and estimation procedure are fixed and identical to the simulations reported in Table~1 of the main text. The noise specification remains
\[
\varepsilon_i\sim N(0,\sigma_i^2), \qquad \sigma_i=s(x_i-z_i).
\]

\begin{figure}[!t]
    \centering
    \includegraphics[width=.78\textwidth]{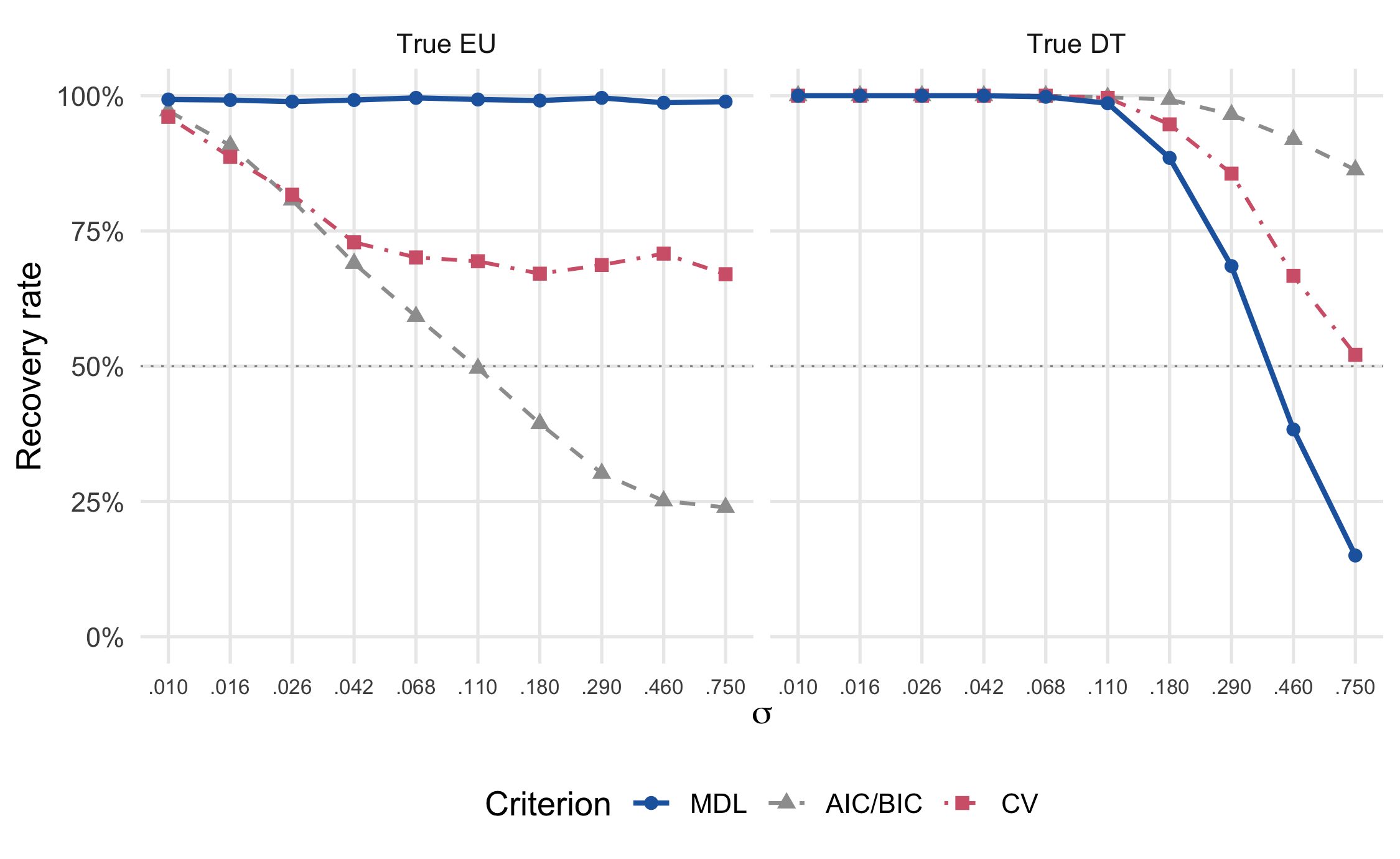}
    \caption{Model recovery across noise levels in the introductory risk simulation. Each point is based on 1{,}000 simulated datasets for the corresponding true model and value of $s$.}
    \label{fig:oa_intro_sigma}
\end{figure}

As expected, all methods recover the true model almost perfectly when noise is very low. As $s$ grows, however, the 2 true models generate different patterns. When EU is true, MDL remains essentially at 100\% recovery throughout the range, while AIC/BIC and CV decline quickly. At $s=0.042$, AIC/BIC recover EU in 69\% of simulations and CV in 73\%; by $s=0.110$, AIC/BIC recovery is 50\%, while CV stabilizes around 70\%. When DT is true, all methods remain near perfect up to moderate noise levels, but MDL eventually declines faster because it penalizes DT for its greater complexity and therefore shifts toward EU as the models become harder to distinguish. Even so, MDL recovery is still 89\% at $s=0.180$.

\FloatBarrier

\subsection{Application 2: risk-preference simulations}
\label{sec:oa_risk_sim_robustness}

Application 2 uses the one-parameter EU and DT restrictions defined in Section~\ref{sec:oa_fixed_sigma}. Each simulated dataset contains 50 lotteries. The balanced design uses payoffs $\{0,.25,.5,.75,1\}$ and probabilities $\{.1,.3,.5,.7,.9\}$. The narrow-payoff design replaces the payoff set with $\{.3,.4,.5,.6,.7\}$, while the narrow-probability design replaces the probability set with $\{.32,.34,.37,.40,.42\}$. For each design and true model, the free parameter is drawn uniformly from $[0.2,0.8]$, and certainty equivalents are generated with $s=0.1$ in the baseline simulations reported in the main text.

The robustness check varies the range of the payoff and probability dimensions separately. In the payoff-range exercise, the probability set is held fixed at $\{.1,.3,.5,.7,.9\}$ and the 5 payoff levels are generated by radially expanding around the midpoint $.5$:
\[
x_j(a)=.5+a(x_j^{\max}-.5), \qquad x^{\max}=\{0,.25,.5,.75,1\}.
\]
Thus $a=1$ gives the balanced payoff set, while $a=.4$ gives the narrow-payoff design used in the paper. In the probability-range exercise, the payoff set is held fixed at $\{0,.25,.5,.75,1\}$ and the probability levels are generated by the analogous construction around the fixed point of the one-parameter Prelec weighting function, $p_0=1/e$:
\[
p_j(a)=p_0+a(p_j^{\max}-p_0), \qquad p^{\max}=\left\{0,\frac{p_0}{2},p_0,\frac{p_0+1}{2},1\right\}.
\]
The value $a=.1$ gives
\[
\{.331,.349,.368,.399,.431\},
\]
which is close to the narrow-probability design used in the paper, and smaller values give even more concentrated probability designs. The robustness check uses the common range scale
\[
a\in\{.025,.040,.063,.100,.158,.251,.400,.631,1\}
\]
for the payoff exercise. The probability exercise uses the same scale except that the final value is set to $.9$ rather than $1$, avoiding the endpoints $p=0$ and $p=1$, where the probability-weighting derivatives entering COMP are not well behaved. At each value of $a$, 1{,}000 datasets are simulated for each true model, both models are re-estimated, and recovery is the fraction of simulated datasets in which the criterion selects the data-generating model. The parameter ranges and estimation procedure are fixed and identical to the simulations reported in the Application 2 simulation table in the main text. The exercise is run at 3 noise levels: $s=.1$, matching the main text, and the additional values $s=.05$ and $s=.2$.

\begin{figure}[!t]
    \centering
    \includegraphics[width=.92\textwidth]{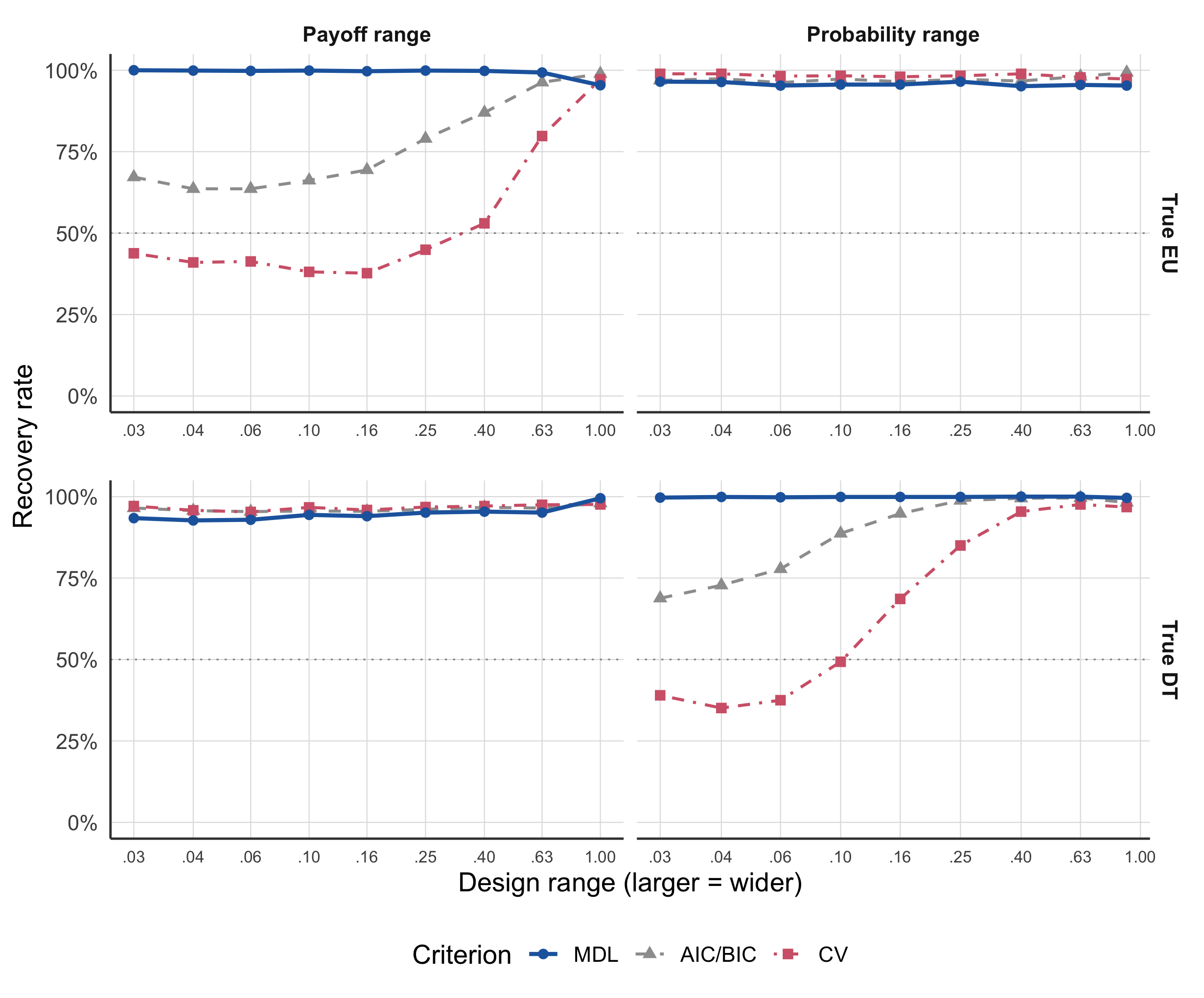}
    \caption{Model recovery across payoff and probability ranges in the risk-preference simulations (Application 2), with $s=0.1$ as in the main text. Each point is based on 1{,}000 simulated datasets for the corresponding true model and design range.}
    \label{fig:oa_application2_design_range}
\end{figure}

\begin{figure}[!t]
    \centering
    \includegraphics[width=.92\textwidth]{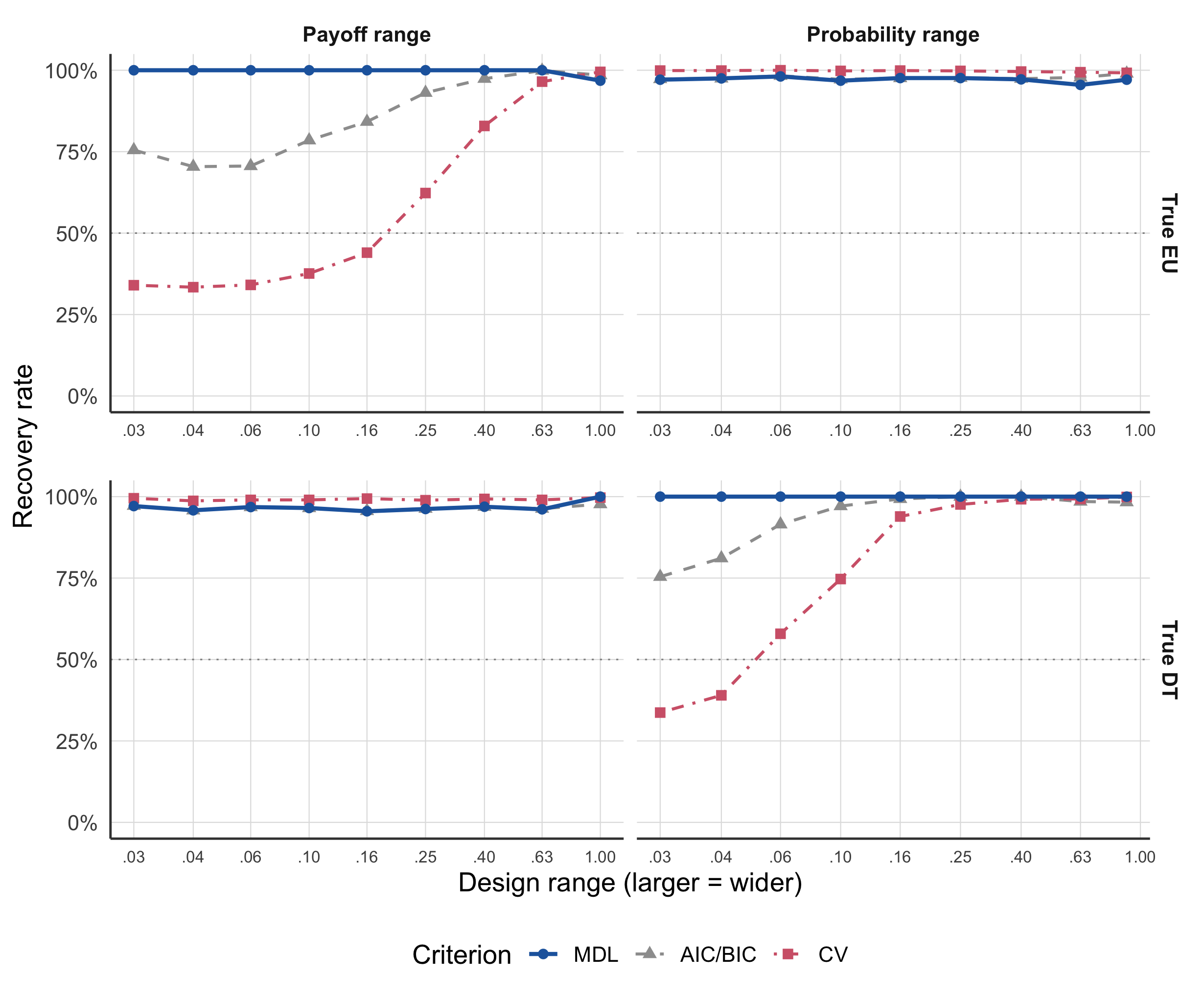}
    \caption{Model recovery across payoff and probability ranges in the Application 2 risk-preference simulations, with lower noise, $s=.05$. Each point is based on 1{,}000 simulated datasets for the corresponding true model and design range.}
    \label{fig:oa_application2_design_range_s050}
\end{figure}

\begin{figure}[!t]
    \centering
    \includegraphics[width=.92\textwidth]{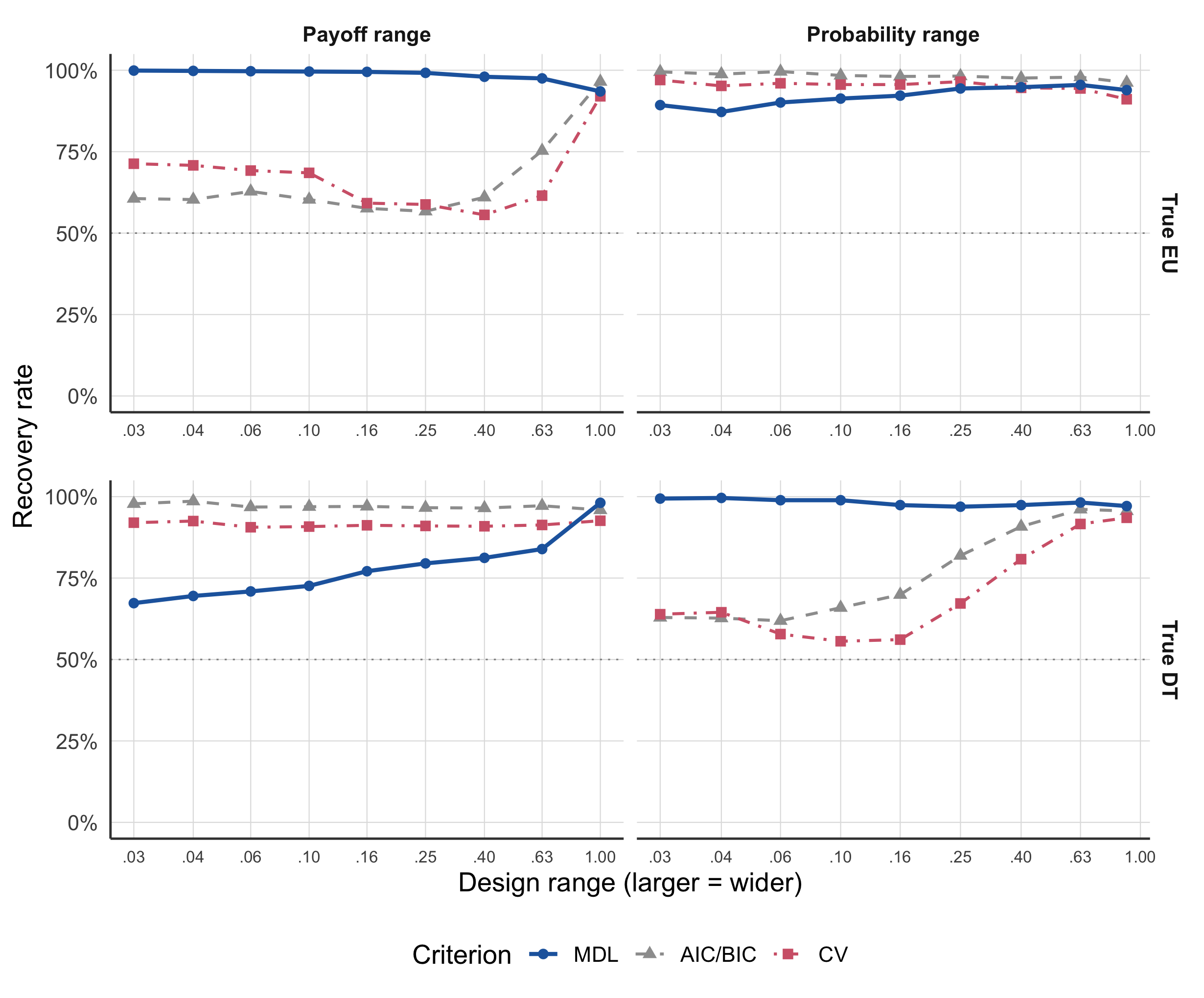}
    \caption{Model recovery across payoff and probability ranges in the risk-preference simulations (Application 2), with higher noise, $s=.2$. Each point is based on 1{,}000 simulated datasets for the corresponding true model and design range.}
    \label{fig:oa_application2_design_range_s200}
\end{figure}

Figures~\ref{fig:oa_application2_design_range}--\ref{fig:oa_application2_design_range_s200} report recovery rates for these design-range simulations. The columns indicate which part of the experimental design is varied: the left column compresses or expands the payoff range while holding probabilities fixed at $\{.1,.3,.5,.7,.9\}$, and the right column compresses or expands the probability range while holding payoffs fixed at $\{0,.25,.5,.75,1\}$. The rows indicate the data-generating model. Thus the upper-left panel displays how often each criterion recovers EU-generated data as the payoff range changes, while the lower-right panel displays how often each criterion recovers DT-generated data as the probability range changes. Figure~\ref{fig:oa_application2_design_range} uses the baseline noise level from the main text, while Figures~\ref{fig:oa_application2_design_range_s050} and~\ref{fig:oa_application2_design_range_s200} repeat the exercise with lower and higher noise.

The Application 2 simulation table in the main text shows that AIC/BIC and CV fail to recover EU reliably when the payoff dimension is compressed, and fail to recover DT reliably when the probability dimension is compressed, while MDL performs well across all designs. The broader design-range analysis shows the same pattern, with the threshold shifting as noise changes. When noise is lower, the same failures appear only at narrower design ranges. When noise is higher, they begin to appear at wider ranges. The main qualification is the lower-left panel of Figure~\ref{fig:oa_application2_design_range_s200}, where MDL also loses accuracy for DT-generated data. This is the conservative MDL error: at high noise and fairly compressed payoff ranges, MDL sometimes selects the simpler EU model over the more complex DT model. Even in this case, MDL has higher recovery rates than the alternatives in the upper-left and lower-right panels of Figure~\ref{fig:oa_application2_design_range_s200}.

\FloatBarrier

\subsection{Application 3: time-preference simulations}
\label{sec:oa_time_sim_robustness}

Application 3 uses the hyperbolic and constant-sensitivity models defined in Section~\ref{sec:oa_fixed_sigma}. Present values are simulated at delays $t\in\{1,\ldots,10\}$. The baseline fixes $\gamma=0.2$ for hyperbolic discounting and $b=1.4$ for constant sensitivity, then chooses $\alpha=0.219$ and $a=0.114$ so that both models imply $\lambda(10)=0.3$. Observations are generated as $y_t=\lambda(t)+\varepsilon_t$, with $\varepsilon_t\sim N(0,0.05^2)$, and model recovery is computed over repeated simulated datasets.

The robustness check varies the noise level and the strength of the discounting pattern. For hyperbolic discounting, $\gamma\in\{.026,.051,.101,.2,.395,.780,1.539,3.039,6\}$; for constant sensitivity, $b\in\{.604,.714,.845,1,1.183,1.4,1.657,1.96,2.319\}$. Both grids are evenly spaced on a logarithmic scale and include the baseline calibrations $\gamma=0.2$ and $b=1.4$; the latter also includes the exponential case $b=1$. At each value, the other parameter is adjusted so that $\lambda(10)=0.3$. The exercise uses $\sigma\in\{.01,.05,.15\}$, with 1{,}000 simulated datasets for each combination of data-generating model and parameter values.

\begin{figure}[!t]
    \centering
    \includegraphics[width=\textwidth]{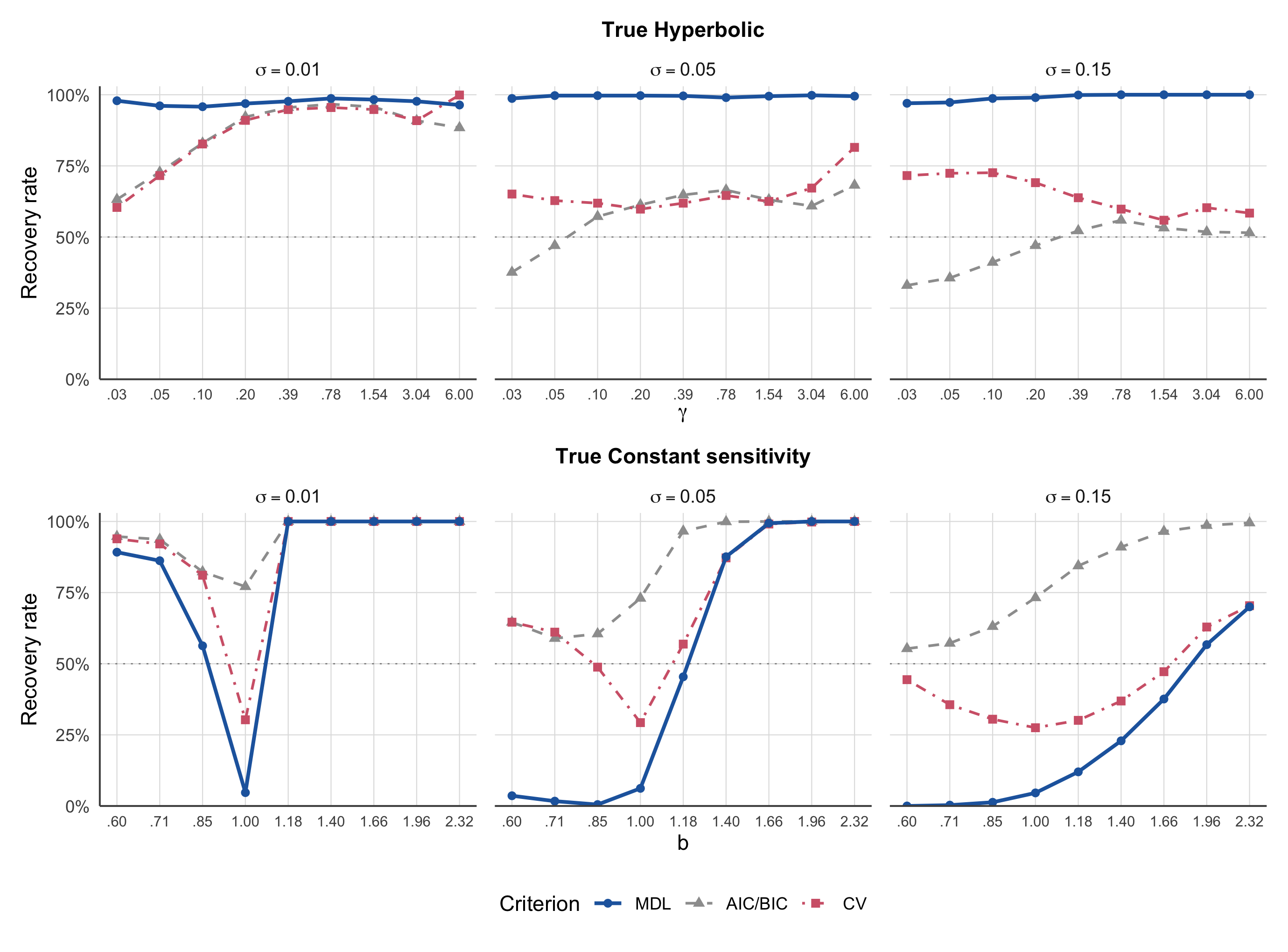}
    \caption{Model recovery across curvature and noise levels in the time-preference simulations (Application 3). Parameter values are evenly spaced on a logarithmic scale. Each point is based on 1{,}000 simulated datasets for the corresponding true model, curvature value, and noise level.}
    \label{fig:oa_time_joint_robustness}
\end{figure}

Figure~\ref{fig:oa_time_joint_robustness} shows that MDL recovers hyperbolic discounting almost perfectly throughout the parameter space, while AIC/BIC and CV become less reliable as noise increases. When constant sensitivity is true, MDL recovers it less often than the other criteria. For $b\leq1$, its discounting patterns can be approximated closely by the less complex hyperbolic model, which MDL therefore tends to select. For $b>1$, recovery increases with $b$ for all criteria, although more gradually for MDL, particularly at higher noise levels.

\FloatBarrier
\clearpage

\section{Selten's Axioms and Compression Gain}
\label{sec:oa_selten_axioms}

This section verifies the claims about the axioms of \citet{Selten1991} made in Section 3.1 of the paper. For a fixed outcome-space size $I\geq2$ and $C=I$, compression gain can be written as a function of Selten's hit rate $r$ and area $a$:
\begin{equation}
g_I(r,a) = r-\frac{\log[1+(I-1)a]}{\log I}. \label{eq:oa_compression_gain_axioms}
\end{equation}
This follows from $I\tilde a=1+(I-1)a$. Define
\[
h_I(a)=\frac{\log[1+(I-1)a]}{\log I},
\]
so that $g_I(r,a)=r-h_I(a)$.

\paragraph{Axioms 1--3 and 5.}
Compression gain is strictly increasing in the hit rate because
\[
\frac{\partial g_I(r,a)}{\partial r}=1>0,
\]
and strictly decreasing in the area because
\[
\frac{\partial g_I(r,a)}{\partial a} =-\frac{I-1}{[1+(I-1)a]\log I}<0.
\]
It therefore satisfies Axioms 1 and 2. Equation~\eqref{eq:oa_compression_gain_axioms} is continuous on $[0,1]^2$, establishing Axiom 3. Finally,
\[
g_I(0,0)=0 \qquad\text{and}\qquad g_I(1,1)=1-\frac{\log I}{\log I}=0,
\]
so compression gain treats the two trivial theories equally, as required by Axiom 5.

\paragraph{Axiom 4.}
Axiom 4 requires the comparison between $(r_1,a_1)$ and $(r_2,a_2)$ to depend only on the differences $r_1-r_2$ and $a_1-a_2$. The function $h_I$ is strictly concave because
\[
h_I''(a) =-\frac{(I-1)^2}{[1+(I-1)a]^2\log I}<0.
\]
It follows that
\[
h_I(1/2)-h_I(1/4)>h_I(3/4)-h_I(1/2).
\]
Choose $q\in(0,1)$ strictly between these two quantities. Consider two comparisons. The first compares $(q,1/2)$ with $(0,1/4)$, while the second compares $(q,3/4)$ with $(0,1/2)$. In both comparisons, the first theory has a hit-rate advantage of $q$ and an area disadvantage of $1/4$. Axiom 4 therefore requires the two comparisons to produce the same ranking. However, in the first comparison,
\[
g_I(q,1/2)-g_I(0,1/4) =q-[h_I(1/2)-h_I(1/4)]<0,
\]
whereas in the second,
\[
g_I(q,3/4)-g_I(0,1/2) =q-[h_I(3/4)-h_I(1/2)]>0.
\]
Thus, despite identical differences in hit rate and area, the theory with the higher hit rate and larger area is ranked lower in the first comparison but higher in the second. Because the ranking depends on the initial area rather than only on these differences, compression gain violates Axiom 4.

\paragraph{Axiom 6.}
For any two hit-rate--area combinations $(r_1,a_1)$ and $(r_2,a_2)$ and any $\alpha\in(0,1)$,
\begin{align*}
    &g_I\bigl(\alpha r_1+(1-\alpha)r_2,
              \alpha a_1+(1-\alpha)a_2\bigr)
      -\alpha g_I(r_1,a_1)-(1-\alpha)g_I(r_2,a_2)\\
    &\qquad=
      \alpha h_I(a_1)+(1-\alpha)h_I(a_2)
      -h_I\bigl(\alpha a_1+(1-\alpha)a_2\bigr).
\end{align*}
By strict concavity of $h_I$, this expression is negative whenever $a_1\neq a_2$, rather than zero as Axiom 6 requires. For example, pooling the two trivial hit-rate--area combinations with equal weights gives
\[
g_I\left(\frac12,\frac12\right) =\frac12-\frac{\log[(I+1)/2]}{\log I} <0 =\frac12g_I(0,0)+\frac12g_I(1,1),
\]
where the strict inequality follows from $(I+1)/2>\sqrt I$ for $I>1$. Thus, compression gain also violates Axiom 6.

\clearpage
\bibliographystyle{apalike}
\bibliography{entropy}